# The climate impact of ICT: A review of estimates, trends and regulations

December 2020.

Charlotte Freitag, and Mike Berners-Lee. Small World Consulting (SWC) Ltd.
charlie, mike @sw-consulting.co.uk

Kelly Widdicks, Bran Knowles, Gordon Blair and Adrian Friday. School of Computing and Communications, Lancaster University.
k.v.widdicks, b.h.knowles1, g.blair, a.friday @lancaster.ac.uk

## Executive Summary

The aim of this report is to examine the available evidence regarding ICT's current and projected climate impacts to stimulate policy developments and governance in the ICT sector.

We examine peer-reviewed studies which estimate ICT's current share of global greenhouse gas (GHG) emissions to be **1.8-2.8% of global GHG emissions**. This corresponds with the baseline estimates for policy making in Europe (i.e. 'more than 2%'). We identify sources of variability in these estimates due to: 1) use of different data sources; 2) variation in the recency of the data used; and 3) varying approaches to boundaries of the analysis, specifically what is considered to be 'in the scope' of ICT's footprint and what elements of its carbon footprint are considered. Our findings indicate that published estimates all **systematically underestimate the carbon footprint of ICT**, possibly by as much as 25%, by failing to account for all of ICT's supply chains and full lifecycle (i.e. emissions scopes 1, 2 *and fully inclusive* 3). Adjusting for truncation of supply chain pathways, we estimate that **ICT's share of emissions could actually be as high as 2.1-3.9%.** We also note the lack of public availability of the data behind many of these estimates.

There are pronounced differences between available projections of ICT's future emissions. **These projections are dependent on underlying assumptions that are sometimes, but not always, made explicit.** Key differences pertain to analysts' answers to the following questions, which have important implications for policy:

*Are energy efficiency improvements in ICT continuing?* While European policy assumes efficiency continues unabated at least through 2050, some experts warn that efficiency improvements in processor technologies are reaching a limit (cf. Moore's Law) and could slow after 2025. If efficiency improvements to date have offset the impact of rising demand, then a limit on further efficiencies threatens soaring future emissions. Conversely, if efficiencies have been driving disproportionately greater demand and hence growth in emissions, then a decline in efficiencies could help to reduce emissions.

*Are energy efficiencies in ICT reducing ICT's carbon footprint?* Throughout at least seven decades of continuous efficiency gains in ICT, the GHG emissions of the sector have risen steadily. This may suggest that efficiencies spur greater demand at a pace that significantly undermines, if not exceeds, the efficiency savings gained.

*Are ICT's emissions likely to stabilise due to saturation?* Some experts argue that energy consumption by ICT will be capped by natural limits to the number of devices people will want to own and the amount of time they have in a day to use them. And yet, technological innovation, not least those resulting from major investment by governments around the world are likely to produce novel applications for ICT that drive additional energy demand. Note that the areas with the heaviest investment are seldom user-mediated (i.e. machine-to-machine communication from machines that run continuously in the background), so a saturation argument would not apply.

*Is data traffic independent of ICT emissions?* While data traffic is not directly proportional to emissions, video streaming provides a clear example of a larger historical pattern that rising data traffic peaks require additional internet infrastructure to meet demand, which then allows further data-intensive services to exist, ultimately leading to higher emissions. Netflix's agreement with the EU to reduce their traffic and ease the load on the network – allowing network provision for homeworkers during the Covid-19 pandemic – further evidences that data traffic is not independent of ICT infrastructure growth and its emissions.

*Is ICT enabling carbon savings in other industries?* The central premise of the new European Green Deal is that ICT saves more emissions through the efficiencies it enables in other sectors than it produces through its own energy consumption; ICT has the huge potential to deliver savings and has shown its value for enabling activities that would otherwise be restricted during the Covid-19 pandemic. However, ICT's net effect on global emissions depends on the extent to which ICT substitutes more traditional, carbon-intensive activities rather than being offered *in addition* to them, thereby increasing our global carbon footprint. Furthermore, ICT-enabled efficiency improvements in other industries might lead to greater demand that more than offsets any efficiency gains and therefore lead to increased emissions in the wider economy.

*Will renewable energy decarbonise ICT?* The ICT sector is leading the way in the shift to more renewable energy consumption which has helped the sector reduce its emissions. This shift is needed along with increases in renewable energy capacity as all sectors will have to replace fossil fuel with energy from renewable sources. Yet, renewable energy itself has a significant carbon footprint, embodied in its infrastructure and supply chains and also for the time being, there are practical constraints to the use of renewables. A limit to absolute energy consumption is most likely needed in addition to higher shares of renewable energy to achieve ambitions for carbon-neutrality of data centres, as per the European Commission's recent commitment for 2030.

Whatever position they take on these crucial questions, analysts agree that **ICT will not reduce its emissions without a major concerted effort involving broad political and industrial action**. Under business as usual, the most optimistic projection sees ICT's emissions staying stable at the current level. Yet, the global economy needs to reduce its emissions by more than 42% by 2030, more than 72% by 2040, and more than 91% by 2050 in order to meet the Paris Agreement goal of staying within 1.5°C, and it remains to be seen whether the ICT sector itself must reduce at a rate on par with the global economy or is justified in reducing at a slower pace. This depends on the balance between the carbon savings that are enabled by ICT and the carbon costs of ICT in enabling those savings. **There are three reasons to believe that ICT's emissions are going to increase barring a targeted intervention**:

- Firstly, we note that **historically, ICT-enabled efficiency improvements have gone hand in hand with increases in energy consumption and GHG emissions both within the ICT sector and in the wider economy.** While it cannot be *proven* that ICT efficiency gains lead to rebounds in emissions that outweigh any savings, there are so many circumstances in which reductions in inputs per unit of output lead to a net increase in inputs that this has to be a significant risk; and one that is often underappreciated. If this dynamic does apply, the only ways to harness emission savings from efficiencies may be through a constraint on consumption (as like Covid-19 has temporarily imposed), or a carbon constraint (such as a carbon tax or a cap on emissions).
- Secondly, **current studies make several important omissions surrounding the growth trends in ICT**. Blockchain is generally excluded from calculations, and Internet of Things (IoT) devices are sometimes partly included but their effect on complementary growth in energy consumption by data centres and networks is not. These trends alongside Artificial Intelligence (AI) do offer opportunities for efficiency gains, but there is *no evidence* to suggest these create GHG savings that outweigh the additional emissions these technologies would cause. Moreover, while associated carbon footprints might be acceptable in scenarios where these technologies are applied toward realising greater carbon reductions in the wider economy, it is certainly not the case that the majority of the innovation in these trending areas is for the purposes of yielding carbon reductions, and so could be purely additional.
- Thirdly, **there is significant investment in developing and increasing uptake of Blockchain, IoT and AI**. All three represent key market opportunities, provide a range of claimed public benefits and are further believed by some to enable up to 15% reductions in global emissions. While significant if achieved, this falls well short of the reductions needed to meet climate change targets. There is a risk that these technologies might also contribute to increases in emissions through stimulating increased carbon-intensive activities such as 'Proof of Work' algorithms and training ever more complex machine learning models.

In light of these considerations, it seems risky, at the very least, to assume that ICT will, by default, assist in the attainment of climate targets.

Recently, some large technology corporations have pledged to voluntarily reduce their carbon footprint and counteract emissions through offsets. Not all pledges are ambitious enough to meet net zero targets by 2050, and so far, there are no mechanisms for enforcing sector-wide compliance.

**Without a global carbon constraint, we contend that a new regulatory framework is required to introduce an ICT-specific constraint on carbon or energy consumption** – ensuring that the full impacts of ICT are considered systematically and drawing consistent boundaries for analysis that are fully inclusive of supply chains. This needs to be backed by the availability of objective and high-quality data reviewed on an annual basis that can be openly and repeatedly inspected. ICT has clearly demonstrated significant potential for year-on-year efficiency improvement and better

exploitation of green energy, but without a cap and inclusion of the full lifecycle of ICT, it looks likely that carbon savings through efficiency improvements will continue to be outpaced by rising demand through rebound effects rather than achieve net zero from ICT efficiency gains. **We contend that this should be the default assumption in the absence of strong evidence of a change in the dynamics of efficiency and growth, capable of causing ICT to buck its historical trend**.

However, we also contend **that if a global carbon constraint were introduced, efficiencies within and enabled by ICT would be even greater enablers of productivity and utility than they are today.** A global carbon constraint is therefore a significant opportunity for the ICT sector.

# Report Contents

# 1. Introduction

The Information and Communication Technology (ICT) sector has seen a massive growth in the last 70 years. With large parts of the economy not yet digitised and emerging economies entering the market, ICT is set to grow even further over the next decade [The Economist 2020]. At the same time, there has been an increasing awareness of the potential environmental effects of ICT, particularly on climate change [e.g. Kaapa 2017, Guardian Environment Network 2017, Belkhir 2018, Tarnoff 2019, BBC iPlayer 2020].

The impact that the ICT sector has on climate change can be expressed as a carbon footprint: that is, an estimate of the amount of greenhouse gases (GHG) released because of a product or activity from all its lifecycle stages. This includes embodied emissions (the GHG emissions released from the extraction of raw materials required, the manufacturing process and transport to the business or user), use phase or operational emissions (from energy use and maintenance) and end-of-life emissions (emissions after disposal).

This report looks at the evidence on the climate change impacts of ICT now and in the future, both through its own GHG emissions and through its effect on the wider society and economy (Section 2). Central to this is the question of efficiency, and whether efficiency gains may reduce emissions over time or if they are more than offset by 'rebound effects'. In this context, we take a broad view of rebound effects to include any increase in emissions due to the introduction of ICT or the efficiencies it enables (see Section 1.1 *Jevons Paradox* for an example of a rebound effect). We also explore important trends in ICT (namely: big data, data science and Artificial Intelligence (AI); the Internet of Things (IoT); and Blockchain) that could provide both opportunities and risks for global emissions (Section 3), as well as relevant government and industry policies (Section 4).

In the study, we have adopted a broad definition of ICT to include all types of data centres, networks and user devices (see Appendix A for Methodology). While there are limitations to our study in terms of the literature review scope and the uncertainties of carbon calculations (see Appendix A.6), we are confident we have captured the main debates regarding this topic, and contribute to those debates through our focus on GHG emissions. We specifically focus on GHG emissions rather than electricity consumption as the former drives climate change and the latter does not capture important factors surrounding ICT's environmental impact (see Appendix A.6).

Given the significance of the topic, there are surprisingly few studies analysing the environmental impact of ICT and they are often characterised by a lack of interrogatability, potential for conflict of interest, a limited scope that leaves out growing ICT trends and an underestimation of ICT's carbon footprint because significant proportions of total emissions are omitted. We have therefore extended our data collection process to include consultations with the lead authors of the main studies included in this review in order to better assess their ICT emission estimations. We endeavoured to understand both the criticisms levelled at each study and how they are countered. We also looked carefully at potential motivations, the quality and transparency of data underpinning the work, and the resources that each study had at its disposal.

Whilst we found broad agreement on the size of ICT's current carbon footprint, there are a range of different views with regards to ICT's future role in climate change – both in terms of ICT's own carbon footprint and its effect on the wider economy's emissions. We discuss the arguments and assumptions underpinning these different views and their policy implications in the following sections.

## 1.1 Jevons Paradox

In 1865, William Stanley Jevons predicted that as the UK's use of coal became more efficient, it would make coal more attractive and thereby would increase demand for coal rather than reduce it [Jevons 1865]. *Jevons Paradox* refers to a situation in which an efficiency improvement leads to an even greater proportionate increase in total demand, with the result that resource requirement goes up rather than down, as is often assumed. There is evidence that Jevons Paradox applies beyond coal [e.g. Alcott 2005, Sorrell 2009, Schaffartzik et al. 2014]. An example is the increased energy efficiency of new forms of lighting (such as electric lighting compared to gas lighting) which allowed lighting to be used more widely – increasing the total energy consumption from lighting. Another demonstration is the fact that electric trains are vastly more efficient than steam trains, let alone horses, yet the carbon footprint of land transport has continuously risen over the time period that these technological advances took place due to expanded use [Berners-Lee and Clark 2013].

While Jevons Paradox is linked with efficiency as the principal driver of rebound effects, the paradox is frequently linked more broadly to a wide range of socio-economic drivers leading to a perverse increase rather than decrease in input demand. Macro-economic models suggest that this 'backfiring' or *rebound effect* leads to savings being cancelled out completely on average and even *adding input* demand relative to previous levels through a variety of mechanisms.[1] At the global level, efficiency improvements in almost every aspect of life have gone hand in hand with rising energy demand and rising emissions.

It is sometimes argued that without the efficiency improvements, demand would have increased even further; this assumes that demand would rise independently of efficiency. It is also argued by some [e.g. GeSI 2015; UK Energy Research Council 2007] that rebound effects are less than 100% of the efficiency savings, but this often results from an incomplete consideration of rebound pathways, especially macro-economic effects. To assess the full impact of rebound effects, all parts of the economy and a longer timescale need to be considered. The only way to feasibly do this is to analyse the combined effect of all global efficiency gains in all sectors and to track this against global energy use. This analysis yields a total energy rebound averaging 102.4% over the past 50 years (i.e. the annual global growth in energy use) [Berners-Lee and Clark 2013]. Despite the increasing utility per unit of energy, the world's energy use is increasing. The same holds true for emissions. Over the last 170 years, $CO_2$ emissions have been rising at 1.8% per year (with only temporary deviations on either side of that trajectory) [Berners-Lee and Clark 2013] alongside the growth of ICT, vast efficiency gains in ICT and other technological advances in other industries.

---

[1] Jevons Paradox and rebound effects are explored in more detail in Berners-Lee and Clark's book *The Burning Question* (2013). Briefly, they argue that when we improve energy efficiency, the available energy becomes more productive and therefore more valuable, leading to increased use. This is because any energy saved bounces back as additional energy elsewhere, either because: 1) efficiency makes the use of the resource cheaper (e.g. lighting, cloud storage of more data than with traditional file storage), 2) the savings are spent on other activities with a carbon footprint, 3) lower resource use leads to lower prices which increases demand for the resource elsewhere, or 4) knock-on effects in other areas of the economy (e.g. when video conferencing enables forming relationships with people on the other side of the world, leading to more air travel to visit them). Resource use can also be displaced into another country (e.g. when burning of fossil fuel domestically is restricted to lower the country's emissions but fossil fuel is continued to be extracted for exports to other countries with fewer environmental concerns).

In terms of the ICT industry, it has been argued that it is through its increasing efficiency that computational power has risen and ICT has been able to become so important in society; the energy consumption of early computers would have been prohibitive for the scale of expansion we have seen over the last decades [Aebischer and Hilty 2015]. An analysis of dematerialisation by Magee and Devezas [2017] found evidence that, in the ICT industry, efficiencies in the material needed for a single product lead to either increasing performance or reduced prices and that this inadvertently leads to increases in demand, resulting in an increase in absolute material consumption. Silicon is one example as it holds a special place in information storage, transmission and computing. Other examples of rebound effects in ICT are provided by Gossart [2015], Galvin [2015] and Walnum and Andrae [2016]. Galvin [2015] estimates that rebound effects in ICT's energy use could range between 115% and 161% based on eight case studies, as efficiency is more than offset by increases in demand.

In addition to efficiencies within the ICT industry, ICT-delivered efficiencies can also have far-reaching effects in other industries – in what we will call *Global Rebounds*. In recent years, ICT has increasingly expanded into other sectors. Common examples include video conferencing technologies or online shopping which could reduce the need to physical travel or reading news on a smartphone. These have the potential to both decrease and increase environmental impact. Where these new technologies evolve to be more energy intensive than their alternatives (e.g. high-quality video streaming), where they are used *in addition* rather than as a substitute (e.g. e-books being used alongside paper books), or where they allow intensified activity or growth in other industries because they are cheaper, more productive or more convenient (e.g. more regular checking of news on a smartphone than with traditional newspapers leading to increased need for news production), the impact of the economy as a whole in terms of energy use, resource use or GHG emissions can increase [Court and Sorrell 2020].

In a systemic review of the direct and economy-wide impact of e-materialisation (such as e-publications, e-games, e-music etc.) on energy consumption, Court and Sorrell [2020] found that studies systematically neglect rebound effects. Most studies assume substitution of old technology with the new digital system where this assumption is not always justified, leading to overestimates of energy savings. Assumptions around the lifetime, the number of users, efficiency of user devices and the replacement of travel lead to a wide range of predictions from 90% decreases to 2000% increases in energy consumption. They conclude that there is no conclusive evidence suggesting significant current or future energy savings from e-materialisation. There is another aspect to efficiency: psychological spillovers through moral licensing where people feel that they have done their part for the environment when increasing efficiency and then go on to have an increased environmental impact elsewhere [Sorrell et al. 2020] - but this is out of scope for this report.

The net effect of ICT depends on the balance of impacts it has both through its own emissions and the effects it has on the wider economy. The economy-wide effects of ICT are difficult to quantify, but in the absence of solid evidence, it would at the very least be risky to assume that the Jevons Paradox and other rebound effects (e.g. time rebounds [Börjesson Rivera et al. 2014]) do not apply to ICT's direct and economy-wide impact.

## 2. Estimating the Carbon Footprint of ICT

In this section, we provide the following:
- a broad overview of the estimates for ICT's carbon footprint before 2015;
- an in-depth analysis of three major peer-reviewed studies of ICT's own greenhouse emissions;
- an overview of the arguments and assumptions underpinning the different estimates for ICT's future carbon emissions and ICT's impact on the wider economy;
- a summary of what this analysis means for global climate targets.

### 2.1 ICT's carbon footprint

**Historically, ICT emissions have grown continuously alongside global emissions.**

Several studies prior to 2015 have estimated the carbon footprint of ICT (summarised in Figure **2.1** see also Table B.**1** in Appendix B.1 for a summary). These show an increase in ICT's carbon footprint over time, even without these studies considering the full life cycle emissions. This growth in ICT's emissions has coincided with consistent growth in our total global carbon footprint [Ritchie and Rose 2019].

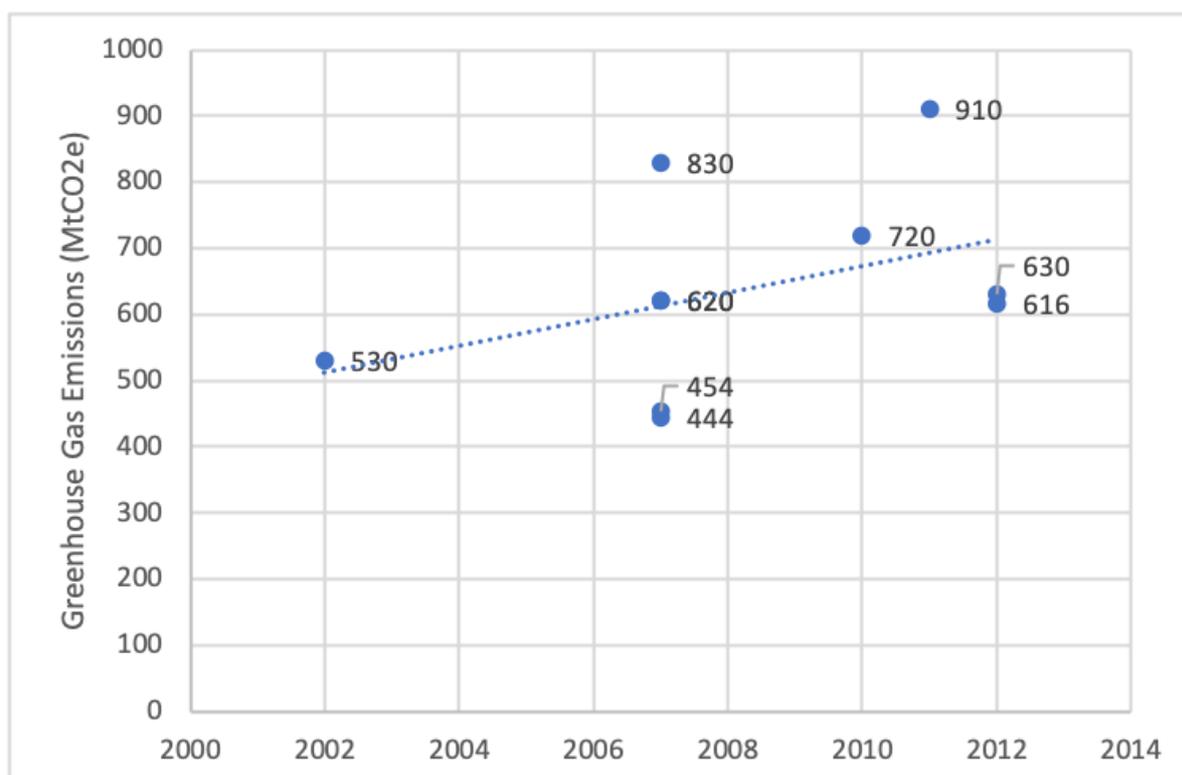

*Figure 2.1 Estimates of ICT's carbon footprint from studies published before 2015. The linear best fit line shows the increase in emissions with time, although the growth is not necessarily linear.*

Scientific debate over ICT's emissions has intensified in the last five years. We therefore focus on research since 2015 – especially studies by three main research groups led by Andrae [Andrae and Edler (A&E) 2015, Andrae 2019a, 2019b, 2019c], Belkhir [Belkhir and Elmeligi (B&E) 2018] and Malmodin [Malmodin and Lundén (M&L) 2018; Malmodin in personal communication].[2] We summarise the arguments here and direct the reader to Appendix B where we analyse these studies in more depth and provide an overview of reports deemed out of scope for this review.

### 2.1.1 ICT's current carbon footprint

**ICT is estimated to form ca. 1.8-2.8% of global GHG emissions in 2020**

Estimates of ICT's emissions in 2020 (see Figure **2.2**) vary between 0.8 and 2.3 $GtCO_2e$. The highest estimates (A&E 'worst case') put ICT's share of global GHG emissions around 6.3%, but Andrae now believes that the A&E 'best case' scenario of around 1.5% is more realistic for 2020 [personal communication]. B&E's estimates are higher at 1.9-2.3%, especially considering they omit TVs in their total estimate. Malmodin's estimates sit in between the others at 1.9% of global emissions. When adjusting for differences in scope (see Table B.**4** in Appendix B.2.2), these studies point towards a footprint of 1.0-1.7 $GtCO_2e$ for ICT, TVs and other consumer electronics in 2020; this is 1.8-2.9% of global GHG emissions. We stress that this estimate carries some uncertainty but gives us a reasonable idea of the impact of ICT. Across studies, roughly 23% of ICT's total footprint is from embodied emissions, yet the share of embodied emissions for user devices specifically is ca. 50%. This is because, unlike networks and data centres, user devices are only used for *parts* of the day and use less electricity, but are exchanged often, especially in the case of smartphones.[3]

---

[2] A&E [2015] estimate ICT's emissions for every year 2010-2030, B&E [2018] for 2007-2040 and M&L [2018] for 2015. Malmodin has also provided additional estimates for 2020 to us in personal communication.

[3] Electricity consumption of user devices and domestic equipment has decreased over the last 15-20 years driven by legislation and public procurement policy such as the EU ERP directive and EnergyStar [Preist, personal communication]. However, efficiency improvements will not be able to reduce embodied emissions drastically. While production processes are becoming more efficient, the manufacturing footprint of smartphones is increasing because of more advanced integrated circuits, displays and cameras [Malmodin, personal communication]. With a large share of their footprint coming from their manufacture, extending smartphones' lifetime is the best way to reduce their footprint. Most studies reviewed here assume an average lifetime of 2 years, partly driven by phone contracts that promise users the newest models [B&E]. There are some signs, though, that this might be increasing slightly. For example, the NPD [2018] reported that in the US, the average use has increased to 32 months in 2017 up from 25 months in 2016. Legislation encouraging repair e.g. the EU Waste Electrical and Electronic Equipment (WEEE) Directive, can help, alongside business models centering around service rather than product provision or selling repairable products to markets in the Global South [Preist, personal communication].

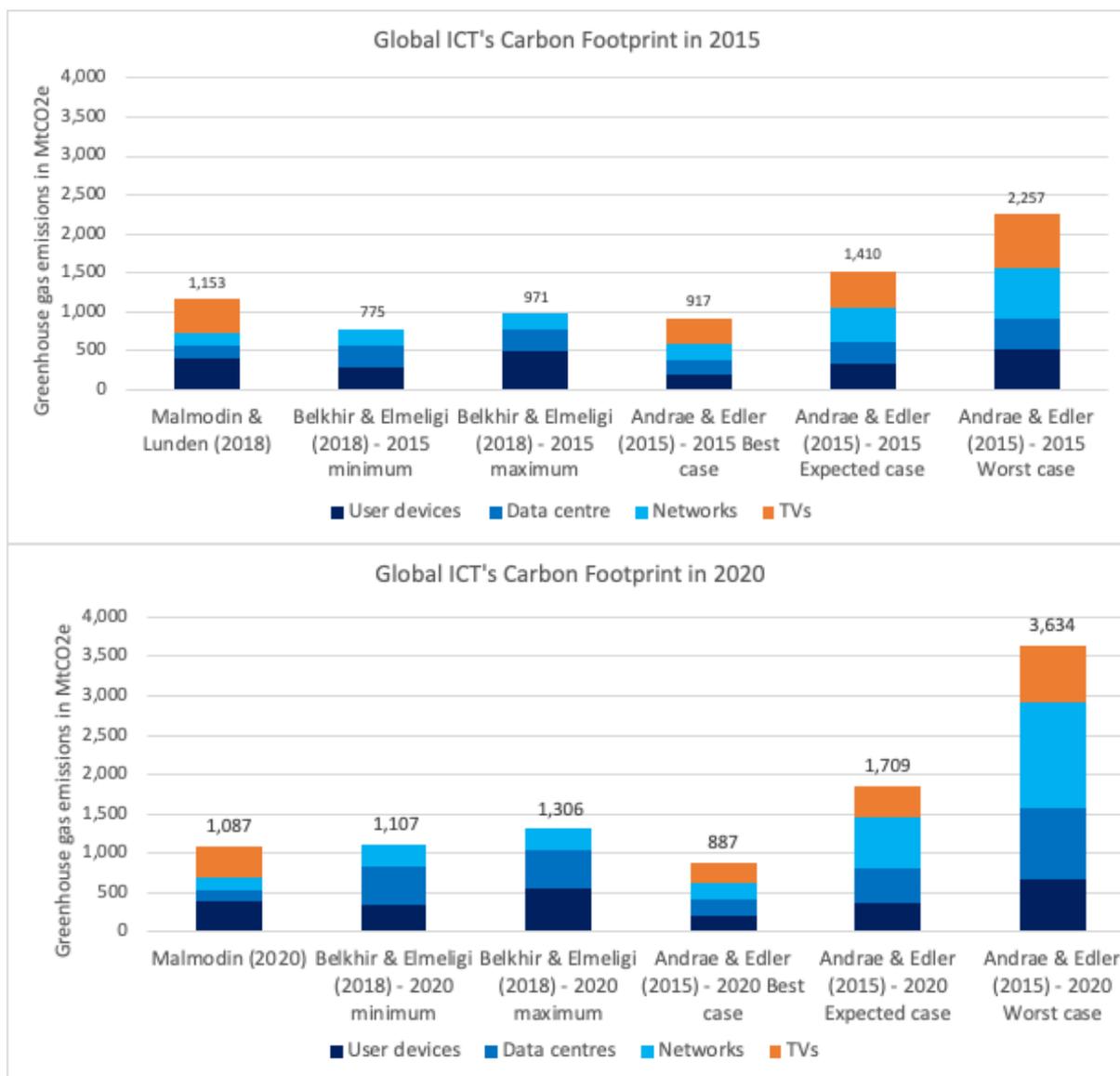

*Figure 2.2 Estimates for global ICT's carbon footprint in a) 2015 and b) 2020. Note that for M&L's estimates, TV includes TV networks and other consumer electronics (**Error! Reference source not found.** in Appendix B.2.1), whereas for A&E's estimates, only TVs themselves and TV peripherals are included. B&E did not include TVs. M&L's original estimates for the ICT and E&M sector includes paper media which we have excluded here. A breakdown of these estimates can be found in Table B.3 in Appendix B.2.2.*

**There are important differences in how analysts arrived at these estimations**
There is a lack of agreement about which technologies ought to be included in calculations of ICT's GHG emissions – particularly TV (see Appendix B.2.1). All studies include data centres, networks and user devices as the three main components of ICT, but there are pronounced differences of opinion regarding the proportional impact of each. A comparison of the different proportions in 2020 estimates (excluding TV) is provided below (Figure **2.3** Proportional breakdown of ICT's carbon footprint, excluding TV. A&E's Best Case is displayed because more recent analysis by the lead author suggest that this scenario is most realistic for 2020. Note that Malmodin's estimate of the share of user devices is highest; this is mostly because Malmodin's network and data centre estimates are lower than those of the other studies.).

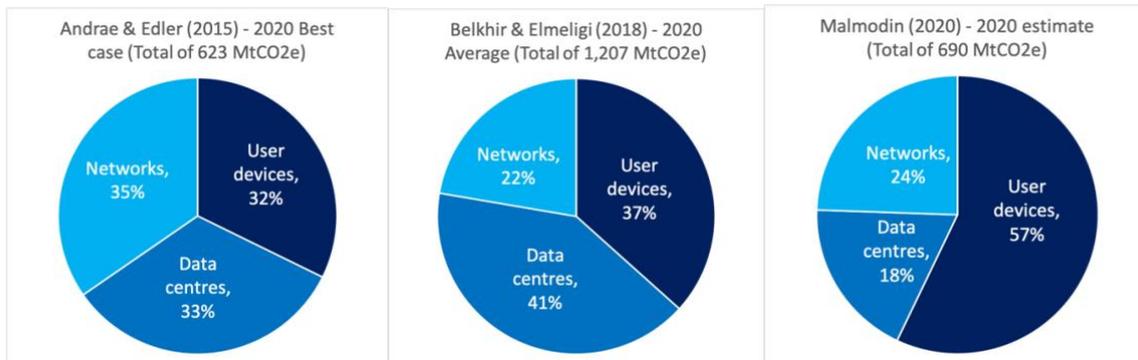

*Figure 2.3 Proportional breakdown of ICT's carbon footprint, excluding TV. A&E's Best Case is displayed because more recent analysis by the lead author suggest that this scenario is most realistic for 2020. Note that Malmodin's estimate of the share of user devices is highest; this is mostly because Malmodin's network and data centre estimates are lower than those of the other studies.*

Regarding data centres, Belkhir himself noted that his projection of 495 MtCO$_2$e for data centres in 2020 is overestimated [personal communication]. Recent evidence by Masanet et al. [2020] of 205 TWh total energy use in 2018 seems to converge with Malmodin's estimate of 127 MtCO$_2$e in 2020. Assuming a global electricity mix at 0.63 kgCO$_2$e/kWh, Masanet et al.'s estimate comes to ca. 129 MtCO$_2$e – higher than A&E's best case estimate of 217 MtCO$_2$e.

**Studies systematically underestimate the carbon footprint of ICT due to truncation error**

Malmodin's studies are the most comprehensive as they include operator activities and overheads (e.g. offices and vehicles used by data centre and network operators), as well as considering the full lifecycle emissions of equipment (i.e. from production, use, to disposal) rather than just production energy (A&E) or only material extraction and manufacturing energy (B&E).

However, A&E, B&E and M&L all follow Life Cycle Analysis (LCA) methodology which is unable to include the infinite number of supply chain pathways of a product, thereby incurring 'truncation error' in their carbon accounting. They also do not consider the full carbon footprint of electricity used to run ICT equipment. This can be rectified by combining LCA with Environmentally Extended Input Output (EEIO) methodologies – taking into ICT's scope 1, 2 *and* 3 emissions – and adjusting the carbon intensity factor of electricity. When these omissions are adjusted for truncation error (see [Appendix F](#)), the carbon footprint for ICT, including TVs and other consumer electronics, rises to 1.2-2.2 GtCO$_2$e (2.1-3.9% of global GHG emissions) in 2020 with ca. 30% coming from embodied emissions and 70% from use phase emissions. We stress once more that these are rough estimates with a significant degree of uncertainty.

### 2.1.2 ICT's future carbon footprint
**There is broad agreement by analysts in the field on certain key assumptions.**
- The world's carbon footprint needs to decrease to avoid climate catastrophe;
- Data traffic is continuing to grow;
- Energy demand by ICT is increasing;
- Demand for data centres and network services will increase;

- The shift to smartphones is decreasing emissions from PCs and TVs;
- Using more renewable energy would reduce ICT emissions;
- ICT could reduce emissions in other sectors but not by default and only under certain conditions (contrasting to GeSI's [2015] SMARTer 2030 claims);
- ICT has the potential to increase its own emissions and facilitate rising emissions in other sectors.

**Opinions are more divided regarding future trends in emissions.**
From 2015 to 2020, B&E's and A&E's estimates of ICT emissions have increased due to an increase in data traffic and the number of user devices (see Figure **2.2**). In contrast, Malmodin's estimates have decreased slightly – mostly for data centres (by 10%), due to an increased adoption of renewable energy, and networks (by 8%), due to decreases in overheads, despite increases in their electricity consumption.

Malmodin [personal communication] argues that: GHG emissions from ICT have stabilised for now; ICT and Entertainment and Media (E&M) sector growth is starting to decouple from GHG emissions; and that ICT could even halve its 2020 emissions by 2030 through renewable energy transformation and collective effort [Malmodin 2019], to 365 MtCO$_2$e in 2030 [Malmodin, 2020]. In contrast, B&E and A&E believe that emissions from ICT will continue to grow (see Figure **2.4**). We explore the differences of opinion between the three in more detail in Appendix B.3.

All analysts think that, theoretically, it would be possible for ICT to decrease its emissions with broad political and industry action – but Malmodin is more optimistic that this will happen than B&E and A&E. A recent Ericsson report [2020] based on M&E claims that ICT's emissions could be reduced by 80% if all its electricity came from renewable sources.

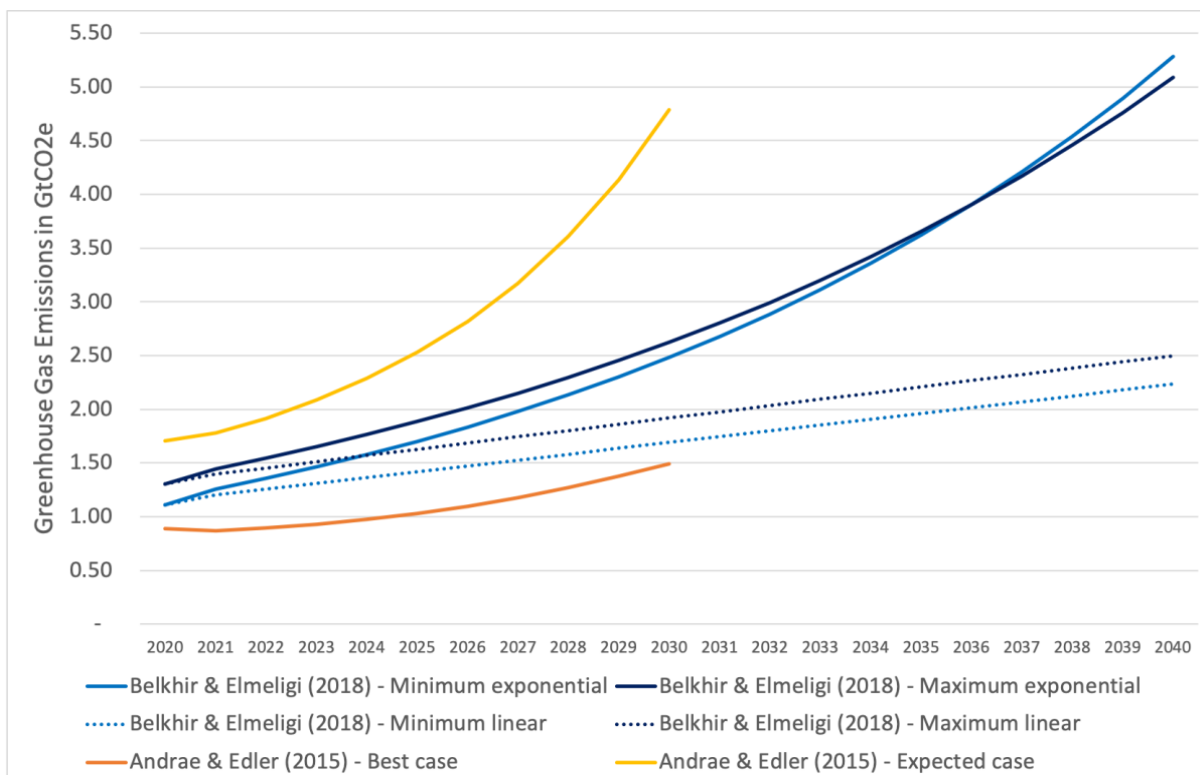

*Figure 2.4 Projections of ICT's GHG emissions from 2020. B&E judge their exponential scenario as most realistic while the linear growth scenario is more conservative and reflects the impact of mitigating actions between now and 2040. M&L [2018] did not make concrete estimates beyond 2020, but Malmodin suggests ICT's carbon footprint in 2020 could halve by 2030 – offering a 2030 estimate of 365 MtCO$_2$e in a recent techUK talk [Malmodin, 2020].*

### Differences in predictions could be due to age of data used.

The data underlying A&E's and B&E's work is somewhat older[4] considering ICT's fast pace of development, meaning their projections are potentially based on historical trends that might no longer apply, such as the assumed exponential growth of energy consumption by data centre and networks. In contrast, M&L might better capture recent changes in emission trends given their estimates are based on data measured directly from industry.[5] M&L also have the most inclusive scope in terms of ICT equipment, lifecycle stages and supply chain emissions considered (see Appendix B.2.1).

However, this access to industry data inevitably comes at the price of a lack of data interrogatability. Part of Malmodin's data was obtained by ICT companies under confidentiality agreements, preventing others from reviewing the original data and the model's assumptions and calculations. There are also potential risks of conflicts of interest as both authors work for network operators.[6] arguably makes M&L's paper open to concerns that claims are less reliable due to selective reporting and assumptions that cannot be properly assessed. We are not suggesting that they cannot be trusted, but the lack of transparency makes independent data and analysis difficult, and transparency is necessary for important policy decisions. As employees of Huawei, A&E also have potential for conflict of interest, but their study is transparent about their data sources, calculations and assumptions. B&E have no obvious conflict of interest and they use only peer-reviewed and publicly available sources.

---

[4] A&E uses some data from 2011 for data centre and networks, while B&E uses data from 2008 for data centres and from 2008-2012 for networks.
[5] M&L's estimates are based on 2015 data; Malmodin's more recent estimates provided in personal communication are based on data from 2018 onwards.
[6] Malmodin works for Ericsson. Lundén works for Talia.

Due to the trade-off between data interrogatability and up-to-date data, it is impossible to judge which study makes the most reliable predictions about ICT's *future* emissions based on methodology alone. It is possible, however, to examine their arguments and the underlying assumptions in order to assess which projection is more likely.

## 2.2 ICT's future carbon footprint: unpacking the studies' assumptions

In the key studies reviewed here, there is disagreement on whether or not:
- energy efficiencies in ICT are continuing;
- energy efficiencies in ICT are reducing ICT's carbon footprint;
- ICT's carbon footprint will stabilise due to saturation in ICT;
- data traffic is independent of ICT emissions;
- ICT will enable emissions savings in other industries;
- renewable energy will decarbonise ICT.

These assumptions have a critical influence on what we can conclude about ICT's role in climate change. We therefore explore the arguments on both sides of the assumptions next to shed some light on the most likely path of ICT's future emissions. In doing so, we draw on several other much-cited sources and our consultation with key experts.

### 2.2.1 Are energy efficiency improvements in ICT continuing?
*Yes, there is scope for energy efficiency improvements in ICT to continue*
There has been a long history of ICT equipment becoming more efficient (and thus cheaper and more productive) with time. Moore's Law allowed the ICT industry to exponentially increase chips' performance, speed and reduce their power consumption. The exponential improvements of processors has kept the exponential growth in demand partly in check in terms of energy consumption.

While M&L acknowledge that Moore's Law has slowed down since 2012, they note that there is usually a time lag before the effects are felt outside of research labs – therefore arguing that efficiencies are continuing for now. Masanet et al. [2020] argue that there is scope for further efficiency improvements in data centres through: improvements in server virtualisation; efficiency gains in servers, storage devices and data centre cooling technology; and the move towards large data centres that are more energy efficient due to efficiencies of scale and the ability to invest in AI to optimise energy use.

For efficiency improvements in user devices, there is evidence of carbon savings from TVs: older, more energy-intensive CRT and plasma TVs have been replaced by more efficient LED TVs; and TV sales have dropped due to users now watching video on laptops and smartphones (B&E, Malmodin). However, smart TVs could change this trend if they become a popular way to access streamed media [Preist, personal communication].

*No, energy efficiency improvements are slowing down and are not fully utilised*
Efficiency improvements might be coming to an end – a view echoed by some of the experts we have consulted (e.g. Peter Garraghan, Belkhir, Andrae). As transistors have shrunk in size and increased in speed, they have begun to heat up; this led to manufacturers putting a speed limit on processing in 2004. The problem now is 'quantum entanglement' where transistor layers become so thin that electrons jump between them, making transistors increasingly unreliable [Waldrop 2016]. Other avenues may exist for improving efficiencies (e.g. decreasing semiconductor use stage power and nanophotonics [Andrae 2020]), but possibly not on the same time scales [Simonite 2016] or with the same size of efficiency gains.

If processor efficiencies are reaching a limit, data centres' power consumption will likely rise as increasing demand will no longer be counterbalanced by increasing efficiency. Despite some remaining scope for further efficiency improvements, Masanet et al. [2020] note that there *are* limits to efficiency improvements and that energy demand will not stabilise by itself – arguing that urgent policy action and investment are needed to limit increases in energy use driven by increasing demand. Furthermore, efficiencies in ICT do not always guarantee replacement of the older, less efficient equipment,[7] and new devices[8] or user habits[9] may conflict with replacement gains.

### 2.2.2 Are energy efficiencies in ICT reducing ICT's carbon footprint?
*Yes, energy efficiencies in ICT can reduce ICT's carbon footprint*
Malmodin argues that so far, efficiency improvements are continuing, and data centre emissions are expected to stay at 1% of global electricity and at the same level of emissions as in 2015 in the next five years. Furthermore, Masanet et al. [2020] reported that data centres' operational energy consumption has increased only marginally from 194 TWh in 2010 to 205 TWh in 2020 despite global data centre compute instances increasing by 550% over the same time period – showing the effectiveness of efficiencies in ICT. Masanet et al. [2020] also note that these efficiency improvements would be able to offset a doubling of data centre demand relative to 2018; beyond that point, energy demand will rise rapidly. This is in line with what Belkhir [personal communication] believes, although he is less optimistic about the remaining scope for efficiency improvements.

*No, energy efficiencies drive growth in ICT's carbon footprint due to rebound effects*
As highlighted above, ICT has seen rapid and continuous efficiency gains. Yet increases in demand for more computations and the number of ICT-enabled devices per person have outpaced these energy efficiency improvements, resulting in ICT's energy consumption, and therefore ICT's carbon footprint, growing year-on-year. This pattern fits with the rebound effect described by Jevons Paradox (Section 1.1) whereby an efficiency improvement leads to an even greater proportionate increase in total demand, meaning total resource requirements rise rather than decrease, as is often assumed. While Jevons Paradox has not been proved to apply within the ICT industry, it is risky to assume it does not apply given historical evidence of ICT emissions consistently rising despite significant efficiencies in the sector (Section 2.1).

---
[7] For example, the development of 5G networks while 2G, 3G and 4G networks still exist.
[8] Some new ICT devices like smart watches and smart speakers are used by people in addition to smartphones and laptops. Court and Sorrell [2020] also highlight the issue of incomplete substitution of e-materialisation trends like e-news or e-books.
[9] For example, multiple user devices in the home have led to a third of UK households watching separate video content simultaneously in the same room once a week [Ofcom 2017] where people may have watched content using the same TV before (see Appendix C).

It would be surprising if rebound effects in ICT – and Jevons Paradox in particular – were to end in the future without a foundational change [Hilty et al. 2011]. There is a theoretical alternative scenario (the reverse of the Jevons Paradox rebound effect) where stalled energy efficiency growth leads to a plateau in ICT emissions due to prohibitive costs as increasing demand cannot be counterbalanced by efficiency improvements any longer. There is little precedent for this in prior work.

### 2.2.3   Are ICT's emissions likely to stabilise due to saturation?
*Yes, the world will become saturated with ICT and this will stabilise ICT's emissions*
The studies reviewed here all agree that the number of smartphones is increasing. According to Cisco [2020], there will be 5.7 billion mobile subscribers by 2023 – 71% of the world population. However, within a few years, every person on earth might have a smartphone and the total number might not further increase [Malmodin in personal communication]. There is some evidence suggesting that the average lifetime of smartphones is increasing too [NPD 2018], which will decrease the yearly embodied carbon associated with people replacing their smartphones. In addition, Malmodin argues that there is a limited time per day that people can be using their phones, theoretically capping energy consumption. The same pattern of saturation could be true for other ICT equipment.

*No, innovation will prohibit saturation in ICT*
In general, ICT companies have a strong incentive to prevent saturation from happening as this would cut their income growth. There is economic pressure for them to create new technologies for individuals and organisations to buy. An example of this is the increase in IoT devices which require little person time and can operate in the background, driving both embodied and use phase emissions from the production of billions of IoT devices, the networks allowing them to communicate and from data centres that analyse the IoT data (see Section 3.2). Other important trends (Section 3) such as AI analytics would also escape this natural saturation. The history of ICT does not provide precedents for a saturation effect; it is therefore unlikely to occur without active intervention. Furthermore, there is still scope for more ICT infrastructure growth beyond smartphones before this innovation cycle even begins, e.g. for data centres in the Global South [Preist, personal communication].

### 2.2.4   Is data traffic independent of ICT emissions?
*Yes, ICT emissions are largely independent of data traffic*
The amount of data traffic on the internet at a given time does not correspond with simultaneous increases in ICT's emissions. Instead, network operators plan capacity for peak data traffic [Sandvine 2014], meaning emissions from ICT are fixed regardless of the amount of data traffic until growth in peak capacity is required. In M&L's view, data traffic is not directly proportional to emissions due to efficiency gains and use of renewable energy in data centres and networks that allow them to process increasingly more data with similar emissions. M&L (reiterated in [Ericsson 2020]) believe the energy consumption of ICT is rather linked to the number of users and time spent using ICT because of the energy consumption of user devices and access equipment like modems and routers, and that data traffic growth is slowing down to a more linear than exponential growth (see Appendix B.2.5 for more details).

*No, data traffic drives ICT growth and the associated emissions*

A&E and B&E both agree that data traffic is a driver in ICT growth and emissions. Growth in the internet's infrastructure capacity allows for new data-intensive services and applications; these offer more affordances to users, driving demand for the services and therefore further infrastructure growth [Preist et al. 2016]. Peak data traffic is one driver for this infrastructure growth due to increased demand for data-intensive services; other influences include ensuring technology is always accessible to all users [Preist, personal communication].

Video streaming is a particularly prominent driver in data traffic (see Appendix C). Netflix have just agreed with EU regulators to reduce their traffic and ease the load on the network, allowing network provision for homeworkers during the Covid-19 pandemic [Sweney 2020]. Belkhir [personal communication] pointed out that this agreement between Netflix and EU regulators makes it difficult to argue that data traffic is independent of ICT infrastructure growth and therefore that data traffic has little effect on emissions.

### 2.2.5 Is ICT enabling carbon savings in other industries?

*Yes, ICT is enabling carbon savings in other industries*

In their report SMARTer 2030, the Global eSustainability Initiative [GeSI 2015], which represents ICT companies, claim that ICT could save 9.1 $GtCO_2e$ in 2020 and 12.08 $GtCO_2e$ in 2030 in other industries such as health, education, buildings, agriculture, transport and manufacturing – mostly due to better efficiency. This would allow a 20% reduction of global $CO_2e$ emissions by 2030, holding emissions at 2015 levels and decoupling economic growth from emissions growth. Relative to their estimate of ICTs own emissions of 1.27 $GtCO_2e$ in 2020 and 1.25 $GtCO_2e$ in 2030, GeSI [2015] argue that ICT is net carbon negative and that governments and businesses should invest *more* into ICT. According to them, already in 2015, ICT saved 1.5 times its own emissions. There is also a strong argument that ICT will accelerate the use of renewable energy in the grid and hence lead to decarbonisation of the energy supply.

*No, ICT drives carbon emissions in other industries due to rebound effects*

The GeSI [2015] report is sponsored by several large ICT companies and there is a lack of transparency in their analysis, raising concerns over possible conflict of interest. So far, there is little evidence that these predictions have come true. History has shown us that growth in the global economy and its carbon footprint has continuously risen, even with ICT creating efficiencies in other industries. It is risky to assume that further ICT-enabled efficiencies will suddenly start to create significant carbon savings in the wider economy without governance and intervention. Rather, it is more likely that ICT enables emission increases in other sectors *because it enables efficiencies*, leading to growth in the very areas into which ICT delivers those efficiency gains – including growth in industries that are already carbon-intensive [Preist, personal communication]. By efficiencies here, it is important to note that we go beyond just energy-specific efficiencies as described by Jevons Paradox; rather, we take into account ICT's emission impacts and rebound effects more widely [cf.

Börjesson Rivera et al. 2014] and refer to *any* potential route for rebound ICT brings to our society.[10]

While GeSI [2015] mention rebound effects, this is only in the appendix and given very limited treatment. Their estimate of an increase of global emissions by 1.37 GtCO$_2$e due to rebound effects is not included in overall calculations for emission savings by ICT and is almost certainly a serious underestimation. This is highlighted by their example of video conferencing. GeSI [2015, p. 69] estimating that *"E-Work technologies like videoconferencing could save around 3 billion liters of fuel."* by cutting workers' commutes. It is difficult to quantify the exact balance of ICT-enabled savings and increased emissions, but one clue is that while video traffic has been expanding rapidly to the extent that it is one of the main contributors of internet traffic [Cisco 2020], emissions from flights were simultaneously increasing (save for pandemics) [Graver et al. 2019]. Therefore, ICT only enables efficiencies in other industries if it completely substitutes more traditional carbon-intensive activities rather than being offered in addition to them.

### 2.2.6 Will renewable energy decarbonise ICT?
*Yes, renewable energy will decarbonise ICT*

Whilst the exact share of renewable energy used for the ICT sector is not known, some ICT operators generate renewable energy on-site and the ICT sector overall is a major purchaser of renewable energy – leading the way for a global shift to this energy source. In a recent Ericsson blogpost building on Malmodin's work, Lövehagen [2020] claims that ICT's carbon footprint could be reduced up to 80% if all electricity came from renewable energy. Renewable energy has a much lower carbon footprint than fossil fuel energy at ca. 0.1 kgCO$_2$e/kWh. Compared to 0.63 kgCO$_2$e/kWh for the global electricity mix,[11] a switch to 100% renewable energy would reduce emissions by ca. 86%.

*No, renewable energy is not a silver bullet*

With unlimited growth in energy demand, even the relatively small carbon footprint from renewable energy compared to fossil fuel would add up significantly. Additionally, there might be limits to the amount of renewable energy that can be generated with present technology, such as the availability of silver which is used in photovoltaic panels.[12] While investments into renewable energy currently have the effect to reduce the price of renewable energy for other sectors, as soon as there are limits to the amount of renewable energy that can be generated, any additional energy used by ICT will take energy away from other purposes.

There are also practical constraints on the extent that renewable energy can be used to power ICT equipment. Even data centres that are powered by 100% renewable energy usually have fossil fuel-powered backups for unexpected demand increases. Powering networks with renewable energy is a lot harder due to their decentralised nature [Belkhir and Elmeligi 2018] and powering user devices depends largely on the greening of national grids – a trend that is ongoing in the UK but still far from complete. Thus, while a shift to more renewable energy is crucial, it does not provide an unlimited supply of energy for ICT to expand into without consequences.

---

[10] For example, consider how ICT has made it far easier to book flights online, contributing to the growth of the aviation industry.

[11] Both figures are based on SWC's EEIO model which draws on official data from the UK government's Department for Business, Energy and Industrial Strategy.

[12] An average solar panel requires ca. 20g of silver [Apergis and Apergis 2019]. There are currently 2.6bn solar panels in the world generating a total of 865 TWh [IEA 2019a]. From 2019 to 2020, 135 TWh of solar energy was added. The manufacture of these requires 52,000 tons of silver. Worldwide, 27,540 tons of silver are being mined in 2020, and the amount increases by ca. 2% every year [IEA 2019a]. On this trajectory, solar panels would use 100% of global silver supplies in 2031 leaving none for electric car batteries and other uses.

### 2.2.7 Six common narratives for ICT's role in climate change.

The assumptions from the studies and unpacked in this section can be summarised into six narratives of ICT's future role in climate change: four around future trends in efficiency and demand and their effect on ICT's own emissions, and two on ICT's effect on emissions in the wider economy; see Appendix D for a description of these narratives.

A) **Assumptions about demand for ICT**

|  | increases less than or in line with efficiency | increases more than efficiency |
|---|---|---|
| **Assumptions about efficiency — continues** | 'Efficiency saves ICT'<br>Emissions decline or stabilise<br>*e.g. Malmodin, Masanet* | 'Rebounds in ICT'<br>Emissions increase |
| **Assumptions about efficiency — stops** | 'Rebounds stalled'<br>Emissions stabilise | 'Growth without efficiency'<br>Emissions increase rapidly<br>*e.g. Andrae, Belkhir* |

B) **ICT's effect on emissions in the wider economy**

| Enablement | Global Rebounds |
|---|---|
| ICT allows for efficiency improvements in other sectors and thereby enables emissions savings that are bigger than its own emissions and bigger than any rebound effects.<br><br>ICT's net effect is to reduce emissions in the world.<br><br>*e.g. GeSI* | The efficiency improvements enabled by ICT in other sectors lead to system growth. Rebound effects are larger than the efficiency gains.<br><br>ICT's net effect is to increase emissions in the world. |

*Figure 2.5 Narratives of ICT's role in climate change and the critical assumptions underlying these. A) ICT's carbon footprint. B) ICT's effects on emissions in the wider economy. The proponents of each narrative are in italics. Efficiency is here defined as GHG emissions per equivalent ICT use. This includes Moore's Law but also higher renewable energy use, energy efficiency of the infrastructure etc.*

## 2.3 Summary of ICT's carbon footprint

**To meet climate change targets, the ICT sector needs to drastically decrease its own emissions and deliver vast savings in other sectors.**

Despite some variability in estimates, research studies reviewed here agree that ICT is responsible for several percent of global GHG emissions and that its footprint has grown until recently. The world needs to reduce its GHG emissions in order to stay within 1.5°C warming [IPCC 2018]. If the ICT sector should decrease its emissions in line with other parts of the economy, it would have to: reduce its $CO_2$ emissions by 42% by 2030, 72% by 2040 and 91% by 2050 (see Figure **2.6**)[13] and net zero by 2050 [ITU 2020]; or deliver equivalent savings in other sectors *in addition* to the savings these sectors will have to deliver themselves to meet these targets, making sure that rebound effects do not offset these savings.

**Under business as usual, increases in emissions are likely. Major concerted effort would be needed to reduce emissions.**

All the analysts we spoke to agree that in order to decrease ICT's emissions - even assuming emissions have stabilised – a strong and unified effort would be needed (Section 4). Without this effort, even if ICT's emissions were to stay stable at the 2020 level over the next decades, the relative share of ICT's emissions in global emissions would increase to more than a third as other sectors reduce their emissions in line with 1.5°C warming (see **Error! Reference source not found.**).

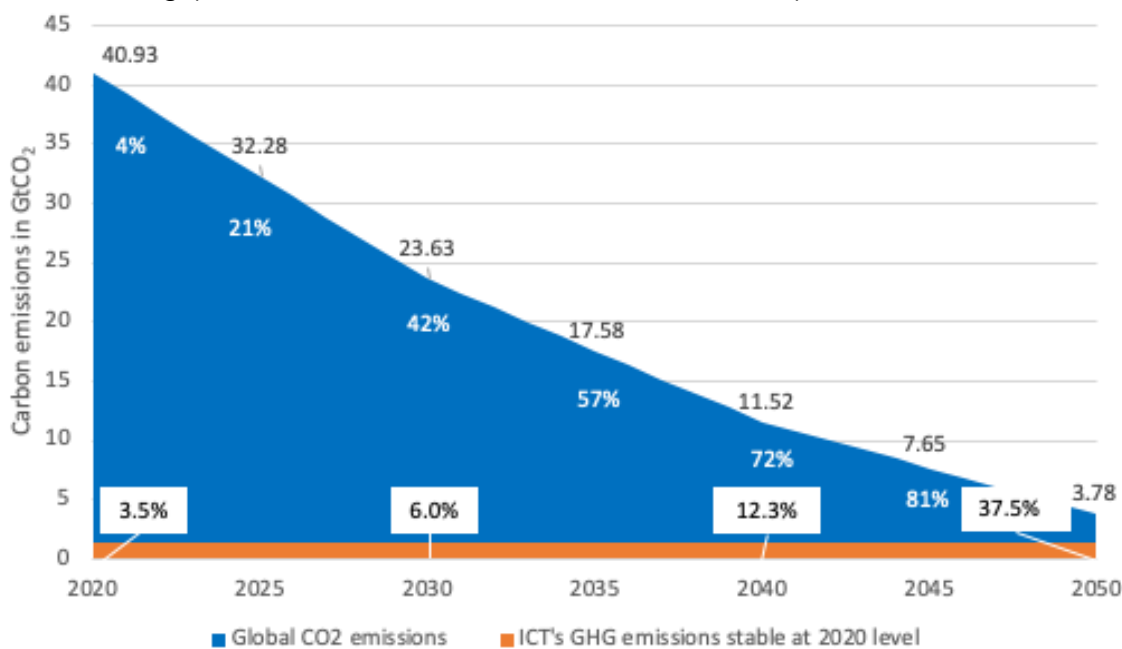

*Figure 2.6* ICT emissions, assuming the 2020 level (adjusted for truncation error) remains stable until 2050, and global $CO_2$ emissions reduced in line with 1.5°C under scenario SSP2-19. White numbers indicate global $CO_2$ cuts needed relative to 2010 and labels at the bottom indicate ICT's share of global $CO_2$ emissions in percent. We

---

[13] Global $CO_2$ emission cuts to 2050 needed to stay within 1.5°C warming by 2100 are based on modelling by Baskerville-Muscutt [2019] based on the Shared Socio-Economic Pathway 2 as outlined by the International Institute of Applied Systems Analysis, available at: https://tntcat.iiasa.ac.at/SspDb/dsd?Action=htmlpage&page=40. This is the 'middle of the road' or average scenario for the trajectory the world will follow. Cuts are relative to global $CO_2$ in 2010. Note that this is CO2 only, assuming ICT emissions are mostly CO2 as a large part if electricity and there are no agricultural components. The comparison to $CO_2$ emissions was chosen because reliable budgets do not exist for GHG emissions at this point.

*assume most of ICT's emissions are from CO₂ because a large proportion of its footprint is from electricity consumption and there are no agricultural components. The comparison to CO₂ emissions was chosen because reliable budgets do not exist for GHG emissions at this point.*

**There are three reasons to believe that ICT's emissions are higher than estimated and that they are going to *increase*.**

*Reason 1: Rebound effects have occurred since the beginning of ICT, and they will likely continue without intervention.*

Even if efficiency improvements are continuing (see Section 2.2.1), this will not completely counterbalance growth in demand for ICT; in fact, efficiencies might spur further growth in emissions by allowing the ICT sector to grow further due to rebound effects (see Section 2.2.2). We believe that a natural peak in ICT emissions due to saturation of demand is unlikely (see Section 2.2.3). To the extent that ICT enables efficiencies in other sectors, there is the risk that rebound effects more than offset any savings following *Global Rebounds* (see Section 2.2.5). Renewable energy will help decarbonise ICT but is not a silver bullet (see Section 2.2.6).

*Reason 2: Current studies of ICT's carbon footprint make several important omissions surrounding the growth trends in ICT.*

The studies reviewed here make several important omissions of growing areas of ICT, such as Blockchain; and IoT is only partially considered for future trends. This leads to an incomplete picture. Some analysts argue that Blockchain is not part of ICT because it requires *specific* hardware, not regular servers. However, we believe that it should be in scope of ICT as it is an ICT-facilitated algorithm (see Section 3.3); having specific hardware for Blockchain is similar to how graphics-specific services (e.g. online games) require Graphics Processing Units [Preist, personal communication]. M&L include some IoT (see Table B.*2* in Appendix B.2.1) and concluded that the impact of IoT is small. However, this is a small share of all IoT and they only accounted for the connected devices themselves, not the energy consumption IoTs create in data centres and networks (based on the assumption that data traffic and energy are not closely related, see Section 2.2.4). Such trends, as well as AI, could help reduce global carbon emissions, but they will also add to ICT's carbon footprint; we discuss this trade-off for prominent ICT trends in the next section (see Section 3).

*Reason 3: There is significant investment in developing and increasing uptake of Blockchain, IoT and AI.*

Despite questionable evidence that ICT growth trends will save more carbon emissions than it will introduce (see Section 3), Blockchain, IoT and AI are seeing increased investment and uptake. As we explore in Section 4, the European Commission discuss these trends as a way to spur economic growth and yield emission reductions; yet, they expect ICT will only enable 15% reductions which is insufficient for meeting climate change targets (see Section 4.1). Some large technology corporations are setting their own carbon pledges which might help reduce the emissions from ICT's growth trends, however these pledges are often not ambitious enough to meet net zero emissions by 2050 (see Section 4.2). Until ICT corporations become net zero, any investment in the ICT industry will be associated with an increase in emissions.

**With a global carbon constraint, ICT will be a vital sector to ensure transition to a net zero world.**

If a global carbon constraint was introduced, we could be certain that rebound effects would not occur, meaning that productivity improvements through ICT-enabled efficiencies both within the ICT sector and the wider economy would be realised without a carbon cost. Under these conditions, ICT would be a key means by which productivity is maintained or increased despite the carbon constraint, and therefore ICT's role in enabling the whole economy can be expected to be even greater than it is today. Given these reasons, under a carbon constraint, ICT's share of global emissions could justifiably be allowed to rise.

# 3. ICT Trends: Opportunities and Threats

Looking towards the future, three known areas of technological innovation may have potentially profound implications (both opportunities and threats) for the carbon footprint of the ICT sector:
- Big Data, Data Science and AI;
- The Internet of Things;
- Blockchain and Cryptocurrencies.

## 3.1 Big Data, Data Science and AI

Big data is one of the most significant trends in technology, made possible by the data store (and computational) capacities of cloud computing. There has been enormous interest in making sense of this large and complex data through data science and AI.

### 3.1.1 Opportunities

**Big data, data science and AI could contribute to a lower carbon 'smart' future.**
There are real opportunities around the use of big data/data science/AI alongside IoT in what has often been referred to as the smart future – encompassing smart grids, cities, logistics, agriculture, homes etc. [e.g. Al-Ali et al. 2017, Minoli et al. 2017, Reka and Dragicevic 2018, Saleem et al. 2019]. Big data/data science/AI could particularly contribute to a lower carbon future, e.g. by finding optimal routes through cities and reducing traffic congestion, or by optimising energy use for heating and lighting in buildings. As these areas rely on IoT, we defer discussion on these opportunities until Section 3.2.

**There is a willingness across industry and academia to apply such technologies for the benefits of society.**
There is a significant move towards data science and/or AI for social good, including applications in health [Raj et al. 2015] and the environment, although this work is in its infancy and generally not migrated to everyday practice. The role of big data in supporting green applications has been discussed, specifically in the areas of energy efficiency, sustainability and the environment [Wu et al. 2016]; and the field of computational sustainability is emerging, using technologies such as AI in support of the United Nations (UN) sustainable development goals [Gomes et al. 2019]. There is also an emerging research community looking at the role of such technologies in supporting environmental sciences as they seek a deeper understanding of our changing natural environment.[14]

### 3.1.2 Threats

**The world's data is doubling every two years.**
Data has been described as 'the new oil' [James 2019] given its commercial impact – yet as data storage and data centres grow to meet demand for big data, this description could have a double meaning due to the environmental impact of data. The storage and analysis of data can help solve complex world problems, but there are concerns over the resources required to facilitate data science and AI technologies, especially the carbon footprint of the underlying data centres (see

---

[14] See for example research in Toronto, Exeter and the Centre of Excellence in Environmental Data Science, a joint initiative between Lancaster University and UK CEH programme called 'data science for social good'.

Section 2). By this year (2020), it is expected that the total size of the world's digital data will be around 44 trillion gigabytes of data [IDC 2014]. AI and data science are therefore an important trend that drives growth in data storage and processing[15] and in data centres, which some experts argue leads to an increase in ICT's carbon footprint (Section 2.2.4).

**Emissions associated with processing this data are increasing due to growing computational complexity.**
Data science and AI offer additional threats *over and above* the potential growth of data centre emissions. The dominant of the two is AI with its potentially computationally complex underlying algorithms when operating on big data, especially around machine learning and deep learning. Researchers have estimated that 284,019 kg of $CO_2e$ are emitted from training just *one* machine learning algorithm for Natural Language Processing, an impact which is five times the lifetime emissions of a car [Strubell et al. 2019]. Whilst this figure has been criticised as an extreme example of the footprint of model training,[16] the carbon footprint of model training is still recognised as a potential issue in the future given the trends in computation growth for AI [Biewald 2019]: AI training computations have in fact increased by 300,000x between 2012-2018 (an exponential increase doubling every 3.4 months) [Amodei and Hernandez 2019].

### 3.1.3 Threat Mitigation
**Sustainability needs more consideration in ethical guidelines of AI.**
Due to this growth of computation, Schwartz et al. [2019] argue the need for 'Green AI' that focuses on increasing efficiencies of AI computations rather than the current focus on what they describe as 'Red AI', i.e. accurate AI computations without consideration of resource costs or efficiencies. Sustainability is currently one of the least represented issues associated with ethics guidelines in AI [Jobin et al. 2019], though a framework and 'leaderboard' to track the energy consumption and carbon emissions of machine learning research have recently been offered in the hope that this will encourage energy efficiency to be considered in such algorithms [Henderson et al. 2020]. Energy efficiencies are required in this field of ICT and opportunities to introduce efficiencies may exist, such as addressing the processing requirements of AI algorithms by using idle PCs as a distributed supercomputer [Folding@Home 2020]. However, we reiterate the concerns of an efficiencies-focused endeavour without a carbon or consumption constraint due to the potential of rebound effects (see Section 2.2.2).

## 3.2 The Internet of Things
The Internet of Things (IoT), where 'Things' represent everyday internet-connected objects from wearable technologies through to appliances, cars and other transport vehicles. This has led to a substantial and ongoing growth of the internet as documented below.

---

[15] Data processing will be the larger contributor to ICT's energy use, as simply storing data is environmentally cheap in comparison [Preist, personal communication].

[16] A more typical case of model training may only produce around 4.5 kg of $CO_2$ [Biewald 2019].

### 3.2.1 Opportunities

**IoT technologies can enable efficiencies within and outside of the ICT sector.**

IoT applications are often viewed as 'smart technology', especially when combined with data science/AI in ways that optimise energy usage. Smart cities aim to provide better public services and resource use at a lower environmental cost [Mohanty et al. 2016], e.g. location-based services from smart city IoT sensing and data analysis can reduce transportation pollution through more efficient driving routes [Bibri 2018]. Govindan et al. [2018] also investigate how such developments can support smarter logistics including reducing energy requirements. As mentioned in Section 2.2.5, ICT has the potential to decarbonise energy supply and a combination of IoT with grid technology has real potential to to support the management of the resultant Smart Grid, e.g. by dealing with intermittency of renewable supply [Collier 2015]. IoT deployments have been tested in schools with the aim of raising awareness of energy consumption and promoting sustainable behaviours based on IoT sensing data [Mylonas et al. 2018], and IoT has also been harnessed to enable energy efficiencies within ICT, e.g. by using IoT sensing data to reduce the required air conditioning for data centres [Liu et al. 2016]. These few examples highlight the breadth of IoT opportunities to reduce GHG emissions, as long as the IoT applications substitute more traditional carbon-intensive activities rather than act alongside them.

### 3.2.2 Threats

**IoT enablement comes at a cost of rapidly rising numbers of devices, device traffic, and associated emissions.**

Despite these opportunities, the sheer number of IoT devices and the associated data traffic is growing significantly. Innovation in IoT is expected to create a fivefold increase from 15.41 billion internet-connected devices in 2015 to 75.44 billion in 2025 [Statista Research Department 2016]. Cisco estimate Machine-to-Machine (M2M) connections will grow from 6.1 billion in 2018 to 14.7 billion by 2023 (a Compound Annual Growth Rate (CAGR) of 19%), representing 1.8 M2M connections per member of the global population in 2023 [Cisco 2020]. The majority of these connections is expected to be formed by IoT in the home for automation, security and surveillance (48% of connections by 2023), yet connected cars (30% CAGR between 2018-2023) and cities (26% CAGR) are the fastest-growing sectors of IoT application [Cisco 2020].

**IoT's carbon footprint is under-explored, but will have significant implications for embodied emissions.**

Whilst the footprint of IoT is uncertain and often unexplored in studies of ICT carbon emissions (Section 2.3), it has been estimated that the energy footprint of IoT semiconductor manufacturing alone might be 556 TWh in 2016 and increase 18-fold to 722 TWh in 2025 [Das 2019].[17] Assuming a global electricity mix of 0.63 MtCO$_2$e/TWh, this would be a total of 424 MtCO$_2$e in 2016 and 6,125 MtCO$_2$e in 2025 for the manufacture and use of the semiconductors; this is without emissions from the entire IoT device, associated sensors and the emissions in data centres and networks that IoT communicate with. It is also worth noting that the introduction of IoT could lead to an initial rise in obsolescence for other non-ICT products, as society makes the transition to an IoT-focused life (e.g. replacing a working kettle with an internet-connected kettle).

---

[17] This does not include other aspects of embodied carbon in IoT, such as material extraction and transport, or sources of GHG emissions other than electricity. This also does not consider energy use of running systems, although Das [2019] estimates that this would be a lot smaller than the embodied carbon in manufacturing, at perhaps 118 TWh in 2016 and decreasing to only 1 TWh in 2025 as we see more energy efficient technologies. This study has been questioned as vastly overestimated, however, by Malmodin [personal communication].

### 3.2.3 Threat Mitigation

**Lower energy IoT systems are a way forward, but may lead to energy-intensification and fuel greater emissions overall.**

Researchers are already looking to create lower energy IoT systems, considering both devices [Kaur and Sood 2015] and communication technologies. One focus is on Low Power Wide Area Networks (LPWANs) [Raza et al. 2017] to reduce the energy requirements of M2M communication, but at a trade-off of lower bandwidth. There is an associated field of study referred to as 'Green IoT' [e.g. see Shaikh et al. 2017, Arshad et al. 2017, Alsamhi et al. 2019, Solanki and Nayyar 2019], which focuses on ensuring that IoT's own efficiencies and environmental costs are considered as we move towards a smarter society and environment. Yet we should be careful of IoT applications which could lead to rebound effects. For example, smart home technologies have the potential to reduce energy consumption (e.g. through remote-controlled heating or lighting), but could perhaps lead to *"energy-intensification"* once adopted through offering new services (e.g. pre-heating homes, continuously running security systems) or intensifying current services (e.g. internet connectivity, audio/visual entertainment) [Wilson et al. 2017] – the latter adding to ICT's carbon footprint through additional user devices and data traffic.

## 3.3 Blockchain and Cryptocurrencies

Blockchain is an example of a decentralised algorithm designed to avoid a centralised authority or central point of failure. Blockchain allows for potentially important new uses, e.g. for decentralised financial systems. Cryptocurrencies are the most popular application for Blockchain, with Bitcoin being the biggest cryptocurrency available today.

### 3.3.1 Opportunities

**Blockchain could offer some opportunities for reducing carbon, but there are no emissions-reducing applications of these technologies yet.**

A decentralised electronic currency could offer a real disruption in the management of market transactions and in the possibility of handling decentralised energy exchanges [Enerdata 2018], although there are no real examples of demonstrable emissions savings yet. Kouhizadeh and Sarkis [2018] discuss the potential of Blockchain technologies to enhance sustainability in the supply chain, for example by supporting transparency in the early stages of supply chain management (e.g. vendor selection and evaluation); this work, however, is speculative at this stage, leading to researchers offering directions to further explore adoption of Blockchain in this domain [Saberi et al. 2019].

### 3.3.2 Threats
**The energy consumed by single cryptocurrency is equivalent to that of entire nations.**

Blockchain is underwritten by energy: the algorithm, if based on 'Proof of Work', creates high levels of replication and redundant computation [Mora et al. 2018].[18] Energy consumption can also increase through escalation of the 'mining arms race' due to improving risk sharing for Proof of Work Blockchains [Cong et al. 2019]. Focusing specifically on cryptocurrencies, one study indicates that Bitcoin's annual electricity requirements of 68.7 TWh in 2020 are equivalent to powering 7 million US households [Digiconomist 2019], associated with a footprint of 44 $MtCO_2$.[19] Due to the inefficiency of transactions, a single transaction could be ca. 750 kWh, enough to power 23 households for one day [Digiconomist 2019], or 473 $kgCO_2e$.[19] Bitcoin currently has a market dominance of 64% of all cryptocurrencies [CoinMarketCap 2020]. Under the assumption that other cryptocurrencies have the same carbon intensity as Bitcoin, the carbon footprint of all cryptocurrencies would be ca. 69 $MtCO_2e$, 0.1% of global emissions. Another study estimated the Bitcoin network consumption at 2.55 gigawatts (GW) of electricity in 2018 (a value which is nearly as much as Ireland at 3.1 GW), but that this could rise to 7.67 GW in the future (making it comparable to Austria at 8.2 GW) [de Vries 2018]. Other researchers argue an annual electricity consumption of 48.2 TWh and annual carbon emissions ranging from 23.6-28.8 $MtCO_2$ for Bitcoin in 2018 [Stoll et al. 2019]. Stoll et al. also estimated that other cryptocurrencies would add another 70 TWh in 2018, bringing the total carbon footprint to ca. 73 $MtCO_2e$ in 2018.

### 3.3.3 Threat Mitigation
**Fiscal policy intervention may be needed to mitigate energy consumption of decentralised algorithms.**

Alternatives to Proof of Work exist that reduce the number of resources required for Blockchain, e.g. Proof of Stake reduces computation and Byzantine protocols remove consensus mining [Monrat et al. 2019, Saleh 2020]. Carbon offset mechanisms for Blockchain also exist, such as SolarCoin whereby solar energy producers are rewarded with a free SolarCoin for each MWh of solar-based electricity they produce [Howson 2019]. Renewable energy can also be used to power these technologies and it is argued to form 73% of Bitcoin's mining [Bendiksen and Gibbons 2019].[20] However, de Vries [2019] does not think Bitcoin can be sustainable due to: 1) the seasonality of hydropower in Sichuan, China (a region which supposedly supports nearly half of global mining capacity [Bendiksen et al. 2018]) meaning energy is required from alternative sources such as coal; and 2) the e-waste associated with mining machines[21] once they reach their end-of-life, estimated at an annual 10,948 metric tons (comparable to Luxembourg at 12 kt) assuming Koomey's efficiencies law [Koomey et al. 2011]. Despite being the most popular use of Blockchain technology, there are, and will continue to be, Blockchain applications beyond Bitcoin and cryptocurrencies. To mitigate the energy consumption of Blockchain technologies and applications, Truby [2018] has proposed a series of fiscal policy options.[22]

---

[18] The methodology and assumptions behind the Mora et al. [2018] projections of Blockchain's future energy use have been questioned by Masanet et al. [2019], but Proof of Work is widely accepted to be energy-intensive.

[19] Based on a global average electricity intensity of 0.63 $kgCO_2e$/kWh. This is likely an underestimate since Bitcoin is likely powered by a higher share of coal than the global average [Stoll et al. 2019].

[20] Note that CoinShares Research who published the report run a cryptocurrency investment fund, so there is a potential conflict of interest.

[21] If the cryptocurrency collapses, mining machines cannot be repurposed as a generic data centre since they are so specialised [Preist, personal communication].

## 3.4   Summary

**If unchecked, ICT trends could drive exponential growth in GHG emissions.**
The three trends discussed above lead to growth in the three core areas of ICT (see Figure 3.1).[23] Whilst we have discussed the trends independently above, it's important to note that these trends are in fact interlinked. For example, IoT involves collecting more sensing data, requiring more data analytics and adding to the issue of big data, data science and AI, with the potential to further increase ICT's emissions. Such growth trends will also be facilitated through innovations in the ICT infrastructure, e.g. the move from 4G to 5G cellular networks would enable faster, data-intensive network transmissions for IoT devices - allowing for even more sensing data to be collected and analysed for big data, data science and AI. If not restrained, these above trends all have potential to help drive further exponential growth, unlikely to be outweighed by the ICT-enabled carbon reductions in other sectors.

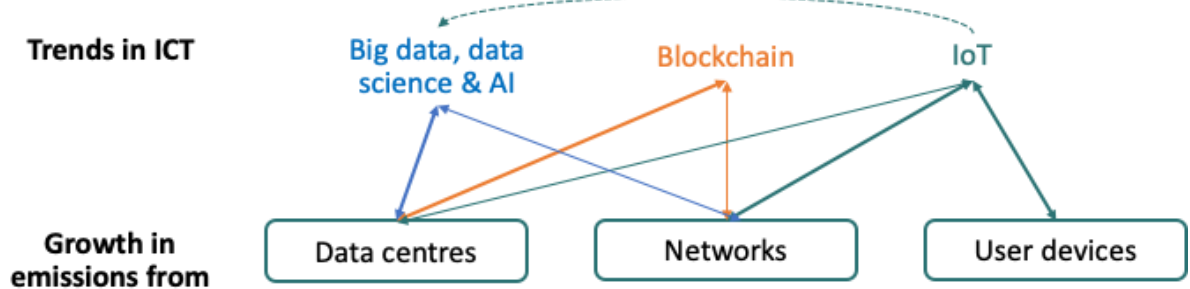

*Figure 3.1* *The threats that trends in ICT have on emissions from data centres, networks and devices. Note that stronger lines depict prominent threats, weaker lines depict secondary threats, and the dotted lines depict the links between the trends.*

**Covid-19 has shown a consumption constraint that could disrupt these trends.**
As many activities have been restricted or avoided during the pandemic, ICT has shown the significant benefits and value it can bring to society – allowing families to communicate, people to work from home, and conferences to be held online. Under these circumstances, ICT serves as a substitution rather than an addition to our regular activities. Coinciding with this, there has been a temporary drop in carbon emissions.[24]  The key question is what society will do when the Covid-19 crisis is over. Will the world: embrace some of these new ways of living and working *instead* of their traditional counterparts and reap the carbon benefits, return to our old ways, or create a mix of the two?

---

[22] For example, introducing a customs duty or excise tax on imports of miners' verification devices based on its energy consumption [Truby 2018].
[23] Note that in this section we expand 'user devices' to 'devices' to include embedded devices.
[24]  A recent study in Nature estimates that daily global $CO_2$ emissions temporarily decreased by 17% in early April relative to 2019 levels, largely due to changed transport and consumption levels, and that 2020 annual emissions could decrease by 4% if restrictions remain in place until the end of 2020 and 7% if restrictions end in June relative to 2019 [Le Quéré et al., 2020]. However, this is negligible if it does not lead to lasting changes after the pandemic.

**There are important policy decisions to be made which determine the future of ICT's carbon footprint.**

There is an increasing awareness of the impacts of ICT, but we note the need to expand our awareness to the full range of narratives and their underlying assumptions (see Section 2.2). We also note that ICT and its trends can bring a lot of value to many people worldwide. We are very much at a crossroads in terms of the choices we make, and there are some positive signals.[25] Without a global carbon constraint, avoiding unsustainable growth trends in ICT becomes a debate of what we should prioritise in the ICT sector, what problems can and should be solved using computing, and who can access the required ICT resources for such solutions – supporting valued use of ICT (for example, for uses that lead to carbon reductions in the economy) whilst constraining consumption and minimising the ICT sector's carbon footprint. An example of such prioritisation in practice is the recent Netflix agreement with EU regulators to reduce its bit rate to ease the burden on internet traffic during the Covid-19 outbreak, enabling more people to work online from home [Sweney 2020]. This in turn places the spotlight on policy makers and governance structures at all levels, including in industry, governments, and academia. We look at this important issue in the next section.

---

[25] For example, in AI research, there have been calls for the EU to incentivise AI applications which are *"socially preferable (not merely acceptable) and environmentally friendly (not merely sustainable but favourable to the environment)"*, recognising the need for a methodology to assess these characteristics [Floridi et al. 2019].

# 4. Current Policy Developments and Governance in ICT

In this section, we provide the following:
- an overview of European policy developments in ICT;
- a summary of ICT emissions self-regulation in top technology companies.

## 4.1 European Policy and ICT

Europe is leading the world in implementation of and experimentation with climate policy [Delbeke and Vis 2016]; therefore, our analysis focuses on documentation produced by the European Commission (EC). Note that while the UK had been involved in the formulation of the European Green Deal, they may be backing away from some of these commitments post-Brexit.

**ICT is a central pillar of Europe's climate strategy.**

Under the European Commission's (EC's) Green Deal, Europe is committed to becoming carbon neutral by 2050, and climate neutral later this century[26] [European Commission 2018a]. ICT features prominently in policymaking around the climate: 1) because of recent efforts to lead the world in a sustainable, human-centric approach to innovation [European Commission 2019d], and 2) to drive down GHGs across the economy.

**European ICT emissions policy emphasises efficiency, renewables and circular waste.**

The EC's official figures put ICT's current share of global GHG emissions at more than 2% [European Commission 2020c], and a study commissioned by the EC anticipates that *"the energy consumption of data centres and telecommunication networks will grow with an alarming rate of 35% and 150% respectively over 9 years"* (from 2018) [PEDCA 2015]. Rather than seeking to directly affect this consumption trend, the clear policy focus is mitigating the impacts associated with rising consumption, specifically through improved efficiency and renewable energy. Three fundamental assumptions are evident in this approach:
- there is scope for energy efficiency improvements in ICT to continue, at least through 2050 (Section 2.2.1);
- energy efficiencies in ICT can reduce ICT's carbon footprint (Section 2.2.2);
- renewable energy will decarbonise ICT (Section 2.3.6).

As discussed in Section 2.3, there are strong arguments against each of these premises which may impede successful decarbonisation of ICT unless simultaneously curbing demand or adding a global carbon constraint. However, publicly facing policy statements do not attend to these counter-assumptions.

Data centres are a particular focus of European policy. The EC has committed to carbon-neutral data centres by 2030, through a mixture of continued efficiency improvements, transitioning toward reliance on renewable energy sources, and developing methods of reusing the heat that servers generate [Kayali et al. 2020]. This is an ambitious proposal, as currently there is no indication that data centre emissions are decreasing despite continuous efficiency improvements (see Section 2.1). The EC also does not specify whether this must be achieved through on-site renewables or can include purchasing of offsets (see Appendix H for why this matters).

---

[26] The EC use the term 'carbon neutral' to refer to no net emissions of carbon dioxide, and the term 'climate neutral' to refer to no net emissions of GHG emissions. This is different from the way most ICT companies use the term 'carbon neutral', which includes all GHG emissions, see Appendix H.

Other noteworthy policy covers e-waste, which is recognised by the World Economic Forum as the fastest growing category of waste [World Economic Forum 2019]. As part of Europe's New Circular Economy Action Plan, the EC plans to put forward a 'Circular Electronics Initiative' by the end of 2021 to improve the lifespan, repairability and recyclability of ICT products [European Commission 2020d]. This initiative would help decrease the embodied carbon of ICT but would be partly offset if the total number of devices continues to increase (i.e. innovation will prohibit saturation in ICT; see Section 2.2.3).

Except for this Circular Electronics Initiative, which will likely include a reward scheme for consumers who recycle their old devices [European Commission 2020a], the Green Deal is notable for its lack of clear incentivisation or enforcement mechanisms regarding decarbonisation of ICT. It may be believed that efficiency naturally improves as technology advances (e.g. through Moore's Law), and/or that market forces will compel industry to drive these improvements, as there is no discussion of either penalties to be applied or assistance to be offered to the sector toward achieving carbon neutrality by 2050. Also not provided within the Green Deal are estimates of the emissions reductions needed within the ICT sector to meet this ambition, which may be incompatible with continuing growth expected of ICT's electricity consumption (see Section 2.1).

**Europe seeks to supercharge enablement through significant investment in ICT.** While policies clearly acknowledge ICT's share of global emissions and commit to reducing them as much as possible, the primary thrust of Europe's climate strategy is the use of ICT to enable emissions savings in other industries. An EC commissioned report states vaguely that ICT *"probably saves more energy than it consumes"* [PEDCA 2015]. The wording of the Green Deal, however, is unambiguous: *"Digital technologies are a critical enabler for attaining the sustainability goals of the Green deal in many different sectors"* [European Commission 2019a]. This includes various initiatives and major funding schemes intended to foster innovation in and uptake of AI, IoT and Blockchain (see Appendix F).

The Green Deal does not provide a detailed roadmap for how these technologies will deliver against these goals, nor figures regarding expected savings to be achieved from these technologies. These are undoubtedly difficult to estimate, but as there is no evidence in the multi-decade history of ICT-driven efficiency savings that enablement works for reducing overall emissions (see Section 2.2.5), in the absence of an intervention such as the introduction of a global carbon constraint, claims of the feasibility of this strategy should be approached with skepticism. As a baseline, staying below 1.5°C warming would require the global economy to reduce by 42% by the year 2030, including the ICT sector (see Section 2.3); so if ICT's emissions do not shrink by 42% by 2030, then it would have to enable reductions in other sectors – beyond the 42% that other industries will have to cut anyway – to compensate for this shortfall.

This may prove a delicate balancing act. To facilitate this work, complete and accurate estimates of ICT's footprint need to be captured regularly, alongside careful accounting of the emissions ICT is driving or saving in other sectors, with sector targets adjusted accordingly to ensure regional and global targets are met.

We note the competing policy priorities of the EC. Europe faces pressures to remain competitive in the global technology market and seeks to lead the way in rapidly growing technologies that would otherwise be capitalised by Asian and US competitors [Palmer 2019]. By stimulating innovation in these areas, Europe seeks to maintain both the health of its economy and the health of the planet. In the current policy environment, economic growth would likely further spur consumption and therefore emissions.

## 4.2 Self-regulation within the ICT industry

**Companies need net zero carbon targets that cover supply chain emissions**.
Several big ICT companies have recently announced carbon pledges to self-regulate their emissions (e.g. Amazon, Apple, BT, Microsoft, Sky). These pledges fall into three main categories (see Appendix G for detail): 1) carbon neutral (least ambitious); 2) net zero; and 3) carbon negative (most ambitious). To limit global warming to 1.5°C [IPCC 2018], we will need to reach net zero emissions by 2050 globally [Pineda and Faria 2019]. Companies should aim for net zero or, even better, carbon negative, to make this possible; carbon neutral targets are not enough because they do not cover supply chain emissions (see Appendix G.5 for more details). Yet only a few firmly aim to be net zero (e.g. Microsoft, Sky, Amazon, BT) and only Microsoft aims to be carbon negative [Smith 2020].

**Carbon offsetting requires truly additional carbon removal methods**.
Companies need to prioritise reducing the total emissions as much as possible [Stephens 2019] – only then should the rest of their emissions be offset by permanent, verifiable and additional carbon removal methods. For a company's emissions to be truly offset, the same amount of carbon that the company emits needs to be removed from the atmosphere (e.g. through afforestation, reforestation, planting seagrass, taking in landfill gas), not simply avoided.[27] However, only 2% of offsets result in truly additional removals [Cames et al. 2016]. Furthermore, some offsetting projects may not be permanent. For example, where forests or peatlands are used to sequester carbon, these carbon stores must be protected from fires or logging – otherwise the carbon removals are negated. Efficiency enablement cannot count as offsetting because it is hard to show that any enabled savings are not negated by rebound effects (see Section 2.2.5).

**Only some renewable energy helps to cut emissions.**
Some companies also claim, or aim for, power provision from 100% renewable energy without specifying whether they aim to cut emissions (see Appendix G.4). Companies need to detail which type of renewable energy they use (e.g. biofuels, solar, wind, hydro), and what proportion of their renewable energy comes from on-site renewable power generation, Power Purchasing Agreements (PPAs) and Renewable Energy Guarantees of Origin (REGOs), as these differ in their additionality (see Appendix H). For a company to claim they are 100% renewable, they should source 100% of their energy through PPAs, on-site renewables and investment in off-site projects but not unbundled REGOs, because the latter cannot claim additionality. Renewable energy projects should not be considered a removal but rather a scope 2 reduction (see Section 2.2.6).

---

[27] An example of an avoided emission offset is an area of forest that is protected from logging; the amount of carbon that would have been released if the forest was cut down is counted as offset. However, there needs to be some certainty that it would have been removed if it had not been purchased, otherwise these offsets cannot be considered additional. Even genuine 'avoided' emissions may end up 'leaking' out at another point in the system, (e.g. a protected area of forest may just lead to more logging somewhere else in the world) [Childs 2019].

**The new ITU standard encourages ICT companies to become net zero by 2050.**
In collaboration with GSMA, GeSI and SBTi, the International Telecommunication Union (ITU) [2020], a UN agency focused on the ICT industry, released a new standard in February 2020 that aims to reduce ICT's GHG emissions by 45% by 2030, and net zero by 2050, in line with limiting global warming to 1.5°C.[28] The 'voluntary' standard comes with reduction targets for each ICT sub-sector for the next decade.[29] Data centre operators adopting the science-based target will need to reduce emissions by at least 53%, mobile network operators by 45% and fixed network operators by 62% [GeSI 2015]. The targets have been approved by the SBTi and require companies to set targets for scope 1 and 2 emissions and some supply chain scope 3 (see Appendix G.2). Most of these reductions between 2020 and 2030 are expected to come from a shift to more renewable and other low-carbon energy. The targets are less ambitious than pledges by some individual companies such BT, Sky and Microsoft which commit to reach net zero by 2030 or 2040 already, but they send a strong signal that the world needs net zero and science-based targets and provide a template that policy makers could adopt.

## 4.3   Summary

**Regulation and changes to climate strategies are required to ensure the ICT sector keeps in line with the 1.5°C limit.**
Estimations of ICT-enabled emissions savings in other sectors fall short of what is required for meeting agreed targets, and there is a risk that ICT's expansion into other sectors could increase those sectors' emissions (see Section 4.1). This fundamentally calls into question the presumed role of efficiency within climate strategy. There is clear need to detail sector by sector the savings ICT is expected to produce – reflecting careful balancing of sector footprints within the contexts of regional and global targets – along with developing a detailed roadmap toward delivering on those expectations. As part of these calculations, the full impacts of ICT need to be considered systematically – particularly using methods that allow for objective, high-quality, up-to-date data and analysis (i.e. rectifying the issues of current estimates, see Section 2.1). Policy-enforced carbon caps on global emissions, or carbon pricing for all industries, would help avoid the risk of Global Rebounds; but without a global carbon constraint, policies will be needed to enforce credible and ambitious carbon pledges within the ICT sector (see Section 4.2). We have outlined below five criteria specifically for ICT sector targets, all of which will need to pervade the ICT sector and be subjected to tough, well-resourced, and independent scrutiny:

---

[28] The scope of ITU's recommendation includes 'mobile networks, fixed networks, data centres, enterprise networks, and end-user devices, but excludes ICT services.'
[29] Sub-sectors are defined as per other ITU documentation, specifically clauses A2 to A6 of ITU-TL.1450 (https://www.itu.int/ITU-T/recommendations/rec.aspx?id=13581).

1. Targets should be inclusive of scope 1, 2 and 3 emissions;
2. Reduction trajectories should be in line with IPCC recommendations for limiting warming to 1.5°C;
3. Where transition to renewable energy is part of the decarbonisation pathway, a careful test should be applied that the renewables are provably additional;
4. Emissions offsets need to pass tests of *permanence*, *verifiability* and *additionality;*
5. Where 'net zero' or 'carbon neutral' targets are announced, these should be disaggregated into an emissions reduction component and an offsetting component so that offsets are not allowed to replace reduction responsibilities;
6. Emission reduction targets should not be replaced by enablement claims due to the risk of rebound effects.

**Self-regulation may not be enough to cut $CO_2$ emissions.**
With growing awareness of the climate emergency, public pressure may be enough to get more ICT companies to announce net zero emissions by 2050. However, there is a lack of net zero pledges thus far. Some companies which have pledged net zero are not on target, or do not have detailed and transparent action plans. Note that this piecemeal approach of individual companies making commitments also comes at a competitive cost for the foreriders, with others gaining financially from being free of such commitments. The way forward for a reduction in ICT's emissions is a sector-wide commitment to net zero that is enforced through incentives and compliance mechanisms, such as procurement clauses that set out carbon criteria and consequences for non-compliance. We flag this as an important issue for the sector but detailed consideration of the form of regulation is beyond the scope of this report. We also note that an ICT-focused net zero commitment is unlikely to limit the emissions from ICT's impact on the wider economy, unless upstream scope 3 emissions are included in the targets.

# 5. Conclusion

As we have explored in this report, there are two central issues for the ICT industry with respect to the climate emergency: ICT's own carbon footprint; and ICT's carbon impact on the rest of the global economy. There has been surprisingly little research into these questions given their significance in response to climate change. The evidence that does exist needs to be interpreted with awareness of problems arising from the following issues: 1) the age of the data; 2) a lack of data interrogatability; 3) a potential for conflict of interest (especially where researchers are employed by ICT companies, and data and analysis is not freely available); and 4) varying approaches to, and lack of agreement on, the boundaries of the analysis of specifically what constitutes the ICT industry in terms of inclusion in estimates of its carbon footprint (e.g. whether or not growth trends in ICT such as Blockchain are included, how scope 3 emissions in the supply chain are included to avoid truncation error).

Historically we can be sure that four phenomena have gone hand in hand: ICT has become dramatically more efficient; ICT's footprint has risen to account for a significant proportion of global emissions; ICT has delivered increasingly wide-ranging efficiency and productivity improvements to the global economy; and global emissions have risen inexorably.

Looking to the future, our concerns are that this growth in emissions will continue at a time when emissions *must shrink* considerably. All analyses reviewed in this report concur that ICT is not on a path to reduce emissions in line with recommendations from climate science *unless additional steps* are taken by the sector, or legislators, to ensure that this happens. Prevalent policy emphasis on efficiency improvements, use of renewables and circular electronics is likely insufficient to reverse ICTs growth in emissions. There are real concerns that the period governed by Moore's Law is coming to an end, and there is huge investment in trends that can significantly increase the carbon footprint of ICT, including in AI, IoT and Blockchain. Recently there are encouraging signs that some ICT giants may be moving in a positive direction (e.g. through net zero and carbon negative targets that include their supply chains). Our hope is that with the right policy to enforce these commitments, ICT companies will be able to deliver on their pledges and that other industries will follow ICT's example, allowing us to stay within 1.5°C warming.

Based on the evidence available, it is also key that regulators move away from the presumption that ICT *saves more emissions than it produces*. While ICT offers opportunities to enable reductions in GHG emissions in other sectors, evidence does *not support* their ability to achieve the sustained significant carbon savings we require by 2050. And whilst ICT might make lower carbon living possible (as the Covid-19 pandemic demonstrates), this will not in itself help to bring about a cut in carbon. The argument of enablement simply does not exempt the ICT sector from addressing its own emissions, and the sector could certainly do more to understand its enablement and rebound effects. To ensure current technologies have a truly positive impact on the environment, the climate emergency requires a global constraint such as a carbon cap on extraction, a price on carbon emissions, or (as Covid-19 has temporarily engineered) a constraint on consumption, to rule out rebounds in emissions.

Given the historical trends, in the absence of strong evidence that the conditions resulting in those trends have fundamentally altered, there is a significant risk that the efficiency gains sought through increased application of ICT solutions will not only fail to lower emissions, but may in fact spur even higher emissions. At the very least it would seem unsafe to assume that ICT efficiencies bring about carbon savings by default. Efficiency is important and, in a carbon-constrained and perhaps consequently energy-constrained world, efficiency may well become more highly valued than ever before since it will be the only means to grow or even maintain productivity; this would mean the ICT industry could become an even larger part of the global economy as a vital sector for the transition to a net zero world.

## 6. Acknowledgements

This work was developed following discussions at the Royal Society project on Digital Technology and the Planet. The research was partially supported by the DT/LWEC Senior Fellowship (awarded to Blair) in the Role of Digital Technology in Understanding, Mitigating and Adapting to Environmental Change, EPSRC: EP/P002285/1, and by the EPSRC Doctoral Prize (awarded to Widdicks) to enable Widdicks to continue her research in sustainability and digital wellbeing, EPSRC: EP/R513076/1. We thank the experts we consulted for this research as well as those who provided feedback, particularly: Anders Andrae, Lotfi Belkhir, Livia Cabernard, Peter Garraghan, Jens Malmodin, and Chris Preist.

# References

1. Aebischer, B. and Hilty, L.M., 2015. *The energy demand of ICT: a historical perspective and current methodological challenges*, in *ICT Innovations for Sustainability*. Springer. p. 71-103.
2. Al-Ali, A.-R., Zualkernan, I.A., Rashid, M., Gupta, R., and AliKarar, M., 2017. *A smart home energy management system using IoT and big data analytics approach.* IEEE Transactions on Consumer Electronics, **63**(4): p. 426-434.
3. Alcott, B., 2005. *Jevons' paradox.* Ecological economics, **54**(1): p. 9-21.
4. Alsamhi, S., Ma, O., Ansari, M.S., and Meng, Q., 2019. *Greening internet of things for greener and smarter cities: a survey and future prospects.* Telecommunication Systems, **72**(4): p. 609-632.
5. Amodei, D. and Hernandez, D., 2019. *AI and Compute*. https://openai.com/blog/ai-and-compute/ - accessed March 2020.
6. Andrae, A.S., 2019a. *Prediction Studies of Electricity Use of Global Computing in 2030.* Int J Sci Eng Invest, **8**: p. 27-33.
7. Andrae, A.S., 2019b. *Comparison of Several Simplistic High-Level Approaches for Estimating the Global Energy and Electricity Use of ICT Networks and Data Centers.* International Journal, **5**: p. 51.
8. Andrae, A.S., 2019c. *Projecting the chiaroscuro of the electricity use of communication and computing from 2018 to 2030.* Researchgate. net.
9. Andrae, A.S., 2020. *Hypotheses for primary energy use, electricity use and CO2 emissions of global computing and its shares of the total between 2020 and 2030.* WSEAS Transactions on Power Systems.
10. Andrae, A.S. and Edler, T., 2015. *On global electricity usage of communication technology: trends to 2030.* Challenges, **6**(1): p. 117-157.
11. Apergis, I. and Apergis, N., 2019. *Silver prices and solar energy production.* Environmental Science and Pollution Research, **26**(9): p. 8525-8532.
12. Arshad, R., Zahoor, S., Shah, M.A., Wahid, A., and Yu, H., 2017. *Green IoT: An investigation on energy saving practices for 2020 and beyond.* IEEE Access, **5**: p. 15667-15681.
13. Aslan, J., Mayers, K., Koomey, J.G., and France, C., 2018. *Electricity intensity of Internet data transmission: Untangling the estimates.* Journal of Industrial Ecology, **22**(4): p. 785-798.
14. Baskerville-Muscutt, K., 2019. *Setting Science-Based Supply Chain Greenhouse-Gas Emission Targets: Aligning Corporate Climate Action With The 1.5°C Target.* MSc, Durham University.
15. BBC iPlayer, 2020. *Dirty streaming: The internet's big secret*. https://www.bbc.co.uk/news/av/stories-51742336/dirty-streaming-the-internet-s-big-secret - accessed March 2020.
16. Belkhir, L., 2018. *How smartphones are heating up the planet.* The Conversation. https://theconversation.com/how-smartphones-are-heating-up-the-planet-92793 - accessed March 2020.
17. Belkhir, L. and Elmeligi, A., 2018. *Assessing ICT global emissions footprint: Trends to 2040 & recommendations.* Journal of Cleaner Production, **177**: p. 448-463.
18. Bendiksen, C. and Gibbons, S., 2019. *The Bitcoin Mining Network: Trends, Average Creation Costs, Electricity Consumption & Sources. December 2019 Update.* CoinShares Research. https://coinsharesgroup.com/research/bitcoin-mining-network-december-2019 - accessed March 2020.
19. Bendiksen, C., Gibbons, S., and Lim, E., 2018. *The Bitcoin Mining Network.*

CoinShares Research.
20. Berners-Lee, M., 2011. *How bad are bananas?: the carbon footprint of everything.* Greystone Books.
21. Berners-Lee, M. and Clark, D., 2013.*The Burning Question: We can't burn half the world's oil, coal and gas. So how do we quit?* : Profile Books.
22. Berners-Lee, M., Howard, D.C., Moss, J., Kaivanto, K., and Scott, W., 2011. *Greenhouse gas footprinting for small businesses—The use of input–output data.* Science of the Total Environment, **409**(5): p. 883-891.
23. Bibri, S.E., 2018. *The IoT for smart sustainable cities of the future: An analytical framework for sensor-based big data applications for environmental sustainability.* Sustainable cities and society, **38**: p. 230-253.
24. Biewald, L., 2019. *Deep Learning and Carbon Emissions.* Towards Data Science, Medium. https://towardsdatascience.com/deep-learning-and-carbon-emissions-79723d5bc86e - accessed March 2020.
25. Börjesson Rivera, M., Håkansson, C., Svenfelt, Å. and Finnveden, G., 2014. Including second order effects in environmental assessments of ICT. Environmental Modelling & Software, 56, pp.105-115.
26. Cabernard, L., 2019. *Global supply chain analysis of material-related impacts in ICT (MRIO approach).* http://www.lcaforum.ch/portals/0/df73/DF73-04_Cabernard.pdf - accessed March 2020.
27. Cabernard, L., Pfister, S., and Hellweg, S., 2019. *A new method for analyzing sustainability performance of global supply chains and its application to material resources.* Science of the Total Environment, **684**: p. 164-177.
28. Cambridge Dictionary, 2020. Available from: https://dictionary.cambridge.org/ - accessed March 2020.
29. Cames, M., Harthan, R.O., Füssler, J., Lazarus, M., Lee, C.M., Erickson, P., and Spalding-Fecher, R., 2016. *How additional is the clean development mechanism? Analysis of the application of current tools and proposed alternatives.* Öko-Institut. https://ec.europa.eu/clima/sites/clima/files/ets/docs/clean_dev_mechanism_en.pdf - accessed March 2020.
30. Childs, M., 2020. *Does carbon offsetting work?* Friends of the Earth. https://friendsoftheearth.uk/climate-change/does-carbon-offsetting-work - accessed March 2020.
31. Cisco, 2020. *Cisco Annual Internet Report (2018–2023).* https://www.cisco.com/c/en/us/solutions/collateral/executive-perspectives/annual-internet-report/white-paper-c11-741490.pdf - accessed February 2020.
32. CoinBundle Team, 2018. Consensus Algorithms. Securing Blockchain Transactions. Medium. https://medium.com/coinbundle/consensus-algorithms-dfa4f355259d#a76e - accessed March 2020.
33. CoinMarketCap, 2020. *Global Charts.* https://coinmarketcap.com/charts/ - accessed March 2020.
34. Collier, S.E., 2016. The emerging enernet: Convergence of the smart grid with the internet of things. *IEEE Industry Applications Magazine*, *23*(2), pp.12-16.
35. Cong, L.W., He, Z., and Li, J., 2019. *Decentralized mining in centralized pools.* National Bureau of Economic Research.
36. Coroama, V.C., Hilty, L.M., Heiri, E., and Horn, F.M., 2013. *The direct energy demand of internet data flows.* Journal of Industrial Ecology, **17**(5): p. 680-688.
37. Coroama, V.C., Moberg, Å., and Hilty, L.M., 2015. *Dematerialization through*

*electronic media?*, in *ICT Innovations for Sustainability*. Springer. p. 405-421.
38. Court, V., and Sorrell, S., 2020. *Digitalisation of goods: a systematic review of the determinants and magnitude of the impacts on energy consumption.* Environmental Research Letters, **15**(4): p.043001.
39. Das, S., 2019. *Global Energy Footprint of IoT Semiconductors.* http://www.lcaforum.ch/portals/0/df73/DF73-09_Das.pdf - accessed March 2020.
40. de Vries, A., 2018. *Bitcoin's growing energy problem.* Joule, **2**(5): p. 801-805.
41. de Vries, A., 2019. *Renewable Energy Will Not Solve Bitcoin's Sustainability Problem.* Joule, **3**(4): p. 893-898.
42. Delbeke, J. and Vis, P., 2016. *EU Climate Policy Explained.* https://ec.europa.eu/clima/sites/clima/files/eu_climate_policy_explained_en.pdf - accessed March 2020.
43. Department for Business, Energy and Industrial Strategy, 2019. *Greenhouse gas reporting: conversion factors 2019.* UK Government. https://www.gov.uk/government/publications/greenhouse-gas-reporting-conversion-factors-2019 - accessed March 2020.
44. Digiconomist, 2019. *Bitcoin Energy Consumption Index. (2019).* https://digiconomist.net/bitcoin-energy-consumption - accessed February 2020.
45. Enerdata, 2018. *Between 10 and 20% of electricity consumption from the ICT\* sector in 2030?* https://www.enerdata.net/publications/executive-briefing/expected-world-energy-consumption-increase-from-digitalization.html - accessed March 2020.
46. Ercan, M., Malmodin, J., Bergmark, P., Kimfalk, E., and Nilsson, E., 2016. *Life cycle assessment of a smartphone.* in *ICT for Sustainability 2016*. Atlantis Press.
47. Ericsson, 2019. *Exponential data growth – constant ICT footprints.* https://www.ericsson.com/en/reports-and-papers/research-papers/the-future-carbon-footprint-of-the-ict-and-em-sectors - accessed March 2020.
48. Ericsson, 2020. *A quick guide to your digital carbon footprint – Deconstructing Information and Communication Technology's carbon emissions.* https://www.ericsson.com/en/reports-and-papers/industrylab/reports/a-quick-guide-to-your-digital-carbon-footprint - accessed March 2020.
49. ETIP SNET, *Integrating Smart Networks for the Energy Transition: Serving Society and Protecting the Environment.* Vision 2050. https://www.etip-snet.eu/etip-snet-vision-2050/ - accessed March 2020.
50. EU Blockchain, 2020. *Observatory and Forum.* Available from: https://www.euBlockchainforum.eu/ - accessed March 2020.
51. European Commission, 2018a. *A Clean Planet for all. A European strategic long-term vision for a prosperous, modern, competitive and climate neutral economy.* https://eur-lex.europa.eu/legal-content/EN/TXT/PDF/?uri=CELEX:52018DC0773&from=EN - accessed March 2020.
52. European Commission, 2018b. *European countries join Blockchain Partnership.* Shaping Europe's digital future. Digibyte. https://ec.europa.eu/digital-single-market/en/news/european-countries-join-Blockchain-partnership - accessed March 2020.
53. European Commission, 2018c. *Digital Innovation Hubs.* https://ec.europa.eu/futurium/en/system/files/ged/digital_innovation_hubs_in_

digital_europe_programme_final2_december.pdf - accessed March 2020.
54. European Commission, 2019a. *The European Green Deal.* https://ec.europa.eu/info/sites/info/files/european-green-deal-communication_en.pdf - accessed March 2020.
55. European Commission, 2019b. *The Internet of Things.* Shaping Europe's digital future. Policy. https://ec.europa.eu/digital-single-market/en/internet-of-things - accessed March 2020.
56. European Commission, 2019c. *Blockchain Unleashed for Climate Action.* Shaping Europe's digital future. Event, Trade Fair and Congress Center of Malaga. https://ec.europa.eu/digital-single-market/en/news/Blockchain-unleashed-climate-action - accessed March 2020.
57. European Commission, 2019d. *On Artificial Intelligence - A European approach to excellence and trust.* https://ec.europa.eu/info/sites/info/files/commission-white-paper-artificial-intelligence-feb2020_en.pdf - accessed March 2020.
58. European Commission, 2019e. *Policy and Investment Recommendations for Trustworthy AI.* Independent High-Level Expert Group on Artificial Intelligence set up by the European Commission. https://www.europarl.europa.eu/cmsdata/196378/AI%20HLEG_Policy%20and%20Investment%20Recommendations.pdf - accessed March 2020.
59. European Commission, 2020a. *Eco-Innovation at the heart of European policies.* Enrivonment. Eco-innovation Action Plan.; Available from: https://ec.europa.eu/environment/ecoap/about-eco-innovation/policies-matters/rewards-recycling_en - accessed March 2020.
60. European Commission, 2020b. *Blockchain Technologies*. Shaping Europe's digital future. Policy. https://ec.europa.eu/digital-single-market/en/Blockchain-technologies - accessed March 2020.
61. European Commission, 2020c. *Supporting the green transition.* https://ec.europa.eu/commission/presscorner/detail/en/fs_20_281 - accessed March 2020.
62. European Commission, 2020d. *Changing how we produce and consume: New Circular Economy Action Plan shows the way to a climate-neutral, competitive economy of empowered consumers.* https://ec.europa.eu/commission/presscorner/detail/en/ip_20_420 - accessed March 2020.
63. FCAI, 2020. *The European Commission offers significant support to Europe's AI excellence.* https://fcai.fi/news/the-european-commission-offers-significant-support-to-europes-ai-excellence - accessed March 2020.
64. Floridi, L., Cowls, J., Beltrametti, M., Chatila, R., Chazerand, P., Dignum, V., Luetge, C., Madelin, R., Pagallo, U., and Rossi, F., 2018. *AI4People—An ethical framework for a good AI society: Opportunities, risks, principles, and recommendations.* Minds and Machines, **28**(4): p. 689-707.
65. Folding@Home, 2020. Folding@Home. https://foldingathome.org/, accessed March 2020.
66. Fryer, E., 2020. *Does streaming really have a dirty secret?* TechUK. https://www.techuk.org/insights/opinions/item/17020-does-streaming-really-have-a-dirty-secret - accessed March 2020.
67. Galvin, R., 2015. *The ICT/electronics question: Structural change and the rebound effect.* Ecological Economics, **120**: p. 23-31.
68. Gartner, G.I. and Green, I., 2007. *The new industry shockwave, presentation at symposium*. in *ITXPO conference, April*.

69. GeSI, 2008. *Smart 2020: Enabling the low carbon economy in the information age.* A report by The Climate Group on behalf of the Global eSustainability Initiative. https://gesi.org/public/research/smart-2020-enabling-the-low-carbon-economy-in-the-information-age - accessed March 2020.
70. GeSI, 2012. *Smarter 2020: The Role of ICT in Driving a Sustainable Future.* A report by Boston Consulting Group on behalf of the Global eSustainability Initiative. https://gesi.org/report/detail/gesi-smarter2020-the-role-of-ict-in-driving-a-sustainable-future - accessed March 2020.
71. GeSI, 2015. *Smarter 2030: ICT Solutions for 21st Century Challenges.* A report by Accenture Strategy on behalf of the Global eSustainability Initiative. http://smarter2030.gesi.org/downloads.php - accessed March 2020.
72. Gilmore, M., 2018. *Final Report.* Expert and Stakeholder Consultation Workshop. Green ICT - Research and innovation activities (2020-2030). European Commission. https://ec.europa.eu/newsroom/dae/document.cfm?doc_id=50342 - accessed March 2020.
73. Gomes, C., Dietterich, T., Barrett, C., Conrad, J., Dilkina, B., Ermon, S., Fang, F., Farnsworth, A., Fern, A., and Fern, X., 2019. *Computational sustainability: Computing for a better world and a sustainable future.* Communications of the ACM, **62**(9): p. 56-65.
74. Gossart, C., 2015. *Rebound effects and ICT: a review of the literature*, in *ICT innovations for sustainability*. Springer. p. 435-448.
75. Govindan, K., Cheng, T., Mishra, N., and Shukla, N., 2018. *Big data analytics and application for logistics and supply chain management.* Elsevier.
76. Graver, B., Kevin Zhang, K., and Rutherford, D., 2019. *$CO_2$ emissions from commercial aviation, 2018.* International Council on Clean Transportation. https://theicct.org/sites/default/files/publications/ICCT_CO2-commercl-aviation-2018_20190918.pdf - accessed March 2020.
77. GSMA, The Enablement Effect, Technical Report, 2019. https://www.gsma.com/betterfuture/enablement-effect, accessed March 2020.
78. Guardian Environment Network, 2017. *'Tsunami of data' could consume one fifth of global electricity by 2025.* The Guardian. https://www.theguardian.com/environment/2017/dec/11/tsunami-of-data-could-consume-fifth-global-electricity-by-2025 - accessed March 2020.
79. Henderson, P., Hu, J., Romoff, J., Brunskill, E., Jurafsky, D., and Pineau, J., 2020. *Towards the Systematic Reporting of the Energy and Carbon Footprints of Machine Learning.* arXiv preprint arXiv:2002.05651.
80. Hewlett, O., 2017. *Ensuring Renewable Electricity Market Instruments Contribute to the Global Low-Carbon Transition and Sustainable Development Goals.* Gold Standard. https://www.goldstandard.org/sites/default/files/documents/gs_recs_position_paper.pdf - accessed March 2020.
81. Hilty, L., Lohmann, W., and Huang, E.M., 2011. *Sustainability and ICT-an overview of the field.* notizie di POLITEIA, **27**(104): p. 13-28.
82. Howson, P., 2019. *Tackling climate change with Blockchain.* Nature Climate Change, **9**(9): p. 644-645.
83. IDC, 2014. The Digital Universe of Opportunities: Rich Data and the Increasing Value of the Internet of Things. EMC Digital Universe with Research & Analysis by IDC. https://www.emc.com/leadership/digital-universe/2014iview/executive-

summary.htm - accessed March 2020.
84. IEA, 2019a. *Solar PV.* https://www.iea.org/reports/tracking-power-2019/solar-pv - accessed March 2020.
85. IEA, 2019b. *Global Energy & CO2 Status Report.* https://www.iea.org/geco/emissions/ - accessed March 2020.
86. IPCC, 2018. *Special Report – Global Warming of 1.5 ºC.* https://www.ipcc.ch/sr15/ - accessed March 2020.
87. ITU, 2020. *ICT industry to reduce greenhouse gas emissions by 45 per cent by 2030.* https://www.itu.int/en/mediacentre/Pages/PR04-2020-ICT-industry-to-reduce-greenhouse-gas-emissions-by-45-percent-by-2030.aspx - accessed March 2020.
88. James, J., 2019. *Data as the new oil: The danger behind the mantra.* The Enterprises Project. https://enterprisersproject.com/article/2019/7/data-science-data-can-be-toxic - accessed March 2020.
89. Jevons, W.S., 1865. *The coal question: Can Britain survive?* In: Flux, A.W. (Ed.), The Coal Question: An Inquiry Concerning the Progress of the Nation, and the Probable Exhaustion of our Coal-Mines. Augustus M. Kelley, New York
90. Jobin, A., Ienca, M., and Vayena, E., 2019. *The global landscape of AI ethics guidelines.* Nature Machine Intelligence, **1**(9): p. 389-399.
91. Kaapa, P., 2017. *It's time to start thinking about our digital carbon footprint.* The Conversation. https://theconversation.com/its-time-to-start-thinking-about-our-digital-carbon-footprint-81518 - accessed March 2020.
92. Kamiya, G., 2020. *Factcheck: What is the carbon footprint of streaming video on Netflix?* Carbon Brief. https://www.carbonbrief.org/factcheck-what-is-the-carbon-footprint-of-streaming-video-on-netflix - accessed March 2020.
93. Kaur, N. and Sood, S.K., 2015. *An energy-efficient architecture for the Internet of Things (IoT).* IEEE Systems Journal, **11**(2): p. 796-805.
94. Kayali, L., Heikkilä, M., and Delcker, J., 2020. *Europe's digital vision, explained.* Politico. https://www.politico.eu/article/europes-digital-vision-explained/ - accessed March 2020.
95. Kennelly, C., Berners-Lee, M., and Hewitt, C., 2019. *Hybrid life-cycle assessment for robust, best-practice carbon accounting.* Journal of cleaner production, **208**: p. 35-43.
96. Koomey, J.G., Berard, S., Sanchez, M., and Wong, H., 2011. *Web Extra Appendix: Implications of Historical Trends in the Electrical Efficiency of Computing.* IEEE Annals of the History of Computing, **33**(3): p. S1-S30.
97. Kouhizadeh, M. and Sarkis, J., 2018. *Blockchain practices, potentials, and perspectives in greening supply chains.* Sustainability, **10**(10): p. 3652.
98. Lannoo, B., Lambert, S., Van Heddeghem, W., Pickavet, M., Kuipers, F., Koutitas, G., Niavis, H., Satsiou, A., Till, M., and Beck, A.F., 2013. *Overview of ICT energy consumption.* Network of Excellence in Internet Science: p. 1-59.
99. Le Quéré, C., Jackson, R.B., Jones, M.W., Smith, A.J., Abernethy, S., Andrew, R.M., De-Gol, A.J., Willis, D.R., Shan, Y., Canadell, J.G. and Friedlingstein, P., 2020. *Temporary reduction in daily global CO 2 emissions during the COVID-19 forced confinement.* Nature Climate Change, pp.1-7.
100. Liu, Q., Ma, Y., Alhussein, M., Zhang, Y., and Peng, L., 2016. *Green data center with IoT sensing and cloud-assisted smart temperature control system.* Computer Networks, **101**: p. 104-112.
101. Lopez Yse, D., 2019. Your Guide to Natural Language Processing (NLP).

Towards Data Science, Medium. https://towardsdatascience.com/your-guide-to-natural-language-processing-nlp-48ea2511f6e1 - accessed March 2020.
102. Lord, C., Hazas, M., Clear, A.K., Bates, O., Whittam, R., Morley, J. and Friday, A., 2015, April. Demand in my pocket: mobile devices and the data connectivity marshalled in support of everyday practice. *In Proceedings of the 33rd Annual ACM Conference on Human Factors in Computing Systems* (pp. 2729-2738).
103. Lövehagen, N., 2020. *What's the real climate impact of digital technology?*. Ericsson. https://www.ericsson.com/en/blog/2020/2/climate-impact-of-digital-technology - accessed March 2020.
104. Magee, C.L. and Devezas, T.C., 2017. *A simple extension of dematerialization theory: Incorporation of technical progress and the rebound effect.* Technological Forecasting and Social Change, **117**: p. 196-205.
105. Malmodin, J., 2020. The ICT sector's carbon footprint. Presentation at the techUK conference in London Tech Week on 'Decarbonising Data', 2020. https://spark.adobe.com/page/dey6WTCZ5JKPu/ - accessed December 2020.
106. Malmodin, J., 2019. *Energy consumption and carbon emissions of the ICT sector.* Presentation given at Energimyndigheten, Stockholm, 19 Dec 2019.
107. Malmodin, J., Bergmark, P., and Lundén, D., 2013. *The future carbon footprint of the ICT and E&M sectors.* on Information and Communication Technologies: p. 12.
108. Malmodin, J. and Lundén, D., 2018a. *The energy and carbon footprint of the global ICT and E&M sectors 2010–2015.* Sustainability, **10**(9): p. 3027.
109. Malmodin, J. and Lundén, D., 2018b. *The electricity consumption and operational carbon emissions of ICT network operators 2010-2015.* Report from the KTH Centre for Sustainable Communications Stockholm, Sweden.
110. Malmodin, J., Moberg, Å., Lundén, D., Finnveden, G., and Lövehagen, N., 2010. *Greenhouse gas emissions and operational electricity use in the ICT and entertainment & media sectors.* Journal of Industrial Ecology, **14**(5): p. 770-790.
111. Masanet, E., Shehabi, A., Lei, N., Smith, S., and Koomey, J., 2020. *Recalibrating global data center energy-use estimates.* Science, **367**(6481): p. 984-986.
112. Masanet, E., Shehabi, A., Lei, N., Vranken, H., Koomey, J., & Malmodin, J. (2019). Implausible projections overestimate near-term Bitcoin $CO_2$ emissions. Nature Climate Change, 9(9), 653-654.
113. McLaren, D.P., Tyfield, D.P., Willis, R., Szerszynski, B., and Markusson, N.O., 2019. *Beyond 'Net-Zero': A case for separate targets for emissions reduction and negative emissions.* Frontiers in Climate, **1**: p. 4.
114. McMahon, L., 2018. *Is staying online costing the earth?* Technical Report. Policy Connect. https://www.policyconnect.org.uk/appccg/research/staying-online-costing-earth - accessed March 2020.
115. Minoli, D., Sohraby, K., and Occhiogrosso, B., 2017. *IoT considerations, requirements, and architectures for smart buildings—Energy optimization and next-generation building management systems.* IEEE Internet of Things Journal, **4**(1): p. 269-283.
116. Mohanty, S.P., Choppali, U., and Kougianos, E., 2016. *Everything you wanted to know about smart cities: The internet of things is the backbone.* IEEE Consumer Electronics Magazine, **5**(3): p. 60-70.
117. Monrat, A.A., Schelén, O., and Andersson, K., 2019. *A survey of Blockchain from the perspectives of applications, challenges, and opportunities.* IEEE

Access, **7**: p. 117134-117151.
118. Mora, C., Rollins, R.L., Taladay, K., Kantar, M.B., Chock, M.K., Shimada, M., and Franklin, E.C., 2018. *Bitcoin emissions alone could push global warming above 2°C.* Nature Climate Change, **8**(11): p. 931-933.
119. Mylonas, G., Amaxilatis, D., Chatzigiannakis, I., Anagnostopoulos, A., and Paganelli, F., 2018. *Enabling sustainability and energy awareness in schools based on iot and real-world data.* IEEE Pervasive Computing, **17**(4): p. 53-63.
120. NoCash, 2020. *European Commission about AI: „Europe needs to increase its investment levels significantly." Digital Innovation Hubs should provide support to SMEs to understand and adopt AI – at least one innovation hub per Member State.* https://nocash.ro/european-commission-about-ai-europe-needs-to-increase-its-investment-levels-significantly-digital-innovation-hubs-should-provide-support-to-smes-to-understand-and-adopt-ai-at-least-one-innovat/ - accessed March 2020.
121. NPD, 2018. *The Average Upgrade Cycle of a Smartphone in the U.S. is 32 Months, According to NPD Connected Intelligence.* https://www.npd.com/wps/portal/npd/us/news/press-releases/2018/the-average-upgrade-cycle-of-a-smartphone-in-the-u-s--is-32-months---according-to-npd-connected-intelligence/ - accessed March 2020.
122. Ofcom, 2017. *Box Set Britain: UK's TV and online habits revealed*. https://www.ofcom.org.uk/about-ofcom/latest/media/media-releases/2017/box-set-britain-tv-online-habits - accessed March 2020.
123. Olivier, J.G.J. and Peters, J.A.H.W., 2019. *Trends in global CO2 and total greenhouse gas emissions: 2019 report.* PBL Netherlands Environmental Assessment Agency, The Hague; 4004. https://www.pbl.nl/sites/default/files/downloads/pbl-2019-trends-in-global-co2-and-total-greenhouse-gas-emissions-summary-ot-the-2019-report_4004.pdf - accessed March 2020.
124. Palmer, M., 2019. *EU launches €2bn AI and Blockchain fund.* Sifted. https://sifted.eu/articles/eu-2bn-ai-Blockchain-fund/ - accessed March 2020.
125. PEDCA (Pan-European Data Centre Academy), 2015. *Final Report Summary.* European Commission. https://cordis.europa.eu/project/id/320013/reporting - accessed March 2020.
126. Pineda, A.C. and Faria, P., 2019. *Towards a science-based approach to climate neutrality in the corporate sector.* Science Based Targets report. https://sciencebasedtargets.org/wp-content/uploads/2019/10/Towards-a-science-based-approach-to-climate-neutrality-in-the-corporate-sector-Draft-for-comments.pdf - accessed March 2020.
127. Preist, C., Schien, D., and Blevis, E., 2016. *Understanding and mitigating the effects of device and cloud service design decisions on the environmental footprint of digital infrastructure.* in *Proceedings of the 2016 CHI Conference on Human Factors in Computing Systems*.
128. Preist, C., Schien, D., and Shabajee, P., 2019. *Evaluating Sustainable Interaction Design of Digital Services: The Case of YouTube*. in *Proceedings of the 2019 CHI Conference on Human Factors in Computing Systems*.
129. Raj, P., Raman, A., Nagaraj, D., and Duggirala, S., 2015. *Big Data Analytics for Healthcare*, in *High-Performance Big-Data Analytics*. Springer. p. 391-424.
130. Raza, U., Kulkarni, P., and Sooriyabandara, M., 2017. *Low power wide area networks: An overview.* IEEE Communications Surveys & Tutorials, **19**(2): p. 855-873.

131. Reka, S.S. and Dragicevic, T., 2018. *Future effectual role of energy delivery: A comprehensive review of Internet of Things and smart grid.* Renewable and Sustainable Energy Reviews, **91**: p. 90-108.
132. Ritchie, H., and Roser, M. 2019. How have global $CO_2$ emissions changed over time? Our World in Data. https://ourworldindata.org/co2-and-other-greenhouse-gas-emissions#how-have-global-co2-emissions-changed-over-time - accessed March 2020.
133. Saberi, S., Kouhizadeh, M., Sarkis, J., and Shen, L., 2019. *Blockchain technology and its relationships to sustainable supply chain management.* International Journal of Production Research, **57**(7): p. 2117-2135.
134. Saleem, Y., Crespi, N., Rehmani, M.H., and Copeland, R., 2019. *Internet of things-aided smart grid: technologies, architectures, applications, prototypes, and future research directions.* IEEE Access, **7**: p. 62962-63003.
135. Saleh, F., 2020. *Blockchain without waste: Proof-of-stake.* Available at SSRN 3183935.
136. Sandvine, 2014. 1h 2014.Global internet phenomena report. Technical Report.Sandvine Incorporated ULC.
137. Sandvine, 2019. *Global Internet Phenomena Report.* https://www.sandvine.com/hubfs/Sandvine_Redesign_2019/Downloads/Internet%20Phenomena/Internet%20Phenomena%20Report%20Q32019%2020190910.pdf - accessed February 2020.
138. Schaffartzik, A., Mayer, A., Gingrich, S., Eisenmenger, N., Loy, C., and Krausmann, F., 2014. *The global metabolic transition: Regional patterns and trends of global material flows, 1950–2010.* Global Environmental Change, **26**: p. 87-97.
139. Schwartz, R., Dodge, J., Smith, N.A., and Etzioni, O., 2019. *Green AI.* arXiv preprint arXiv:1907.10597.
140. SBTi (Science Based Targets), 2019. *Science-Based Target Setting Manual.* https://sciencebasedtargets.org/wp-content/uploads/2017/04/SBTi-manual.pdf - accessed March 2020.
141. SBTi (Science Based Targets), 2020. *SBTi Criteria and Recommendations.* https://sciencebasedtargets.org/wp-content/uploads/2019/03/SBTi-criteria.pdf - accessed July 2020.
142. Scott, M., 2019. *New rules to crack down on 'greenwash' in corporate clean energy claims.* Ethical Corporation. http://www.ethicalcorp.com/new-rules-crack-down-greenwash-corporate-clean-energy-claims - accessed March 2020.
143. Shaikh, F.K., Zeadally, S., and Exposito, E., 2015. *Enabling technologies for green internet of things.* IEEE Systems Journal, **11**(2): p. 983-994.
144. Shehabi, A., Walker, B., and Masanet, E., 2014. *The energy and greenhouse-gas implications of internet video streaming in the United States.* Environmental Research Letters, **9**(5): p. 054007.
145. Simonite, T., 2016. *Moore's Law Is Dead. Now What?* Technology Review. https://www.technologyreview.com/s/601441/moores-law-is-dead-now-what/ - accessed March 2020.
146. Sky, 2020. *Sky Zero.*; Available from: https://www.skygroup.sky/sky-zero - accessed March 2020.
147. Smith, B., 2020. *Microsoft will be carbon negative by 2030.* Microsoft. https://blogs.microsoft.com/blog/2020/01/16/microsoft-will-be-carbon-negative-by-2030/ - accessed March 2020.

148. Solanki, A. and Nayyar, A., 2019. *Green internet of things (G-IoT): ICT technologies, principles, applications, projects, and challenges*, in *Handbook of Research on Big Data and the IoT*. IGI Global. p. 379-405.
149. Sorrell, S., 2009. *Jevons' Paradox revisited: The evidence for backfire from improved energy efficiency.* Energy policy, **37**(4): p. 1456-1469.
150. Sorrell, S., Gatersleben, B. and Druckman, A., 2020. The limits of energy sufficiency: A review of the evidence for rebound effects and negative spillovers from behavioural change. *Energy Research & Social Science*, *64*, p.101439.
151. Statista Research Department, 2020. *Internet of Things (IoT) connected devices installed base worldwide from 2015 to 2025 (in billions)*. https://www.statista.com/statistics/471264/iot-number-of-connected-devices-worldwide/, accessed March 2020.
152. Stephens, A., 2019. *Net zero: an ambition in need of a definition*. Carbon Trust. https://www.carbontrust.com/news-and-events/insights/net-zero-an-ambition-in-need-of-a-definition - accessed March 2020.
153. Stoll, C., Klaaßen, L., and Gallersdörfer, U., 2019. *The carbon footprint of bitcoin.* Joule, **3**(7): p. 1647-1661.
154. Strubell, E., Ganesh, A., and McCallum, A., 2019. *Energy and policy considerations for deep learning in NLP.* arXiv preprint arXiv:1906.02243.
155. Sweney, M., 2020. *Netflix to slow Europe transmissions to avoid broadband overload.* The Guardian. https://www.theguardian.com/media/2020/mar/19/netflix-to-slow-europe-transmissions-to-avoid-broadband-overload - accessed March 2020.
156. Tarnoff, B., 2019. *To decarbonize we must decomputerize: why we need a Luddite revolution.* The Guardian. https://www.theguardian.com/technology/2019/sep/17/tech-climate-change-luddites-data - accessed March 2020.
157. The Economist, 2020. *How to make sense of the latest tech surge.* https://www.economist.com/leaders/2020/02/20/how-to-make-sense-of-the-latest-tech-surge - accessed March 2020.
158. The Shift Project, 2019. *Climate Crisis: The Unsustainable Use of Online Video. The practical case for digital sobriety.* https://theshiftproject.org/en/article/unsustainable-use-online-video/ - accessed March 2020.
159. The Shift Project, 2019. *Lean ICT: Towards digital sobriety.* https://theshiftproject.org/en/lean-ict-2/ - accessed March 2020.
160. Truby, J., 2018. *Decarbonizing Bitcoin: Law and policy choices for reducing the energy consumption of Blockchain technologies and digital currencies.* Energy research & social science, **44**: p. 399-410.
161. UK Energy Research Council, 2007. *The Rebound Effect: An Assessment of the evidence for economy-wide energy savings from improved energy efficiency.* http://www.ukerc.ac.uk/programmes/technology-and-policy-assessment/the-rebound-effect-report.html - accessed March 2020
162. Van Heddeghem, W., Lambert, S., Lannoo, B., Colle, D., Pickavet, M., and Demeester, P., 2014. *Trends in worldwide ICT electricity consumption from 2007 to 2012.* Computer Communications, **50**: p. 64-76.
163. Vaughan, A., 2019. *Cloud gaming may be great for gamers but bad for energy consumption.* New Scientist. https://institutions.newscientist.com/article/2206200-cloud-gaming-may-be-great-for-gamers-but-bad-for-energy-consumption/ - accessed March 2020.

164. Vereecken, W., Deboosere, L., Simoens, P., Vermeulen, B., Colle, D., Develder, C., Pickavet, M., Dhoedt, B., and Demeester, P., 2009. *Energy efficiency in thin client solutions*. in *International Conference on Networks for Grid Applications*. Springer.
165. Waldrop, M.M., 2016. *The chips are down for Moore's Law.* Nature. https://www.nature.com/news/the-chips-are-down-for-moore-s-law-1.19338 - accessed March 2020.
166. Walnum, H.J. and Andrae, A.S., 2016. *The internet: Explaining ICT service demand in light of cloud computing technologies*, in *Rethinking Climate and Energy Policies*. Springer. p. 227-241.
167. Widdicks, K., Hazas, M., Bates, O., and Friday, A., 2019. *Streaming, Multi-Screens and YouTube: The New (Unsustainable) Ways of Watching in the Home*. in *Proceedings of the 2019 CHI Conference on Human Factors in Computing Systems*.
168. Wiebe, K.S., Bjelle, E.L., Többen, J., and Wood, R., 2018. *Implementing exogenous scenarios in a global MRIO model for the estimation of future environmental footprints.* Journal of Economic Structures, **7**(1): p. 20.
169. Wilson, C., Hargreaves, T., and Hauxwell-Baldwin, R., 2017. *Benefits and risks of smart home technologies.* Energy Policy, **103**: p. 72-83.
170. World Economic Forum, 2019. *A New Circular Vision for Electronics. Time for a Global Reboot.* Platform for Accelerating the Circular Economy (PACE). http://www3.weforum.org/docs/WEF_A_New_Circular_Vision_for_Electronics.pdf - accessed March 2020.
171. Wu, J., Guo, S., Li, J., and Zeng, D., 2016. *Big data meet green challenges: Big data toward green applications.* IEEE Systems Journal, **10**(3): p. 888-900.

# Appendix

# A Methodology

## A.1 Definitions

*Table A.1 Definitions for terms used throughout the report. Unless a reference is provided, these are pulled or adapted from the Cambridge Dictionary [2020] or Berners-Lee [2011].*

| Term | Definition |
|---|---|
| 1.5°C | 1.5 degrees Celsius global warming has far fewer climate-related risks in terms of sea level rise, drought, hot weather and precipitation extremes than 2 degrees Celsius. For this reason, world leaders agreed to limit global warming to well-below 2 degrees Celsius and 'in pursuit' of 1.5 degrees Celsius at the 2015 United Nations Climate Change Conference in Paris [IPCC 2018]. |
| 2G/3G/4G/5G | Second, third, fourth and fifth generation communication technology. |
| Artificial Intelligence (AI) | The study of how to produce machines that have some of the qualities that the human mind has, such as the ability to understand language, recognize pictures, solve problems, and learn. |
| Algorithm [in the context of Blockchain/AI/Natural Language Processing] | A set of mathematical instructions or rules that, especially if given to a computer, will help to calculate an answer to a problem. |
| Augmented reality | Images produced by a computer and used together with a view of the real world. |
| Big data | Very large sets of data that are produced by people using the internet, and that can only be stored, understood, and used with the help of special tools and methods. |
| Bitcoin | A type of cryptocurrency. |
| Blockchain | A decentralised algorithm. In the context of cryptocurrencies: a system used to make a digital record of all the occasions a cryptocurrency is bought or sold, and that is constantly growing as more blocks are added. |
| Cap and trading scheme (for carbon) | A cap is set on the total amount of certain GHGs that can be emitted. Within this cap, companies buy or receive emission allowances, which they can trade with one another. At the end of the year, a company must give up enough allowances that cover all its emissions or face a fine. Any spare allowances can be kept to cover future emissions or sold to other companies. |
| Carbon | A shorthand for all the different global-warming greenhouse gases. |
| Carbon footprint | A best estimate for the full climate change impact of something, including all greenhouse gases, expressed in |
| | carbon dioxide equivalent (the amount of carbon dioxide that would have the same impact as the specific greenhouse gas associated with a thing); the central climate change metric. |
| Carbon intensity | The amount of greenhouse gas emissions associated with an activity. |
| Carbon negative | The process by which an activity sequesters more greenhouse gas emissions than are emitted through said activity. |
| Carbon neutral | Releasing no net greenhouse gas emissions into the atmosphere. Typically achieved by reducing emissions and using offsets to counterbalance any emissions generated. |
| Climate change | Changes in the earth's weather, including changes in temperature, wind patterns, and rainfall, especially the increase in the temperature of the earth's atmosphere that is caused by the increase of particular gases, especially carbon dioxide. |
| Cloud computing | The use of services, computer programs, etc. that are on the internet rather than ones that you buy and put on your computer. |
| $CO_2$ | Carbon dioxide, the most common greenhouse gas. |
| $CO_2e$ | Carbon dioxide equivalent. Different greenhouse gases have different global warming potentials. $CO_2e$ expresses the climate change impact of all greenhouse gases emitted in association with an activity as the amount of carbon dioxide that would have the same climate change impact. |
| Cryptocurrency | A digital currency produced by a public network, rather than any government, that uses cryptography to make sure payments are sent and received safely. |
| Data centre | A place where a number of computers that contain large amounts of information can be kept safely. |
| Data science | The use of scientific methods to obtain useful information from computer data, especially large amounts of data. |
| Data traffic/Internet traffic | The activity of data and messages passing through an online communication system or the number of visits to a particular website. |
| Decarbonising | Reducing the carbon footprint of an activity. |
| Dematerialisation | Reducing the amount of material needed to produce a product. |
| Downstream traffic | Data traffic that is moving in a downstream direction (i.e. being downloaded). |
| Economy-wide impacts of ICT | The impact the ICT industry has on other industries, for example through allowing for efficiencies, providing additional products and/or replacing more traditional technologies, but also allowing intensified activity or growth in other areas of the economy. The effect can be both to increase or decrease impact and those other industries. Differentiated from ICT's impact within the ICT industry. |

|  | The net effect of ICT depends on the impact it has in both areas and their balance. |
|---|---|
| Emissions | A shorthand for greenhouse gas emissions. |
| Entertainment and Media (E&M) sector | A sector category used by Malmodin and colleagues; it covers TV, consumer electronics (such as cameras and audio systems in a car and portable GPS) and print media. |
| Environmentally Extended Input Output (EEIO) analysis | A "top-down" approach for estimating life cycle emissions, capable of capturing impacts from the entire supply chain. See Appendix F for details. |
| Embedded device/system | A computer system that does a particular task inside a machine or larger electrical system, or physical object. |
| Embodied carbon/emissions | The greenhouse gas emissions released from the extraction of raw materials required, the manufacturing process and transport and distribution of a product. It includes a share of all the activities required to take goods and services at the point of sale, but excludes the product use phase. It can be from cradle to factory gate, from cradle to site of use or from cradle to grave – in the latter case, end of life emissions are included. In this report, we assume cradle to point of sale unless otherwise stated. |
| Enablement | The avoidance of emissions in the wider economy through ICT applications, including through improved efficiency. |
| End of life emissions (see lifecycle stages) | Emissions after disposal of a product, after the end of the use phase. |
| Energy footprint | The amount of energy used by a product, activity or industry. |
| Exponential growth | A rate of increase which becomes quicker and quicker as the thing that increases becomes larger. |
| Fossil fuel | Fuels, such as gas, coal, and oil, that were formed underground from plant and animal remains millions of years ago. |
| GB (Gigabytes) | A unit of computer information consisting of 1,000,000,000 bytes. |
| Greenhouse gas (GHG) emissions, or emissions for short | Gases that contribute to global warming, including carbon dioxide ($CO_2$), methane (CH4), nitrous oxide (N2O) and fluorinated gases. |
| ICT's own impact | The impact the ICT industry has in terms of its energy use or GHG emissions through the entire lifecycle of its products, including their manufacture, operation and disposal. Differentiated from the effect ICT has on other industries, that is, economy-wide impacts. |
| Information Communication Technology (ICT) | The use of computers and other electronic equipment and systems to collect, store, use, and send data electronically. |
| Internet of Things (IoT) | Objects with computing devices in them that are able to connect to each other and exchange data using the internet. |

| Life Cycle Analysis (LCA) | The detailed study of the series of changes that a product, process, activity, etc. goes through during its existence and the resulting environmental impact. |
|---|---|
| Lifecycle stages (Material extraction, manufacturing, transport, use phase, end of life) | The stages of resource use and environmental releases associated with an industrial system from the extraction of raw materials from the Earth and the production and distribution, through the use, and reuse, and final disposal of a product. |
| Machine-to-Machine (M2M) communication | The act of sending data between machines or computers. |
| Machine Learning (ML) | The process of computers changing the way they carry out tasks by learning from new data, without a human being needing to give instructions in the form of a program. |
| Moore's Law | The observation by Gordon Moore of Intel Corporation that the cost of a computer chip for a particular amount of processing power will continue to fall by half every two years. |
| Natural Language Processing (NLP) | A field of Artificial Intelligence that gives the machines the ability to read, understand and derive meaning from human languages [Lopez Yse 2019]. |
| Net zero | Having no net climate change impact through greenhouse emissions in a company's value chain. This is achieved by reducing greenhouse gas emissions in the value chain and removing the remaining emissions through additional carbon removals. |
| Network | A number of computers that are connected together so that they can share information. |
| Offset | A mechanism to negative a certain amount of GHG emissions either through avoiding emissions elsewhere (e.g. through protecting a forest from logging) or removing emissions form the atmosphere (for example through natural carbon sequestration, such as peatland restoration or reforestation projects). There is debate whether avoided emissions should count as additional and whether they are permanent. |
| Operational emissions | See use phase emissions. |
| Operator activities | Activities by operators of manufacturing plants, data centres and networks, such as office heating and lighting, business travel, maintenance of equipment, a share of which should be allocated to the lifecycle emissions of equipment using data centre and network services. |
| Proof of Work / Proof of Stake | Types of Blockchain consensus algorithms which are processes in computer science used to achieve agreement on a single data value among distributed systems [CoinBundle Team 2018]. |
| Rebound effect | The way in which micro-actions can be nullified by counter balancing adjustments elsewhere in the global system. Often used to refer to the way increased energy efficiency leads to more energy usage overall. |

| Renewable energy | Energy that is produced using the sun, wind, etc., or from crops, rather than using fuels such as oil or coal. |
|---|---|
| Router | A piece of electronic equipment that connects computer networks to each other, and sends information between networks. |
| Scope 1 emissions | Direct emissions from burning of fossil fuels on site (includes company facilities and vehicles) |
| Scope 2 emissions | Indirect emissions from purchased electricity and gas. |
| Scope 3 emissions | All other indirect emissions in a company's value chain; including upstream emissions in the supply chain (e.g. emissions from purchased goods and services, transportation of these goods to the company, use of leased assets such as offices or data centres, business travel and employee commuting) and downstream emissions (from transportation, distribution, use and end of life treatment of sold products, investments and leased assets). Scope 3 emissions form the majority of a company's emissions. |
| Semiconductor | A material, such as silicon, that allows electricity to move through it more easily when its temperature increases, or an electronic device made from this material. |
| Server | A central computer from which other computers get information. |
| Set top box | An electronic device that makes it possible to watch digital broadcasts on ordinary televisions. |
| Smart technology [e.g. smart grids/cities/ logistics/agriculture] | An object/city/process etc. that is internet-connected and therefore able to make intelligent decisions. |
| Supply chain emissions | Emissions that occur upstream of a company's own operations, including emissions from purchased goods and services, transportation of these goods to the company, use of leased assets such as offices or data centres, business travel and employee commuting. See upstream scope 3 emissions. |
| Truncation error | In the context of carbon accounting, truncation error describes the omission of some proportion of the total carbon footprint by LCAs because this approach is unable to track all the supply chain pathways associated with a thing. That means, they disregard or *truncate* the pathways that individually only contribute a small share of the total, often set to less than 1%, even though these can make up sizeable share of the total if they are all added up. See Appendix F for more detail. |
| Upstream traffic | Data traffic that is moving in an upstream direction (i.e. being uploaded). |
| Use phase/Operational emissions | Emissions associated with the use of a product, mainly from energy use and maintenance. |
| Value chain emissions | Emissions occurring in a company's value chain, both upstream in the supply chain (from manufacture and |

| | transport), from its operations and downstream (from product use by customers). |
|---|---|
| Video-on-demand services | Services (e.g. Netflix, Amazon Prime, BBC iPlayer) that provide a system for watching films or recorded programmes on the internet or television at any time. |
| Virtual reality | A set of images and sounds, produced by a computer, that seem to represent a place or a situation that a person can take part in. |
| Virtualisation/ Server virtualisation | The process of changing something that exists in a real form into a virtual version. |

## A.2  Abbreviations

*Table A.2 Abbreviations used throughout the report.*

| Abbreviation | Term |
|---|---|
| AI | Artificial Intelligence |
| A&E | Andrae and Edler (2015) |
| BEIS | Department for Business, Energy and Industrial Strategy |
| B&E | Bekhir and Elmeligi (2018) |
| CEH | UK Centre for Ecology and Hydrology |
| CRT | Cathode Ray Tube |
| DAC | Direct Air Capture |
| EC | European Commission |
| EEIO | Environmentally Extended Input Output |
| GeSI | Global e-Sustainability Initiative |
| GHG | Greenhouse gas emissions |
| GSMA | Global System for Mobile Communications Association |
| ICT | Information and Communication Technology |
| IoT | Internet of Things |
| ITU | International Telecommunication Union |
| LCA | Life Cycle Analysis |
| LED | Light-Emitting Diode |
| M&L | Malmodin and Lundén (2018) |
| PC | Personal Computer |
| PPA | Power Purchasing Agreements |
| REGO | Renewable Energy Guarantees of Origin |
| SBTi | Science Based Targets initiative |
| SWC | Small World Consulting |
| TV | Television |
| UN | United Nations |

## A.3 Units

Emissions are measured in kilograms (kg), tons (t), kilotons (kt), megatons (Mt) and gigatons (Gt). GHG emissions, for example, are often expressed in $MtCO_2e$ or $GtCO_2e$.

1 Gt = 1,000 Mt; 1 Mt = 1,000 kt; 1 kt = 1,000 t; 1 t = 1,000 kg

Energy consumption is measured in watt-hour (Wh), kilowatt-hour (kWh), megawatt -hour (MWh), gigawatt -hour (GWh) and terawatt -hour (TWh).

1 TWh = 1,000 GWh; 1 GWh = 1,000 MWh; 1 MWh = 1,000 kWh; 1 kWh = 1,000 Wh

## A.4 Scope

For the purposes of this report, we have adopted a broad definition of ICT to include all types of data centres, networks and user devices used for processing, storing, sending and receiving digital information. This includes data centres of all scales (i.e. servers run by companies in cupboard up to large data centres), all major types of networks (telephony, mobile and broadband data, TV), and a wide range of digital end user devices, such as PCs, laptops, tablets, mobile and fixed phones, TVs, displays and gaming equipment (see [Appendix B.2.1](#)). We included all stages of equipment lifecycle, from the extraction of the raw materials, manufacture, transport and use to end of life. For networks and data centres, we included infrastructure (such as the construction and running of the building housing the servers, including cooling, and the digging down of network cable tracks) and operator activities (e.g. business travel, office heating and lighting etc.).

## A.5 Method

For the literature review, we built on our collective knowledge of the literature and carried out additional literature searches using Google Scholar, the ACM Digital Library and the citation information from relevant papers. Note that this was not a systematic literature review. For the main review of ICT's carbon impact ([Section 2](#)), we included peer-reviewed journal articles published from 2015 onwards with the key words outlined below. These key words were also drawn upon to facilitate our analysis of the trends in ICT and their environmental implications ([Section 3](#)). For our policy analysis ([Section 4.1](#)), we focused solely on European Commission documents and websites. Our analysis of industry pledges ([Section 4.2](#)) draw on a survey of annual reports, blog posts and web pages for 18 major ICT companies (Microsoft, Sky, Vodafone, Apple, Amazon, Netflix, Facebook, Tesla, Google, Samsung, Ericsson, Spotify, Huawei, Cisco, Sony, Nintendo, Intel and IBM).

### A.5.1 Key words
- Sustainability
- Energy consumption/ energy
- Carbon emissions
- GHG (Greenhouse gas) emissions
- LCA (Life-Cycle Analysis)
- Efficiency
- ICT/ IT (information technology)/ digital technologies

- User devices
- Internet traffic/ Internet
- Data traffic
- Data centres/ data centers
- Communication networks/ networks
- Big data
- AI (Artificial Intelligence)
- Machine learning
- Data science
- IoT (Internet of Things)
- Smart (home, grid, city)
- Cryptocurrencies
- Blockchain
- Bitcoin
- Video streaming/ video
- YouTube
- Video-on-demand
- TV/ television
- Cloud computing/ services
- Jevons Paradox
- Rebound effect

### A.5.2 Selection of key papers

Articles were selected guided by the following questions:
- Does the paper focus on the energy or carbon impacts of ICT, its major components (e.g. data centres, networks), or its major application areas (e.g. AI, IoT)?
- Does the paper focus on the impact ICT has on energy or carbon consumption in other sectors?

### A.5.3 Consultation with key experts

In addition to this, we consulted with the following leading experts based on their extensive knowledge on the carbon impacts of ICT through video conference calls:
- Dr. Lotfi Belkhir (Associate Professor at W Booth School of Engineering Practice and Technology, McMaster University)
- Dr. Anders Andrae (Senior Expert at Huawei Technologies)
- Jens Malmodin (Senior Specialist at Ericsson)
- Dr. Peter Garraghan (Associate Professor at the School of Computing and Communications, Lancaster University)
- Livia Cabernard (PhD student at the Institute of Science, Technology and Policy, ETH Zurich)
- Prof. Chris Preist (Professor of Sustainability and Computer Systems at University of Bristol)

We discussed their research in relation to ICT's carbon footprint, their opinion of other prominent studies, their response to criticism from the other experts, their view on the future of ICT's emissions and on the trends posing risks and opportunities for ICT's impact on climate change.

### A.5.4   Other sources of information

For this report, we drew on research by Small World Consulting (SWC) Ltd. into sector emissions to adjust estimates by the key studies in [Section 2](#) for truncation error.

SWC developed an environmentally extended input output (EEIO) model (described in detail by Berners-Lee et al. [2011] and Kennelly et al. [2019]) that uses data from the Office of National Statistics on the expenses and GHG emissions from 105 industries in the UK to calculate the carbon intensity per Pound spent. This allows us to model carbon flows in the UK economy and the upstream scope 3 emissions of an industry in its supply chain, by tracking the economic activity stimulated by each sector in other sectors. In contrast to LCAs, SWC's EEIO model tracks 100% of all supply chains associated with a sector. It can be used to estimate the truncation error of LCAs for a particular sector; that is, the percentage of the total emissions that is typically omitted by an LCA. We note that SWC's EEIO model is based on UK emissions data which are not representative of other economies, yet it provides a good–enough estimate to help understand the potential truncation error incurred by LCA estimates ([Appendix F](#)).

For manufacture of ICT equipment, these omissions include radiative forcing, manufacture of buildings and machines, of mining equipment and of transport vehicles and other operator activities and overheads associated with the manufacture of a product. We also know that most LCAs do not include pathways that contribute less than 1% of the total carbon footprint. In total, these excluded pathways make up 40% of the total embodied carbon.

For the operation of ICT equipment, electricity is the most important source of GHG emissions. Based on a hybridised EEIO-LCA model SWC developed from scope 1 and 2 emissions data from BEIS [Department for Business, Energy and Industrial Strategy 2019] and the IEA [2019b], SWC estimates that the carbon intensity of global average grid electricity in 2018 was 0.63 kg$CO_2$e/kWh or Mt$CO_2$e/TWh. The carbon intensity factor for electricity used in most LCAs includes emissions from electricity generation and transmission and distribution losses, but not extraction and transportation of fuel to the plant, the manufacture of equipment used in these processes and operator activities. Based on this, we estimate that LCAs omit 18% of the use phase carbon. Our truncation error mark-up applied to the key studies reviewed here is based on the difference between the specific electricity intensity factor they report and SWC's factor of 0.63 kg$CO_2$e/kWh.

All percentages out of global GHG emissions are based on a total of 57.9 Gt$CO_2$e in 2020. This is based on 55.6 Gt$CO_2$e GHG emissions, including land use change, in 2018 [Olivier and Peters 2019], assuming the growth rate of 2% in 2018 applies to 2019 and 2020. Note that this extrapolation did not consider the impact of Covid-19 on emissions.

## A.6   Limitations

This report is not based on a systematic literature review but rather built on our own knowledge of the sector alongside strategic literature searches aimed at covering the main studies in the field. We have focused on critically analysing the main arguments surrounding the ICT sector's environmental footprint and trends. We have not scrutinised reports about the impact of individual components of ICT (e.g. the carbon impacts of servers alone) or covered the full breadth of research papers within ICT on IoT, Blockchain and AI that do not take an environmental position (e.g. instead focus solely on health, finance, etc.). We have limited our discussion of impacts on the wider economy to the ICT sector's potential to enable efficiencies or drive emissions in other sectors; a full, economy-wide assessment of ICT's environmental impacts globally was deemed out of scope for this study. We are confident, however, that we have captured the main academic debates and the most relevant non-academic publications on the climate change impact of the ICT sector as a whole and the impacts of prominent ICT trends going forward. We call for future work to fully assess the Enablement and Global Rebounds narratives (see Figure 2.5) on the world's economy.

Carbon accounting is a rather imprecise science due to the complexity of the supply chain emissions pathways and issues with how to allocate emissions to a particular product, activity or sector. For each carbon footprint calculation, there is a margin of error. The uncertainty increases even further for projections of future emissions, as these are influenced by the actions of companies, policy makers, individual users and unforeseen events like natural catastrophe and pandemics. There are several unknowns including what changes future innovations might bring or the carbon footprint of activities which are largely undocumented (e.g. the dark web). The carbon footprint of some of the emerging ICT trends are also difficult to calculate, e.g. IoT and Blockchain due to their hidden and distributed nature.

We have tried to make this uncertainty clear throughout the report. The carbon footprints calculations in this report serve as approximations indicating the order of magnitude and important trends in emissions that can guide decision-making about the effects that different courses of action could have on climate change. Furthermore, the lack of coherent standards for carbon accounting leads to different approaches, scopes and assumptions being used by different studies. We have attempted to make these explicit and compare the different methodologies used by the key studies reviewed in Section 2.2.

For reasons outlined above, emission estimates are more uncertain than estimates of electricity consumption. Nevertheless, an assessment of ICT's climate change impact needs to focus on GHG emissions rather than electricity consumption alone because it is emissions that ultimately drive climate change, and electricity consumption itself does not capture the impact of factors such as energy source mix and emissions in the energy generation supply chain. Since most studies focus on ICT's energy consumption, we felt that we could most usefully contribute to the scientific debate by applying our expertise in supply chain emission accounting to clarify some of the complexities around the emission footprint from energy use and other sources of GHG.

# B Estimates of ICT Emissions

## B.1 Historical Estimates of ICT's GHG footprint

*Table B.1 Historical estimates of ICT's GHG footprint. Unless otherwise stated, all estimates include embodied (based on LCAs) and use phase GHG emissions.*
*\*Based on 670 TWh in 2007 and 930 TWh in 2012 [Lannoo et al. 2013] and 0.68 MtCO$_2$e/TWh (SWC estimate).*
*\*\*Based on 655 TWh in 2007 and 909 TWh in 2012 [Van Heddeghem et al. 2014] and 0.68 MtCO$_2$e/TWh (SWC estimate).*

| Study | Year | MtCO$_{2e}$ | Scope for emissions |
|---|---|---|---|
| Gartner [2007] | 2007 | 620 | CO$_2$ emissions only; use phase and emissions for phones, PCs, printers, data centres and networks |
| GeSI [2008] | 2002 | 530 | Desktop PCs and laptops and PC peripherals (monitors, printers), data centres, telecoms networks and devices |
|  | 2007 | 830 |  |
|  | 2020 | 1430 |  |
| Malmodin et al. [2010] | 2007 | 1,150 | Phones, PCs, modems, networks and data centres (630 MtCO$_2$e); TVs, TV peripherals and TV networks (390 MtCO$_2$e); other E&M equipment, including audio devices, cameras and gaming consoles (130 MtCO$_2$e) |
| GeSI [2012] | 2011 | 910 | PCs (desktops, laptops), mobile devices (tablets, smartphones, regular mobile phones), and peripherals (external monitors, printers, set-top boxes, routers, IPTV boxes); fixed and wireless networks (excluding local WiFi networks), data centres (servers, storage and cooling) |
|  | 2020 | 1270 |  |
| Lannoo et al. [2013]* | 2007 | 454 | Emissions from electricity and use phase only; computers, data centres, networks |
|  | 2012 | 630 |  |
| Malmodin et al. [2013] | 2020 | 2,200 | Phones (fixed, mobile), PCs (desktops, laptops), modems, networks and data centres (1,100 MtCO$_2$e); TVs, TV networks and TV peripherals (1,100 MtCO$_2$e); other E&M equipment, including audio devices, cameras and gaming consoles (420 MtCO$_2$e) |
| Van Heddeghem et al. [2014]** | 2007 | 444 | Emissions are use phase electricity only; desktops, laptops, monitors, networks and data centres |
|  | 2012 | 616 |  |
| Malmodin [2019] | 2010 | 720 | Phones (fixed, smartphones, other mobile), tablets, PCs (desktops, laptops), displays, modems, some IoT, networks and data centres |

## B.2 Detailed Review of the Key Studies

This report focuses on reviewing peer-reviewed studies by three main research groups published from 2015 that estimate ICT's carbon footprint from 2015 onward. Here, we include a summary of the studies scope and assumptions (B.2.1), then follow with an overview of estimates (B.2.2) and a detailed review of relevant studies by researchers around Andrae (B.2.3), Belkhir (B.2.4) and Malmodin (B.2.5).

### B.2.1 Overview of scope and methodological differences

Studies on the energy and carbon footprint of IT can be classed as either bottom-up (based on LCAs, energy use reports for certain devices and company reports, combined with data on the number of devices produced and used in a given year and the number of network subscriptions), or top-down (based on national or global statistics and input-output analysis). The latter is often difficult to obtain. Most studies use a bottom-up approach in combination with some top-down data, for example combining LCAs for user devices with global statistics for data traffic, such as from Cisco. Using a combined method is probably the best approach to assess emissions accurately.

The studies reviewed for this study use different methodological approaches. Some only include emissions from electricity (e.g. A&E), presenting a more limited picture, while others also include other sources of GHGs (e.g. B&E; Malmodin's research), such as fossil fuel backup power for data centres, fuels used by vehicles and other sources of emissions in the process of mining.

All the key studies include use phase emissions but studies vary as to the other lifecycle stages considered. On one end of the spectrum, in addition to the use phase, A&E only include production energy, just one aspect of embodied emissions. On the other end, Malmodin's research includes end of life emissions, that is, the emissions associated with waste management. The stages of the equipment lifecycle covered by the different studies in this review as well as the scope and assumptions applied are summarised in Table B.*2*.

*Table B.2 Scope matrix for studies included in this review. T&D = Transmission and distribution losses in electricity grids. Note that Malmodin includes some 'Other digital technologies or trends', specifically: wearables such as smart watches and fitness trackers, smart energy meters, control units, surveillance cameras, public displays, payment terminals and the internet-connected communication device in vending machines.*
*\*Included in the E&M sector estimates, not ICT estimates.*

| Component of ICT sector | Andrae and Edler (2015) | Belkhir and Elmeligi (2018) | Malmodin and Lundén (2018) |
|---|---|---|---|
| **User devices** | | | |
| Smartphones | ✓ | ✓ | ✓ |
| Nonsmart mobile phones | ✓ | ✗ | ✓ |
| Fixed phones | ✗ | ✗ | ✓ |
| Tablets | ✓ | ✓ | ✓ |
| Phablets | ✓ | ✓ | ✗ |
| Laptops/Notebooks | ✓ | ✓ | ✓ |
| Desktop PCs | ✓ | ✓ | ✓ |
| Displays | ✓ | ✓ | ✓ |
| Computer peripherals (e.g. mouse and keyboard) | ✗ | ✗ | ✓ |
| Projectors | ✗ | ✗ | ✓* |
| Cameras | ✗ | ✗ | ✓* |
| Home media players/audio systems/traditional speakers | ✓ | ✗ | ✓* |
| Portable media players, e.g. iPods | ✗ | ✗ | ✓* |
| Smart speakers | ✗ | ✗ | ✗ |
| Smart watches/fitness trackers | ✗ | ✗ | ✓* |
| Headphones/Earphones | ✗ | ✗ | ✓* |
| Game consoles | ✓ | ✗ | ✓* |
| Arcade game machines | ✗ | ✗ | ✓* |
| Customer premises equipment (routers, modems) | ✓ | ✓ | ✓ |
| | | | |
| **Networks** | | | |
| Fixed telephony | ✓ | ✓ | ✓ |
| Mobile | ✓ | ✓ | ✓ |
| Fixed access wired | ✓ | ✓ | ✓ |
| Fixed access WiFi | ✓ | ✓ | ✓ |
| Enterprise networks | ✗ | ✓ | ✓ |
| Lower power, lower bandwidth device networks for IoT | ✗ | ✗ | ✗ |
| | | | |
| **Data centres** | | | |
| Servers | ✓ | ✓ | ✓ |
| Buildings that house servers | ✗ | ✓ | ✓ |
| Cooling | ✓ | ✓ | ✓ |
| Backup power supplies | ✗ | ✓ | ✓ |
| Operator activities, such as offices, business travel, maintenance of equipment | ✗ | ✗ | ✓ |
| | | | |
| **TVs, TV peripherals and TV networks** | | | |
| TVs | ✓ | ✗ | Yes* |
| Set top boxes | ✓ | ✗ | Yes* |
| Aerials | ✗ | ✗ | - |
| Satellite dishes | ✗ | ✗ | Yes* |
| DVD/BD players | ✓ | ✗ | Yes* |
| TV networks | ✗ | ✗ | Yes* |
| >>Cable | ✗ | ✗ | Yes* |
| >>Satellite | ✗ | ✗ | Yes* |
| >>DTT | ✗ | ✗ | Yes* |
| | | | |
| **Other digital technologies or trends** | | | |
| Cryptocurrencies/Blockchain | ✗ | ✗ | ✗ |
| AI/Machine Learning | ✗ | ✗ | ✓ |
| IoT | ✗ | ✗ | ✓ (some) |
| Satellites | ✗ | ✗ | ✗ |
| Radio (device+networks) | ✗ | ✗ | ✗ |
| Embedded devices, e.g. sensors for smart cities, smart home tech, M2M communication | ✗ | ✗ | ✓ (some) |
| Private internet, e.g. for military purposes | ✗ | ✗ | ✗ |
| | | | |
| **Trends considered for future projections** | | | |
| Blockchain | ✗ | ✗ | ✗ |
| Artificial Intelligence/Deep learning/Machine Learning | ✓ | ✗ | ✗ |
| IoT | ✓ | ✗ | ✓ |
| Video | ✓ | ✗ | ✗ |
| | | | |
| **Assumptions** | | | |
| Electricity carbon intensity (kgCO$_2$e/kWh) | Varies by scenario; 0.61 in 2015; 0.6-0.61 in 2020; 0.55-0.65 in 2030 | 0.5 | 0.6 |
| Aspects of electricity likely covered (inferred from number) | generation, well-to-tank and T&D losses | generation only | generation, well-to-tank and T&D losses |
| Use phase included | Yes | Yes | Yes |
| Embodied included (based on LCAs) | Yes | Yes | Yes |
| Embodied carbon included | Production electricity only; no transport or end of life considered and no other sources of GHG other than electricity | Material extraction and manufacturing energy, not transport and end of life | Material acquisition, parts and component production and assembly, transport and end of life |

### B.2.2 Estimates for ICT's GHG Emissions in 2015 and 2020

Table B.**3** below estimates summaries by the key studies for 2015 and 2020.

*Table B.3 Estimates of GHG emissions from the ICT sector in a) 2015 and b) 2020.*

| GHG emissions from ICT in 2015 (MtCO$_2$e) | User devices | Data centres | Networks | Total without TV | TVs | Total with TV |
|---|---|---|---|---|---|---|
| M&L 2015 | 395 | 160 | 180 | 733 | 420 | 1,153 |
| B&E 2015 minimum | 290 | 281 | 204 | 775 | - | N/A |
| B&E 2015 maximum | 485 | 281 | 204 | 971 | - | N/A |
| B&E 2015 average (calculated) | 388 | 281 | 204 | 873 | - | N/A |
| A&E 2015 best case | 186 | 213 | 190 | 589 | 329 | 917 |
| A&E 2015 expected case | 324 | 441 | 287 | 1,052 | 463 | 1,515 |
| A&E 2015 worst case | 514 | 582 | 454 | 1,550 | 706 | 2,257 |
| GHG emissions from ICT in 2020 (MtCO$_2$e) | User devices | Data centres | Networks | Total without TV | TVs | Total with TV |
| Malmodin 2020 | 392 | 127 | 168 | 690 | 400 | 1,090 |
| B&E 2020 minimum | 343 | 495 | 269 | 1,107 | - | N/A |
| B&E 2020 maximum | 542 | 495 | 269 | 1,306 | - | N/A |
| B&E 2020 average | 443 | 495 | 269 | 1,206 | - | N/A |
| A&E 2020 Best case | 201 | 216 | 206 | 623 | 264 | 887 |
| A&E 2020 Expected case | 369 | 448 | 631 | 1,448 | 413 | 1,860 |
| A&E 2020 Worst case | 790 | 1,001 | 1,251 | 3,042 | 711 | 3,634 |

The studies vary in the scope with B&E only including user devices, data centres and networks, A&E including TVs and M&L including other consumer electronics such as cameras and audio systems in a car and portable GPS. In order to make estimates more comparable, we have brought them to the same 'system boundary', by adding Malmodin's estimate for the E&M sector (400 MtCO$_2$e; excluding print media) to B&E's and A&E's estimates (after subtracting emissions from TV from A&E's total estimates, using information provided in their supplementary information). The results are shown below. Considering that Andrae judges his Best case to be most realistic for 2020 [personal communication], the most likely range is 1.0-1.7 GtCO$_2$e for ICT, TVs and other consumer electronics in 2020; this is 1.8-2.9% of global GHG emissions.

*Table B.4 Estimates of GHG emissions from the ICT sector in 2020 after adjusting for scope to include TVs and other consumer electronics.*

|  | MtCO$_2$e | Share of total GHG emissions |
|---|---:|---:|
| Malmodin 2020 | 1,090 | 1.9% |
| B&E 2020 Minimum | 1,507 | 2.6% |
| B&E 2020 Maximum | 1,706 | 2.9% |
| A&E 2020 Best case | 1,023 | 1.8% |
| A&E 2020 Expected case | 1,848 | 3.2% |
| A&E 2020 Worst case | 3,442 | 5.9% |

### B.2.3 Research by Andrae and colleagues

**Approach**

Andrae and Edler [2015], from here on A&E, used a hybrid top-down bottom-up approach to model the production and use phase electricity use of user devices, networks, data centres and TVs between 2010 and 2030. User device emissions are modelled bottom-up from predicted production numbers and estimates for production and use phase energy use derived from LCAs. Estimates for the use phase electricity consumption of data centre and network are based on top-down data traffic trends based on Cisco data and estimates for electricity per data unit from the literature, while estimates for production electricity use are based on a fixed share of total electricity use by networks and data centres (5%, 10% and 15% for best, expected and worst case, respectively) – a method that seems somewhat imprecise. Their model also considers changes in energy efficiency (1% annually in the worst case, 3% in the expected case and 5% in the best case) and in electricity carbon intensity based on projected share of renewables which vary by year and scenario.

**Findings**

A&E's estimates for 2030 vary by a factor of 13, yet all scenarios show an increase relative to 2020 (see Figure B.**1** Andrae and Edler's projections for GHG emissions from ICT by year.). While a growth trend in data traffic underlies the increase in total emissions, the large uncertainty in the size of this trend leads to the wide range of estimates. While the footprint of user devices is becoming less important, partly due to a shift from desktops and laptops to smaller devices like smartphones, and networks and data centres will contribute an increasing share of the total emissions over the next decade, due to the increase in data traffic. A&E argue that this growth in data traffic is driven by the popularity of video streaming, especially over mobile data, and emerging new data-heavy technologies, such as cloud computing. In more recent papers [2019a, 2019b, 2019c], Andrae also argues that AI and deep learning, IoT, Blockchain, virtual and augmented reality, facial recognition, and the rollout of 5G could lead to an explosion of data traffic over the next decade. In addition, IoT devices could increase the production footprint of ICT.

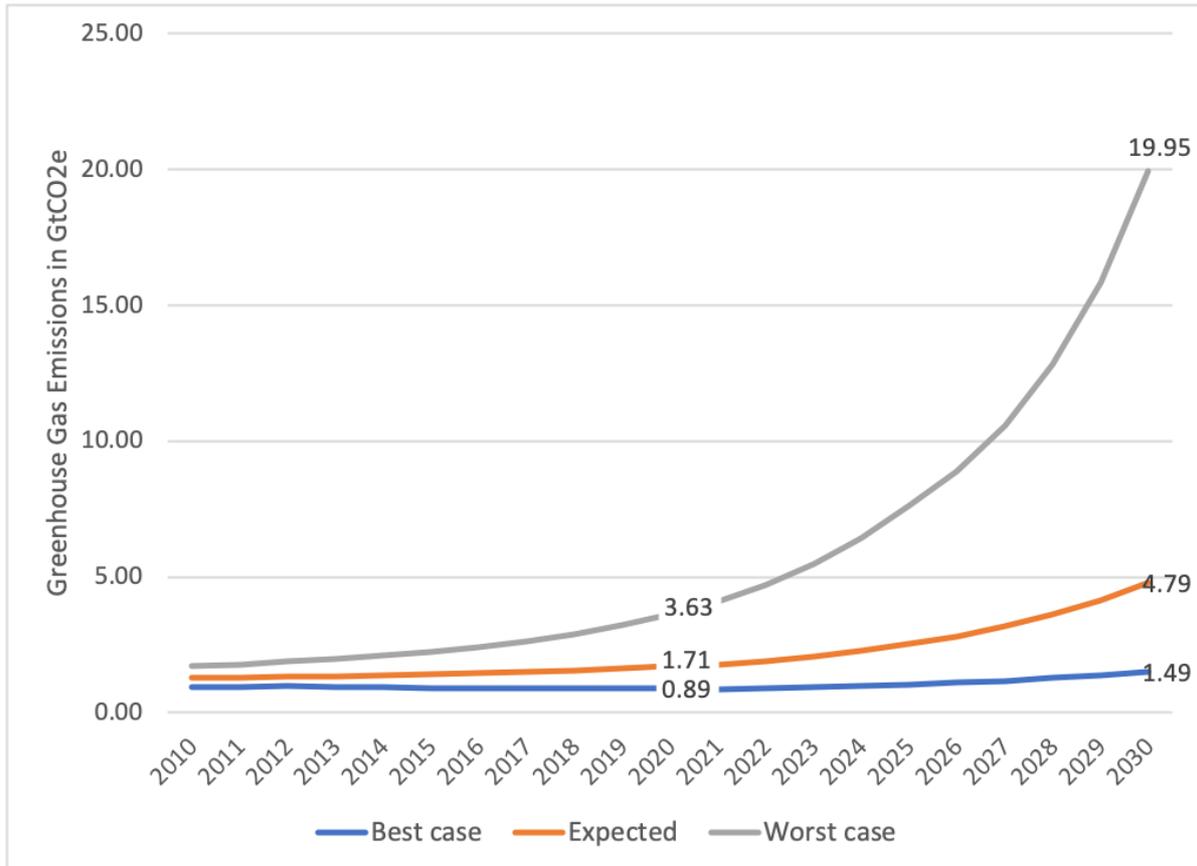

*Figure B.1 Andrae and Edler's projections for GHG emissions from ICT by year.*

A&E's worst case paints a dark picture and has been criticised as unrealistic by B&E and Malmodin [personal communication]. In personal communication, Andrae noted that A&E's study overestimated the carbon footprint of fixed wired and WiFi networks quite much even in the best case but underestimated mobile networks. Thus, for wireless and fixed access, the best case is the most relevant while for mobile, the expected case is the most likely. In 2019c, Andrae also notes that he overestimated production electricity of networks and data centres and that the ratio of production electricity to use phase electricity should be 2% instead of 5% used for the best case. Andrae's revised estimates for 2020 and 2030 (see Table B.**5**) are close to the best case scenario, partly thanks to increasing awareness of ICT's large energy footprint. But while ICT is saving electricity, those savings are used for further expansion, such as in cloud computing. If the trends discussed above take off unexpectedly, Andrae believes that data centre electricity use could be more than 4000 TWh in 2030 [personal communication].

*Table B.5 Andrae's revised estimates for 2020 and 2030 [personal communication], based on Andrae [2019b, 2019c, 2020]. Andrae uses an electricity carbon intensity of 0.55 MtCO$_2$e/TWh in 2020 and 0.54 MtCO$_2$e/TWh in 2030. Consumer devices including WiFi modems and TVs. *Use phase only*

| Year | 2020 | | | 2030 | | |
|---|---|---|---|---|---|---|
| Metric | TWh | Range of MtCO$_2$e | Avg. MtCO$_2$e | TWh | Range of MtCO$_2$e | Avg. MtCO$_2$e |
| Consumer devices* | 600-1000 | 330-550 | 440 | 400-1000 | 216-540 | 378 |
| Networks* | 200-270 | 110-149 | 129 | 330-870 | 178-470 | 324 |
| Data centres* | 290-300 | 160-165 | 162 | 600-1000 | 324-540 | 432 |
| Production of the above | 250-380 | 138-209 | 173 | 180-300 | 97-162 | 130 |
| | **1300-1900** | **715-1045** | **880** | **1500-3200** | **810-1728** | **1269** |

In summary, Andrae believes that ICT's carbon footprint will continue to grow if not for major breakthroughs, albeit at a lower rate for the next few years than previously estimated. While the absolute total will increase, ICT's share of global electricity might stay stable if there are interventions or breakthroughs but this is unlikely under business as usual. Emissions might only reduce if data centres and production facilities are run entirely on renewable energy and if data intensity grows slower than expected [Andrae 2020]. Andrae [2020] notes that there are several potential 'engineering tricks' that might uphold efficiency gains even though Moore's Law has ended, such as decreasing semiconductor use stage power and nanophotonics, but it is unclear to what extent they have already been exploited.

**Limitations and Criticism**

One important difference to the other studies reviewed here is that only production electricity is included but not transport and material extraction, nor other sources of GHG other than electricity. Using the example of smartphones, they estimate between 18.8 (best case) and 37.5 (worst case) kgCO$_2$e in 2010 but with efficiency gains, this decreases to 14.1-35.0 kgCO$_2$e in 2015 and 10.6-33.0 kgCO$_2$e in 2020. The assumption of decreasing embodied emissions for devices is problematic because smartphones are getting bigger and computationally more powerful, counteracting efficiency gains and greening of the electricity grid. In comparison, M&L estimated 45 kgCO$_2$e for the average smartphone embodied footprint and B&E estimated 24.5-45.3kg. Thus, the study likely underestimates the embodied carbon footprint of ICT.

A&E's study has also been criticised by B&E for using a variety of device lifetimes in their scenarios that are not based on the published literature (e.g. 1, 2 and 3 years for smartphones, nonsmartphones, phablets and tablets for worst, expected and best case, respectively), thereby increasing the variance of embodied emission estimates.

In their calculations, A&E use the same number of device units for both the production and operational energy. In a sense, they are calculating how much electricity was used to produce all devices in use in a given year, regardless of when these devices were produced. That assumption is flawed as the same energy efficiency assumptions are applied to all devices used in a given year, even though older devices produced in earlier years will have not benefitted from these efficiency gains. Their figures might therefore underestimate the production energy for user devices further.

However, this paper has to be credited as the most transparent of all the papers reviewed, as the authors lay out clearly their assumptions and calculations in the supplementary information, broken down to the individual device and network type for every year between 2010 and 2030. It is also well-grounded in the previous published literature. The biggest criticism is probably the wide range of projections which leads to a difference between the best and worst case by a factor of 13.4; that said, a high degree of uncertainty does exist especially in such a rapidly developing sector as ICT.

### B.2.4 Research by Belkhir and colleagues

**Approach**

Belkhir and Elmeligi (2018), from here on B&E, used a bottom-up approach for user devices and a top-down approach for data centres and networks and for total projections beyond 2020. The footprint of user devices is calculated by multiplying the number of phones sold in a given year by the lifecycle annual emissions, including embodied carbon spread over the expected lifetime and annual electricity consumption. One of the strengths of this study lies in the systematic review of useful life estimates in the literature for user devices which is more rigorous and provides a smaller range of estimates for user device embodied footprints than A&E more arbitrary useful life estimates. Data centres estimates are based on data from 2008 by Vereecken et al. [2009] and network estimates are based on 2008-2012 data from Van Heddeghem et al. [2014] which B&E projected to grow linearly.

B&E provide a breakdown of emissions for user devices, data centres and networks for each year 2007-2020, but model only the total carbon footprint of ICT from 2020 to 2040 by fitting a linear and an exponential growth curve to their estimates of total emissions from 2007 to 2020.

**Findings**

B&E estimate ICT's footprint in 2020 at between 1.11 GtCO$_2$e (minimum) and 1.31 GtCO$_2$e (maximum). The range is considerably smaller than in A&E and is due to uncertainties in the carbon footprint of user devices, mainly desktops and displays. Data centre and network estimates are the same for both minimum and maximum estimates. The authors suggest that including TVs could add another 435 TWh for operational energy use alone (assuming a global electricity carbon intensity of 0.6 MtCO$_2$e/TWh, this would add 261 MtCO$_2$e), assuming a 2% growth per year in number of TVs.

With regards to projections beyond 2020, both the linear and exponential curve show an increase (see Figure **2.4** in Section 2.1.2). The authors note that exponential growth, which would lead to between 2.48 and 2.62 GtCO$_2$e in 2030 and 5.1 and 5.3 GtCO$_2$e in 2040, is the most realistic, and that growth is highly likely if business as usual continues. These predictions necessarily assume that trends active over the last decade continue for the next two decades and they assume unchecked growth. With an explosion of data traffic driven by trends like AI, Blockchain and IoT and the slowing down of efficiency improvements, there could be an additional jump in ICT's emissions within the next 3-5 years and an overall higher growth in emissions than modelled in their projections. However, while data centres will likely increase in power consumption, they might decrease in GHG emissions, if the trend of powering them with renewable energy continues [Belkhir in personal communication].

The predictions beyond 2020 are limited to totals and are not broken down by component but in their paper, the authors discuss the trend of wireless and mobile communications, cloud-based computing and IoT driving increases in data centres and networks. One of B&E main findings is the disproportionate impact of smartphones, whose footprint they estimate at 125 MtCO2e in 2020.[30] Most of this is due to their embodied footprint which is a concern in combination with their short average lifetime of 2 years. As many online data-heavy activities such as social media and video streaming are accessed by consumers on their mobile phones, the emissions associated with the data centres behind platforms like Facebook can be seen as a knock-on effect of mobile phone usage [Belkhir in personal communication].

**Limitations and Criticism**
B&E included the production and operational energy of ICT for user devices but they only considered the operational energy of data centres and networks, ignoring their embodied carbon because they found it to be negligibly small and excluding operator activities, potentially leading to a slight underestimate of total emissions.

B&E do not consider efficiency improvements in their estimates. For user device footprints this assumption might hold approximately as devices are 'upgraded' with more functionalities and a correspondingly higher footprint. However, for data centres, their estimates are based on extrapolation from 2008 emissions from data centres with a power usage effectiveness of 2, much higher than most modern data centres, without adjusting for efficiency improvements. This might explain why B&E's data centre estimate of 495 MtCO$_2$e in 2020 is at the higher end relative to the other studies discussed here. Belkhir himself noted that their projection for data centres in 2020 is overestimated as efficiency improvements have unexpectedly been able to keep up with growing demand, even though this counterbalancing effect will soon come to an end as efficiency improvements slow down according to Belkhir.

While transparent about their sources, all peer-reviewed articles and publicly accessible industry reports, they did not make available supplementary data with the raw data for the total carbon footprint of ICT (with the exception of figures for 2020, 2030 and 2040) and user devices by year or their calculations. However, in personal communication with the lead author, we were able to get access to the raw data.

---

[30] This is based on 3.6 billion smartphones, including phablets, in 2020. Note that B&E did not include traditional mobile phones or fixed phones in the study's scope. B&E estimate 5.6 billion mobile phones to be in operation in 2030 and 8.7 billion in 2040, although data from Cisco [2020] suggests that the number of mobile phones could rise to 8.3 billion in 2023 already so there might actually be an even steeper rise in carbon emissions from smartphones.

B.2.5  Research by Malmodin and colleagues

**Approach**

Malmodin and Lundén [2018], hereafter M&L, use a hybrid top-down/bottom-up approach and draw on primary industry data from major manufacturers, sales statistics and LCAs for equipment. Emissions of user devices were modelled: 1) bottom-up based on the number of shipped and in-use devices in 2015 and the embodied and use phase emissions per unit estimated in LCAs, and 2) top-down the energy and carbon footprints of 35 major ICT and E&E manufacturers (reported in the supplementary materials) which were extrapolated based on those companies' share of revenue. Embodied footprints of user devices are the most uncertain part of their study.

Network footprints are based on a top-down analysis of network electricity consumption published by the authors [Malmodin and Lundén 2018b] which draw on anonymised operator data covering 70% of subscriptions, which was extrapolated to the global level. Data centre emissions are based on a mix of public and anonymised operator data. The authors note that it was not easy to get hold of primary data from networks and data centres operators because it is considered competitive. The operator data are therefore anonymised; however, they report data collected from public reports in the supplementary materials.

M&L only estimated ICT's global emissions for 2015 but in personal communication, Malmodin has shared his more recent and yet unpublished estimates for 2018 and 2020 with us (see Table B.**6** and Table B.**7**) which follow the same approach as M&L but are based on more recent operator data. We are also drawing on a presentation given by Malmodin at Energimyndigheten in 2019 [Malmodin 2019].

*Table B.6 Breakdown of ICT's carbon footprint as provided by Malmodin [personal communication].*

| MtCO$_2$e | 2015 | 2018 | 2020 |
|---|---|---|---|
| ICT without TV | 730 | 705 | 690 |
| Networks | 182 | 173 | 168 |
| Data centres | 141 | 129 | 127 |
| Enterprise networks | 17 | 16 | 15.5 |
| User devices | 392 | 380 | 375 |
| TVs, TV networks and other consumer electronics | 420 | N/A | 400 |
| ICT with TV | 1,153 | N/A | 1,087 |

*Table B.7 Operational electricity consumption by the ICT industry as provided by Malmodin [personal communication]. The total is adjusted for double counting.*

| Operational TWh | 2015 | 2018 | 2020 |
|---|---|---|---|
| ICT Total | 803 | 836 | 859 |

| Networks | 222 | 243 | 257 |
| --- | --- | --- | --- |
| Data centres | 220 | 225 | 230 |
| Enterprise networks | 25 | 25 | 25 |
| User devices | 343 | 349 | 353 |

**Findings**

Malmodin and colleagues argue that ICT's global emissions have broadly stabilised; they increased slightly from 720 $MtCO_2e$ in 2010 to only 733 $MtCO_2e$ in 2015 and decreased to 690 MtCO2e in 2020 (this includes user devices, networks and data centres). Relative to 2015, emissions from data centres have decreased the most (by 10%) followed by enterprise networks (9%) and networks (8%), even though electricity consumption increased in that time period due to the continued build out of 4G and 5G networks and an increasing number of data centre servers. The reason why the increased total energy use by ICT does not translate into higher carbon footprints is that it is partly offset by a higher share of renewable energy, a slight decrease in network overheads and that embodied carbon has stayed largely the same. The E&M sector (specifically, TVs, TV networks and other consumer electronics) add another 420 $MtCO_2e$ in 2015 [M&L] and 400 $MtCO_2e$ in 2020 (see Table B.**3**), showing a slight decrease too.

M&L further argue that ICT and E&M sector growth is starting to decouple from GHG emissions as ICT use is continuing to grow in terms of number of users and data traffic, albeit slower than previously as the world is moving towards saturation.[31] M&L argue that electricity consumption in data centre and networks does not follow the same exponential growth curve because of efficiency gains by servers. They show that the carbon footprint per subscription and per GB of data in networks has decreased fast since the 1990s and argue that while the number of networks users has increased by a factor of 10 and data traffic has increased 10,000 times between 1995 and 2015 if voice traffic is included, yet ICT's carbon footprint has only tripled during that time. They hold that ICT's footprint does increase with use but is better correlated with the number of users rather than data traffic. M&L acknowledge that Moore's Law has slowed down since 2012/13 but note that there usually a time lag until the effects are felt outside of research labs. In personal communication, Malmodin also noted that so far, efficiency improvements are continuing. The decoupling is also helped by enablement of emission savings in other industries, for example from print media as newspapers increasingly shift online, even though they admit that the effect still has to be seen in other sectors like transport.

These predictions stand in contrast to an earlier study [Malmodin et al. 2013] that projected ICT's footprint at 1.1 $GtCO_2e$ in 2020 plus an additional 1.3 $GtCO_2e$ for the E&M sector, including 680 $MtCO_2e$ for TVs and TV peripherals. This study argued that the increase in emissions would be driven mainly by an increase in the number of devices and therefore network subscriptions and data traffic, which will have been partly counterbalanced by energy efficiency improvements in networks, more efficient

---

[31] Malmodin [2019] argues that data traffic grew 70 times 1995-2000, 15 times 2000-05, 4 times 2010-15, 3 times 2015-20.

TVs, a shift from desktops to laptops and lower standby electricity consumption.

Explaining the discrepancy with earlier predictions, M&L argue that their earlier study was still based on older data which assumed the historical growth of PC and TV sales would continue, whereas more recent research takes into account the peak of PCs around 2011 and a slow decline of TV and tablet sales as well as better power management. The only PC type that is expected to increase is gaming PCs which have a higher carbon footprint. Consumer electronics sales (e.g. cameras, media players) are also declining thanks a move to smaller devices like laptops, tablets and smartphones and in particular the integration of functions into smartphones which *replace* older and less efficient user devices. There is also a shift from traditional storage devices (e.g. memory sticks) to cloud storage.

At the same time, they argue that M2M communication and IoT are only adding a very small footprint. In contrast to earlier studies by Malmodin [e.g. Malmodin et al. 2010, 2013], M&L's study included several IoT devices, including wearables, smart energy meters control units, surveillance cameras, public displays, payment terminals and the internet-connected communication device in vending machines. For 2015, they concluded that their impact on emissions was marginal. However, they did not include other now-common IoTs like smart speakers or any connected devices from other sectors, such as those embedded in vehicles, buildings or IoT used for military, medical, security and industrial purposes, other than those listed above, although they note that these are expected to add to GHG emissions in the future. The authors note that the number of IoT devices might explode and that M2M communication and therefore the number of network subscriptions are likely to increase rapidly in the future too. Already between 2010 and 2015, the number of M2M IoT subscriptions increased from 70 million in 2010 to 350 million in 2015.

A report by Ericsson [2019] presents a 2020 scenario which assumed a large increase in IoT and other new devices, with 500 billion sensors and tags, one billion connectivity boxes in ICT and 27 billion connectivity modules in other sectors, and found that their life cycle emissions were only minimal. The report argues that data traffic is driven more by video than by IoT. However, the methodology underlying this analysis is not provided.

Malmodin [2019] thinks that ICT can halve its emissions by 2030 relative to 2020 through renewable energy transformation and through collective effort. He expects data centres to be 1% of global electricity use even in the future, even though the absolute amount is expected to go up. In a recent Ericsson blogpost building on Malmodin's work, Lövehagen [2020] claims that ICT's carbon footprint could be reduced up to 80% if all electricity came from renewable energy. Importantly, she makes clear that ICT could be both a tool for decreasing or *increasing* global carbon emissions by accelerating carbon-intensive processes, depending on how it is used.

Another reason for stabilisation of emissions for Malmodin is his belief that there are limits to ICT's carbon footprint as smartphone markets saturate and there are a limited number of hours per day that users can use ICT equipment [personal communication]. Malmodin [personal communication] notes that the impact of AI and machine learning is so far very small even though it has been around for a while, and that emissions from AI are unlikely to explode unless training AI becomes more efficient.

**Limitations and criticism**

M&L and Malmodin's follow-up research present the most recent research on global ICT emissions reviewed here and the one with the widest scope in terms of ICT equipment, lifecycle stages and supply chain emissions considered.[32] Unlike Belkhir's and Andrae's studies, the estimates are based on more recent and a wider range of measured data directly from industry rather than older data reported in the academic literature. This is valuable as trends in ICT can change fast.

However, this data is also the biggest weakness, as it is not made public and cannot be scrutinised by the reader. Apart from one LCA for smartphones [Ercan et al. 2016], LCAs for device embodied and use phase carbon are not based on peer-reviewed LCA studies but on data by market research companies like IDC, IHS and Gartner whose reports are not available for free, even through university subscriptions. They are therefore not easily accessible for the reader either.

In personal communication, Malmodin noted that data centre emissions might be overestimated by up to 13 MtCO$_2$e for 2020 if the use of green electricity is taken into account, even though the investigation is still ongoing. For 2015, he notes that M&L should have used 0.63 kgCO$_2$e/kWh instead of 0.6 kgCO$_2$e/kWh. This would change the use phase carbon footprint upward by 5%.

Malmodin has been challenged about his opinion that ICT's carbon footprint has reached a peak and that energy growth is slowing despite data traffic increases, as he believes energy use is largely unrelated to data traffic, and his assumption that energy efficiency gains will continue [Belkhir and Andrae in personal communication]. This debate is examined in more detail in Sections 2.2.1 and 2.2.4.

## B.3  Drivers of change in ICT Future Emissions

A&E predict that over the next decade, user devices will become less important and networks and data centres will become more important for ICT total emissions. This is partly due to: a shift from desktops and laptops to phones; an increase in data traffic through trends like video streaming, cloud computing and emerging new data-heavy technologies; and a higher share of mobile data transmission because of the popularity of smartphones. In more recent papers [Andrae 2019a, 2019b, 2019c], Andrae also argues that trends such as AI and deep learning, IoT, Blockchain, virtual and augmented reality, facial recognition, and the rollout of 5G could lead to an explosion of data traffic over the next decade—increasing the share of data centres in ICT's emissions. While the trend towards smaller devices like smartphones is helping reduce ICT's emissions, we are also adding more devices like smart speakers and IoT.

---

[32] Unlike A&E and B&E, M&L consider all stages of ICT equipment lifecycle, that is material acquisition, parts and component production and assembly, transport and end of life. In addition to 'classic' ICT, they include the entertainment and media (E&M) sector, which includes TVs and other consumer electronics, such as cameras and audio systems in a car and portable GPS, e.g. for use in cars (see Table B.*2*). The study also includes operator activities and overheads, such as offices and business travel used by data centre and network operators, and enterprise networks, which are wireless and wired networks within business buildings that are operated by the company. The network emissions include the embodied carbon of infrastructure, like digging cable ducts and constructing antenna towers spread over its lifetime. ICT used by the financial system is also in scope, including computers, TVs, networks and servers, but not cryptocurrencies with the rationale that mining cryptocurrencies required *specific* hardware, not regular servers. This rationale has been challenged by Belkhir [personal communication] as unreasonable as mining computers and servers use GPUs, which are found in gaming and are therefore within the scope of ICT. AI is included indirectly by covering all data centre emissions with a top-down approach.

B&E point to the rising footprint of smartphones, with a share of 11% of total ICT emissions in 2020. As the number of users and the amount of data-heavy activities like video-streaming and social media on smartphones increases, this in turn contributes to an increase in mobile network and data centre use. Due to their short lifetime and increasing energy efficiency, the vast majority of smartphones' emissions are embodied. They predict that data centres' electricity use is going to rise but that emissions might stabilise if the trend of powering them with renewable energy continues. Network emissions are increasing slowly, and PCs' emissions are decreasing.

In contrast, M&L believe that AI and Machine Learning (ML) will not play a large role, unless training AI becomes more energy efficient because it would not be economical to run. Their assessment of IoT (even though with limited scope – see Section B.2.5) led them to conclude that the impact of IoT is, and will continue to be, minimal in the foreseeable future due to low data volumes; this is despite a possible explosion in the number of devices and network subscriptions. TV and PC emissions are decreasing due to better power management, lower standby power consumption and decreased sales. Their view is that other electronics are also declining – helped by a shift to smartphones which integrate functions such as video streaming, cameras, and portable media players, into one device. They highlight the large energy consumption of user access equipment that is on 24/7, such as modems, routers and set top boxes. They argue that the growth in data traffic is slowing down and that data centres and networks electricity consumption will not grow exponentially alongside the growth of data traffic because of efficiency gains and shifts to renewable energy.

## B.4  Reports out of scope of the review

There have been several reports in recent years on the topic of ICT's emissions, including on behalf of the Global eSustainability Initiative (GeSI), which represents ICT companies. In their report SMART 2020, produced by The Climate Group, they estimated ICT's emissions at 530 $MtCO_2e$ in 2002 and 830 $MtCO_2e$ in 2007 and projected ICT's footprint to rise to 1,430 $MtCO_2e$ in 2020 under business as usual. In a later report compiled by BCG, SMARTer 2020 [GeSI 2012], they estimated emissions in 2011 at 0.91 $GtCO_2e$ and revised their 2020 projection to 1.27 $GtCO_2e$. Their latest report, SMARTer 2030 [GeSI 2015], compiled by Accenture, extends their earlier projections to 2030 with an estimate of 1.25 $GtCO_2e$. The scope is summarised in Table B.**1**; for the 2015 report, it also includes 3D printers. In their reports, they also discuss 'abatement potential' by ICT in other industries whereby ICT could save 9.1 Gt $CO_2e$ in 2020 and 12.08 Gt $CO_2e$ in 2030; we explore these trends in more detail in Section 2.2.5. Another report claiming emission reductions of 2.1 GtCO2e enabled by mobile technology was released by GSMA [2019], which represents mobile operators.

Policy Connect, a London-based thinktank, produced a report *Is Staying Online Costing The Earth?* [McMahon 2018] sponsored by Sony in 2018 that concluded that energy consumption by ICT is not necessarily going to rise due to efficiency gains, renewable energy, a trend to smaller devices and ICT-enabled carbon savings in other industries.

The report *Lean ICT – Towards Digital Sobriety,* produced by The Shift Project [2019b], a Paris-based thinktank, came to a very different conclusion. They projected that ICT emits between 2.1 and 2.3 $GtCO_2e$ in 2020 and between 3.3 and 4.2 $GtCO_2e$ in 2025, including embodied and use phase carbon for PCs, phones, tablets, TVs, some IoT, networks and data centres. The modelling is based on Andrae and Edler's (2015) study but with updated assumptions, such as data traffic and the number of devices used. These estimates lie between Andrae and Edler's expected and worst case and therefore much higher than what the three main experts whose papers we reviewed above believe. The report points to several important trends, such as the impact of video streaming and short-lifespan devices, the underestimation of ICT's emissions by consumers because the underling infrastructure is invisible, and the unequal distribution of data consumption with high-income countries benefitting and thus emitting more than low-income countries.

These reports have not been included in the detailed review as they are not peer-reviewed. In addition, there are potential conflicts of interest where reports are sponsored by ICT companies [e.g. McMahon 2018; GeSI 2015; GSMA 2019]. Policy Connect's report largely relies on M&L's study, which is included in our detailed review, rather than offering original insights. For The Shift Project's [2019b] report*,* new modelling was done based on A&E's study but the findings have been discredited by all of the experts consulted (including Andrae, Belkhir, Malmodin and Preist). In the case of GeSI, the modelling behind the report is not transparent and assumptions are not made clear so it cannot be fully assessed.

# C   Video Streaming

Video streaming has become the dominant driver of data traffic consumption - forming 60% of downstream traffic and 22% of upstream traffic globally in 2018 [Sandvine 2019]. This traffic demand has been driven by adoption of video-on-demand services offered by companies such as Netflix, Amazon Prime and Disney; the popularity of YouTube and the embedding of video clips into other online services (e.g. social media such as Facebook and Twitter); and the use of video for security surveillance and video conferencing.

**If travel is fully replaced by video conferencing, video offers significant carbon savings.**
Online video can most prominently provide opportunities for reduction in travel-related carbon emissions. For example, video conferencing for co-locating a conference can create significant emission reductions from flights [Coroama et al. 2013], creating dematerialisation if the potential of this media is *"actively sought and unleashed"* [Coroama et al. 2015]. Video streaming has shown how useful it is during the Covid-19 outbreak, allowing entertainment during isolation as well as supporting home working. During the pandemic: replacing physical face-to-face meetings, for example, will reduce the travel-based emissions from business flights and peoples' commutes to work; we have also seen academic conferences moving online. However, as highlighted in Section 2.2.5, the rise in video traffic and availability of video conferencing has not yet led to a reduction in air travel [Graver et al. 2019], although this may change following the Covid-19 crisis.

**Video is accelerating data traffic.**
Video is clearly a prominent driver in data traffic which could significantly add to ICT's growth and emissions (Section 2.2.4). For example, higher streaming qualities such as High Definition (HD) and Ultra HD (UHD) can have a *"multiplier effect on traffic"*: 4K (UHD) doubles the bit rate of HD video and multiplies the bit rate of Standard Definition (SD) by nine [Cisco 2020]. Streaming qualities also affect device adoption, e.g. 66% of flat-panel TV sets are expected be UHD in 2023 (doubling the 33% share in 2018) [Cisco 2020], therefore impacting the embodied emissions of video-focused devices as users replace older TVs with newer models. In addition, faster infrastructure (e.g. 5G, fibre to the home) enables applications such as UHD cameras and VR streaming [Cisco 2020], multiple simultaneous streams within households [Widdicks et al. 2019], and now data-intensive gaming activities [Vaughan 2019] – driving the demand for video related network traffic and high performance streaming infrastructure such as content delivery networks and data centres further.

**Changes are required to stop continuous video and internet infrastructure growth.**
The Shift Project [2019a] estimated that 300 MtCO$_2$e was generated in 2018 due to online video and argue that this is comparable to annual emissions of Spain in 2010. These estimates have come under scrutiny [Kamiya 2020] due to arguments that they were based on old data, that energy impacts of the internet are much lower [cf. Shehabi et al. 2014] and that energy intensities of data transmissions are halving every two years [cf. Aslan et al. 2017] – following the 'Efficiency saves ICT' narrative (Section 2.2.2). These arguments also underpin some criticisms by TechUK [Fryer 2020] on a recent documentary BBC iPlayer [2020] *"Dirty Streaming"*, arguing the documentary provides misleading or incorrect information on ICT's environmental impacts.

The Shift Project may overestimate absolute emissions due to the direct processing of video traffic – especially as the energy per bit does improve over time, there is evidence that data traffic, including video, links more potently to growth in infrastructure and capacity. Preist et al. [2016] argue that growth in the internet's infrastructure capacity allows for new data-intensive services and applications (of which video is a part) – offering new affordances to users, in turn driving demand for these services and therefore further infrastructure growth. Peak data traffic is one driver for this infrastructure growth due to increased demand for data-intensive services; other influences include overprovisioning the infrastructure to ensure these services are always available to all users even at peak times [Preist, personal communication]. Growth begets more growth, unless we put a ceiling on absolute demand. In addition, Belkhir [personal communication] highlighted that the agreement between Netflix (a major video streaming service) and EU regulators to ease Netflix's load on the network during the Covid-19 pandemic [Sweney 2020] makes it difficult to argue data traffic is not interlinked with ICT infrastructure growth.

This is where changes in online service design may have a positive impact, e.g. turning off the video for a large portion of YouTube users who are only *listening* to the content [Lord et al. 2015, Widdicks et al. 2019] can have comparable emission reductions to running data centres on renewable energy [Preist et al. 2019] — but much more will need to be done to mitigate the significant growth of video streaming.

# D  Narratives

The assumptions about efficiency improvements and demand for ICT and predictions about ICT's impact on emissions in the scientific literature and non-scientific reports and the media can be summarised in the form of six common 'narratives', as detailed below. Note that Rebounds in ICT, Rebounds stalled and Global Rebounds are theoretical possibilities for which there is some evidence in the scientific literature (Section 1.1); however, they are not commonly discussed in the literature on ICT's emissions.

The first four narratives relate to ICT's own emissions, and the final two relate to ICT's impact on the rest of the global economy. The arguments underlying each narrative are underlined; we explore these in Section 2.2.

**Efficiency saves ICT**

Efficiency improvements are continuing; in combination with a shift towards more renewable energy, this will offset increases in ICT's energy use, stabilising ICT's emissions at the current level or even decreasing it in the future. Emissions are not so much influenced by the increasing data traffic but rather by the number of users, which will naturally level off soon as the world reaches saturation for personal ICT devices.
*E.g. research by Malmodin and colleagues, Masanet and colleagues*

**Growth without efficiency**

The growth in data traffic will lead to increases in network and data centre energy use, while the growth in IoT will lead to increases in embodied device emissions.
A) In combination with efficiency improvements slowing down, this will lead to an exponential growth in ICT's emissions.
B) Even if efficiency improvements continue, they will lead to further emission growth because of Jevon's paradox (Section 1.1), unless emissions are capped.
*E.g. research by Andrae and colleagues, Belkhir and colleagues*

**Rebounds in ICT**

The efficiency improvements enabled by ICT in other sectors lead to system growth within ICT. Under current conditions, rebound effects are greater than 100%. Therefore, the net effect of efficiency through ICT's is a rise in global emissions. If efficiencies continue, ICT's emissions will also increase unless they are deliberately constrained.

**Rebounds stalled**

If efficiency improvements stall (for example because Moore's Law reaches its quantum limit), this will lead to a plateau of emissions because growth requires efficiency gains.

**Enablement**

Because ICT enables carbon savings in other industries, the net effect of ICT is to lower global emissions despite growth in the ICT sector's own footprint.
*E.g. GeSI's SMARTer 2030, GSMA's The Enablement Factor report*

**Global Rebounds**

ICT enables efficiencies in other sectors which lead to growth in the wider economy. Rebound effects are larger than the efficiency gains (i.e. greater than 100%) and lead to an overall increase in global emissions.

# E  Truncation Error

There are two core methodologies for estimating the embodied carbon: the more commonly used Life Cycle Analysis (LCA) and Environmentally Extended Input Output (EEIO) analysis. LCA has potential for greater specificity as it is tailored to specific models, such as an iPhone 11, but inevitably incurs a truncation error; an underestimation arising from LCAs being unable to include the infinite number of supply chain pathways. To illustrate, a factory manufacturing computers will itself use computers to manage the production, a small share of whose embodied carbon needs to be attributed to the factory's output. Most of the literature assessing the embodied carbon in ICT is LCA-based.

EEIO offers a much more generic estimate, based on macro-economic modelling of financial and carbon flows between industrial sectors. It provides estimates of the total carbon emissions resulting from production of different types of goods per unit of monetary value. Whilst lacking specificity (i.e. all goods within broad categories, such as 'manufacture of office machinery and computers', have the same carbon footprint per dollar), EEIO-based estimates have the important advantage of taking account of emissions from all supply chain pathways; they do not incur truncation error.

To get some of the best of both approaches, it is possible to combine LCA and EEIO methodologies by approximating and adjusting for the truncation error incurred by LCAs. This can be done by mapping the LCA's system boundaries onto the EEIO model. Such a hybrid methodological approach stands to have both the specificity of LCA and the system-completeness of EEIO. In this report, we have drawn upon work carried out by SWC (see Appendix A.5.4) to derive adjustment factors for LCAs in different product categories and applied these to LCA-based embodied carbon assessments to derive system-complete estimates.

Based on SWC's EEIO model, we estimate that truncation error causes an omission of ca. 40% of the total embodied carbon and ca. 18% of the use phase carbon. When this is factored in, adjusting each study's specific electricity intensity figures, estimates for 2020 are on average 25% higher. Table E.**1** shows A&E, M&L and B&E LCA carbon estimates without adjustment of truncation error; Table E.*2* shows the adjusted estimates when truncation error is taken into account. This is just an approximation. We reiterate the caveat that SWC's EEIO model is based on UK data which is not representative of the world economy yet we have applied it to A&E's, M&L's and B&E's global estimates for a rough estimate of underestimation. We also note that these studies likely incur truncation errors of different sizes due to their differences in methodology. We have only adjusted for these differences with respect to electricity carbon intensity, but not for embodied emissions. Due to its more inclusive scope, M&L's is likely to have a smaller truncation error and A&E likely has a larger one than the average truncation error assumed here.

*Table E.1 Original estimates of embodied and use phase carbon for 2020. Malmodin's and A&E's estimates include TVs, B&E's estimates do not.*

| Study | Embodied (MtCO$_2$e) | Use phase (MtCO$_2$e) | Total (MtCO$_2$e) | MtCO$_2$e/TWh |
|---|---|---|---|---|
| Malmodin (2020) | 300 | 787 | 1087 | 0.60 |
| B&E - 2020 minimum | 213 | 894 | 1107 | 0.50 |
| B&E - 2020 maximum | 349 | 957 | 1306 | 0.50 |
| B&E - 2020 average (calculated) | 281 | 926 | 1207 | 0.50 |
| A&E - 2020 Best case | 157 | 730 | 887 | 0.59 |
| A&E - 2020 Expected case | 326 | 1534 | 1860 | 0.59 |
| A&E - 2020 Worst case | 1024 | 2610 | 3634 | 0.61 |

*Table E.2 Estimates for 2020 adjusted to include all supply chain pathways.*
*\*Included in the average. A&E's best case was chosen because Andrae (2020) reported that this is the most realistic for 2020. An average calculated for B&E's minimum and maximum estimates was included sinde B&E did not endorse either scenario and we wanted to avoid skewing the average by considering two of their estimates.*

|  | Embodied (MtCO$_2$e) | Use phase (MtCO$_2$e) | Total (MtCO$_2$e) | MtCO$_2$e/TWh |
|---|---|---|---|---|
| Malmodin (2020)* | 500 | 826 | 1326 | 0.63 |
| B&E - 2020 minimum | 355 | 1127 | 1482 | 0.63 |
| B&E - 2020 maximum | 582 | 1206 | 1788 | 0.63 |
| B&E - 2020 average (calculated)* | 469 | 1166 | 1635 | 0.63 |
| A&E - 2020 Best case* | 262 | 781 | 1043 | 0.63 |
| A&E - 2020 Expected case | 543 | 1628 | 2171 | 0.63 |
| A&E - 2020 Worst case | 1706 | 2704 | 4410 | 0.63 |
| **Average** | **410** | **925** | **1335** | **0.63** |

For ICT, TV and other consumer electronics, the adjusted total ranges between 1.2 and 2.2 GtCO$_2$ in 2020 (2.1-3.9% of global GHG emissions, see Table E.**3**) with 30% coming from embodied carbon and 70% from use phase on average.

*Table E.3* Estimates for 2020 adjusted to include all supply chain pathways and brought to the same scope.
*Included in the average.

| | Embodied (MtCO$_2$e) | Use Phase (MtCO$_2$e) | Total (MtCO$_2$e) |
|---|---|---|---|
| Malmodin (2020)* | 500 | 826 | 1326 |
| B&E's Minimum + Malmodin's E&M figure | 457 | 1482 | 1940 |
| B&E's Maximum + Malmodin's E&M figure | 684 | 1562 | 2246 |
| B&E's Average (calculated) + Malmodin's E&M figure* | 571 | 1522 | 2093 |
| A&E's Best without TV + Malmodin's E&M figure* | 296 | 898 | 1194 |
| **Average** | **456** | **1082** | **1538** |

This scope does not include some ICT equipment, such as radios, Blockchain and most IoT. Using EEIO, Livia Cabernard from ETH Zurich [Cabernard 2019, Cabernard et al. 2019] estimates that ICT's embodied carbon footprint (including manufacturing and transporting and covering a wider scope of ICT, specifically computers, mobile phones, TVs, radios, office machinery and all embedded ICT) was ca. 1.1 GtCO$_2$e in 2015. Of the 1.1 GtCO$_2$e embodied emissions, 27% were from computers, 55% from radio, TV and mobile and 18% from ICT embedded in other end products. Over the last few decades, a production shift to China (61% of ICT production in 2015, 45% in 1995) has increased the embodied carbon footprint of ICT because China's electricity use is mainly from coal [Cabernard in personal communication].

Modelling future population growth, efficiency improvements in relevant sectors, such as steel production, and a transition from coal to renewable energy in line with actions taken to limit global warming to 2°C (6°C), Cabernard [personal communication, based on Wiebe et al. 2018] predicts that ICT's embodied carbon footprint could be 1.38 (1.5) GtCO$_2$e in 2020, 1.27 (1.56) GtCO$_2$e in 2025 and 1.16 (1.64) GtCO$_2$e in 2030 for embodied emissions of ICT manufacturing. That means that the estimates for embodied carbon in Table E.**3** only cover a third of ICT's true embodied carbon footprint, because some ICT equipment, such as radio, Blockchain and embedded ICT, is left out of scope.

In summary, in most research on ICT's emissions, ICT's embodied carbon is considerably underestimated. While users are keeping their PCs and smartphones for slightly longer, the manufacturing footprint of smartphones is increasing because of more advanced integrated circuits, displays and cameras [Malmodin in personal communication]. With the number of IoT devices predicted to grow exponentially, the embodied footprint of ICT is likely to increase in the future.

# F European Commission's Investment in ICT

## F.1 Artificial Intelligence

The European Commission has ambitions to significantly increase uptake of AI as a way of *"strengthen*[ing] *the competitiveness of European industry"* [European Commission 2019d]. Specific initiatives and funding streams include Digital Innovation Hubs [European Commission 2018c] to *"provide support to SMEs to understand and adopt AI"* [NoCash 2020], and InvestEU (more on this below). The EC recently funded the 50 million Euro project "ELISE" which aims to *"make Europe competitive by setting up a 'Powerhouse of AI'"* [FCAI 2020]. New Common European data spaces are also being set up to enable greater data sharing [Kayali et al. 2020], which may be used by AIs. The Commission also views AI as a *"driving force to achieve the Sustainable Development Goals"* [European Commission 2019d], so is funding *"competitions and missions for AI solutions tackling specific environmental problems"* [European Commission 2019e]. Examples of how AI is expected to produce emissions reductions in other sectors include *"increasing the efficiency of farming, contributing to climate mitigation and adaptation,* [and] *improving the efficiency of production systems through predictive maintenance"* [European Commission 2019d].

## F.2 Internet of Things

The European Commission has launched the Alliance for Internet of Things Innovation to *"to support the creation of an innovative industry driven European Internet of Things ecosystem"* [European Commission 2019b]. Within the Digitising European Industry initiative, the Commission identifies three foci: 1) *"a thriving IoT ecosystem"*; 2) *"a human-centred IoT approach"*; and 3) *"a single market for IoT"*. The latter is facilitated through the *"European data economy"* initiative, which *"proposes policy and legal solutions concerning the free flow of data across national borders in the EU, and liability issues"* [European Commission 2019b]. The IoT is seen as playing a key role in the European response to climate crisis through providing an infrastructure to enable the 'smart future' (see, e.g. [ETIP SNET]), and for distributing energy consumption across smaller data centres [Gilmore 2018].

## F.3 Blockchain

The European Commission has established the European Blockchain Observatory and Forum to *"accelerate Blockchain innovation and the development of the Blockchain ecosystem within EU and* [to] *help cement Europe's position as a global leader in this transformative new technology"* [EU Blockchain 2020]. Along with AI, Blockchain is the target of the 2 billion Euro joint European Commission and European Investment Fund InvestEU Programme. In addition, the European Blockchain Partnership [European Commission 2018b] commits all member states to *"realising the potential of Blockchain-based services for the benefit of citizens, society and economy"* [European Commission 2020b]. As one of its potential benefits, Blockchain is viewed by the Commission as the kind of 'disruptive technology' required by the climate crisis [European Commission 2019c], and the Commission has outlined five core areas they seek to 'unleash' Blockchain technology in service of climate. Based on the idea that Blockchain can *"improve the transparency, accountability and traceability"* of GHGs, these application areas include areas of clean power, smart transport systems, sustainable production and consumption, sustainable land use, and smart cities and homes.

# G  Carbon Pledges

Company pledges vary in particular with regards to scope of emissions covered (see Table G.1), offsets used and how much emissions are reduced. Note that there are also some companies with plans to cut emissions by a certain percentage but without the sweeping ambitions of the below pledges.

*Table G.1 Emissions covered by Scope 1, 2 and 3.*

| Scope 1 emissions | Direct emissions from burning of fossil fuels on site (includes company facilities and vehicles). |
|---|---|
| Scope 2 emissions | Indirect emissions from purchased electricity and gas. |
| Scope 3 emissions | All other indirect emissions in a company's value chain including upstream emissions in the supply chain (e.g. emissions from purchased goods and services, transportation of these goods to the company, use of leased assets such as offices or data centres, business travel and employee commuting) and downstream emissions (from transportation, distribution, use and end of life treatment of sold products, investments and leased assets). Scope 3 emissions form the majority of a company's emissions. |

## G.1  Carbon Neutral

Carbon neutral means no net release of carbon-dioxide emissions into the atmosphere, through using offsets to counterbalance any emissions generated.[33] For a company to be carbon neutral, they must measure, reduce, and offset their emissions.

There is no set standard on which (if any) scope 3 emissions should be included, how these emissions should be reduced (purchasing credits or on-site renewables, for example), and which offsets are robust and credible.

*Companies pledging carbon neutral: Sky (since 2006), Google (since 2007), Microsoft (since 2012), and Apple (by 2030).*

## G.2  Net Zero

Net zero is defined by the Science Based Targets initiative (SBTi) as a company having no net climate change impact through GHG emissions in a company's value chain. This is achieved by reducing GHG emissions in the value chain and removing the remaining emissions through additional carbon removals. Companies setting a net zero pledge can register with the SBTi if their targets Net zero means no net climate change impact through GHG emissions in a company's value chain, achieved by reducing GHG emissions in the value chain and removing the remaining emissions through additional carbon removals. A net zero target is more ambitious than a carbon neutral target because it specifies that scope 3 emissions need to be included, while for carbon neutral, generally only emissions from business travel and waste are included but not other scope 3 emissions.

---

[33] Note that most ICT companies refer to carbon dioxide and other GHG when they use the term 'carbon neutral'. This is sometimes called 'climate neutral'.

Organisations can register their net zero targest with the Science Based Target initiative (SBTi) which certifies organisations whose emission targets are in line with the IPCC's recommendations that global warming be limited to 2°C, well below 2°C or more recently to 1.5°C [IPCC 2018, Pineda and Faria 2019]. According to the SBTi, for company pledges validated after July 8th 2020, only two thirds of scope 3 emissions need to be accounted for and scope 3 emissions targets 'generally need not be science-based'. A target needs to be set for reducing scope 3 emissions only if they make up over 40% of a company's total emissions [SBTi 2019]. However, the Carbon Trust believes that net zero should apply to 100% of scope 3 emissions [Stephens 2019].

Net zero is also more ambitious than carbon neutral regarding offsets. SBTi specifies that 'offsets must not be counted as emissions reduction toward the progress of companies' science-based targets' [SBTi 2020]. In the way Sky and Microsoft define net zero, any emissions that cannot be reduced need to be removed from the atmosphere rather than being avoided, whereas carbon neutral can use both emission removal offsets and avoided emission offsets [Sky Zero 2020, Smith 2020].

*Companies pledging net zero: Microsoft (by 2030), Sky (by 2030), Amazon (by 2040), BT (by 2045) and Sony (by 2050).*

### G.3   Carbon Negative

Carbon negative (also sometimes described as 'carbon positive' or 'climate positive') is where an activity removes more emissions than it emits across an entire value chain. This goes further than both carbon neutral and net zero. There is currently no official definition or standards for this apart from the ones used by Microsoft for their carbon negative pledge [Smith 2020].

*Companies pledging carbon negative: Microsoft (by 2050).*

### G.4   100% Renewable

100% renewable means that all of a company's power consumptions comes from renewable sources, such as solar, wind and hydro. It does not specify whether and how much the company's emissions should be cut. Unfortunately, often companies do not provide detail on whether they will generate additional renewable energy on-site, or buy renewable energy certificates, including unbundled REGOs which should not be seen as truly 100% renewable, in the authors' opinion (see Appendix I).

*Companies pledging 100% renewable: Netflix (since 2018), Google (2017), Facebook (by 2020), Samsung (by 2020), Microsoft (by 2025), Sky (by 2020), Vodafone (by 2025), Apple (since 2018).*

*Please note: the achievement of these pledges is self-reported and not externally validated.*

## G.5 Why scope matters

Carbon neutral pledges are not enough for the world to limit global warming to 1.5°C because they do not require the company to account for scope 3 emissions, which form the majority of ICT's carbon footprint. Voluntary emissions reductions of company Scope 1 and 2 emissions alone could theoretically be sufficient if every company in the economy played its part. However, without any kind of enforcement or reputational consequences, companies have a competitive advantage if they do not set of meet targets. In addition, companies could lower their scope 1 and 2 emissions by outsourcing carbon-heavy activities to suppliers, which would not decrease overall emissions. By signing up for scope 3 targets, companies take responsibility for their supply chain. They then have an incentive to encourage their suppliers to cut their emissions too, creating a snowballing effect in the economy.

# H Renewable Energy Purchases

## H.1 On-site Generation

On-site generation of renewable energy is the best option for purchasing renewable energy. This is because, with on-site generation, the company carries the set-up costs of the renewable energy project and there are fewer transmission and distribution losses than when electricity is sourced from the grid. It also ensures that additional renewable energy is created which is needed for a successful global transition away from fossil fuels.

## H.2 PPA with Bundled REGOs

Companies can also buy a Power Purchasing Agreement (PPA) for renewable energy. PPAs do not cover the set-up costs of a renewable energy project (unlike on-site generation) but they pay for the cost of the power generation and receive a bundled Renewable Energy Guarantees of Origin (REGOs) that certifies that each 1 MWh of electricity purchased comes from a renewable energy project. REGOs ensure for each unit of renewable energy generated, only one company can claim the environmental benefits, avoiding double counting. PPAs ensure additionality because a company purchasing a PPA pays for additional renewable energy to be generated and fed into the electricity grid. They can also encourage the setup of new renewable energy projects because projects can attract funding from investors more easily with guaranteed buyers of PPAs lined up.

## H.3 Unbundled REGO

Companies also have the option of buying an unbundled REGO without the actual energy to lower their scope 2 emissions. In this case, the environmental benefit of renewable energy gets separated from the energy itself. The company generating renewable energy can sell off the environmental benefit represented by the REGO but without selling the actual energy. The company buying the unbundled REGO can claim lower scope 2 emissions. Unfortunately, unbundled REGOs do not encourage greater power generation from renewable sources because the demand for REGOs is currently vastly outstripped by the supply thus making them very cheap. This means they are an ineffective instrument for investment into renewable projects [Scott 2019] and cannot claim additionality. Because unbundled REGOs cannot claim additionality and the company buying it lays a sole claim to the "greenness", thereby not sharing it with other electricity users, unbundled REGOs raise the carbon footprint of the electricity grid [Hewlett 2017]. The separation also makes it harder to track what renewable energy projects lie behind each certificate. A company wishing to reduce their Scope 2 emissions and become powered by 100% renewables should therefore look towards investing directly into renewable projects and PPAs in which bundled REGOs are purchased, crucially, with the underlying power.